\documentclass[12pt]{article}
\usepackage{relsize}
\usepackage[english]{babel}
\usepackage[utf8]{inputenc}
\usepackage[colorinlistoftodos]{todonotes}
\usepackage{amsthm}
\usepackage{mathtools, nccmath}
\usepackage[margin=1.25in]{geometry}
\usepackage{amsmath}
\usepackage{booktabs}
\usepackage{amssymb}
\usepackage{array}
\usepackage{graphicx}
\usepackage{listings}
\usepackage{tikz}
\usetikzlibrary{arrows.meta,chains,matrix}
\usepackage{enumerate}
\usepackage{enumitem}
\usepackage{bbm}
\usepackage{multirow}

\usepackage[bottom]{footmisc} % Reduce indent in footnotes

\makeatletter
\newcommand\customfootnotesize{%
  \@setfontsize\customfootnotesize{9.5}{10}%
}
\makeatother

\usepackage[T1]{fontenc}
\usepackage{lipsum}
\usepackage{multicol}
\usepackage{changepage}
\usepackage{caption,pdflscape,subfigure,float}
\usepackage[12pt]{moresize}
\usepackage{natbib}
\usepackage{caption}
\usepackage[hidelinks]{hyperref}
\usepackage{xcolor}
\hypersetup{ % remove the box around the link and make the link blue instead 
	colorlinks,
	linkcolor={black},
	citecolor={black},
	urlcolor={red!80!black}
}
\usepackage[]{titlesec}
\newtheorem{theorem}{Theorem}[section]
\newtheorem{lemma}[theorem]{Lemma}
\newtheorem{proposition}[theorem]{Proposition}
\newtheorem{corollary}[theorem]{Corollary}
\newtheorem{assumption}{Assumption}

\newtheorem{definition}{Definition}
\newtheorem{remark}{Remark}
\newtheorem{example}{Example}
\newtheorem{condition}{Condition}
\newtheorem{examplealt}{Example}[example]
\newenvironment{examplecont}[1]{
  \renewcommand\theexamplealt{#1}
  \examplealt
}{\endexamplealt}

\newtheorem{theoremalt}{Theorem}[theorem]
\newenvironment{theoremp}[1]{
  \renewcommand\thetheoremalt{#1}
  \theoremalt
}{\endtheoremalt}
\newtheorem{definitionalt}{Definition}[definition]
\newenvironment{definitionp}[1]{
  \renewcommand\thedefinitionalt{#1}
  \definitionalt
}{\enddefinitionalt}
\usepackage{setspace}
\usepackage{etoolbox}
\makeatletter

\usepackage{etoolbox}
\makeatletter
\newenvironment{mycenter}[1][\topsep]
  {\setlength{\@tempdima}{#1}%
   \begin{list}{}{%
     \setlength{\leftmargin}{\z@}%
     \setlength{\rightmargin}{\z@}%
     \setlength{\itemsep}{\z@}%
     \setlength{\parsep}{\z@}%
     \setlength{\topsep}{\@tempdima}%
     \setlength{\partopsep}{\z@}}%
      \item\relax\centering\fontsize{13.5}{15.5}\selectfont}
  {\end{list}}
\makeatother

\date{\today}
\begin{document}

	\onehalfspacing

\begin{titlepage}
\title{Comparative Statics for Optimal Stopping Problems in Nonstationary Environments\thanks{\scriptsize We thank Dan Barron, Benjamin Bernard, Jeff Ely, Drew Fudenberg, Alessandro Pavan, Mike Powell, Luis Rayo, Eran Shmaya, Bruno Strulovici, St\'{e}phane Villeneuve, and Asher Wolinsky for many detailed comments and suggestions. This project has also benefited from conversations with Arjada Bardhi, Job Boerma, Svetlana Boyarchenko, Henrique Castro-Pires, Isaias Chaves, Harry Pei, Lones Smith, Mete Soner, Boli Xu, and the comments of seminar participants at Northwestern and at the ``Applications of Stochastic Control to Finance and Economics'' workshop at BIRS. All mistakes are our own.}}
\author{Matteo Camboni\thanks{\scriptsize Economics Department, University of Wisconsin-Madison, Madison, WI 53706, camboni@wisc.edu} \: \: \: Th\'{e}o Durandard\thanks{\scriptsize Economics Department, University of Illinois Urbana-Champaign, Urbana, IL 61801,  theod@illinois.edu}}
\date{\today%\\ \vspace{0.7em}
% \href{https://www.dropbox.com/s/d99ef9lga43rgyb/Monitoring_Team_Members.pdf?dl=0}{\Large \textcolor{blue}{\underline{Click here for the latest version}}} 
}
\maketitle

\begin{abstract}
   How do decisions change with the economic environment and with time? This paper studies general nonstationary stopping problems and provides the methodological tools to answer these questions. First, we identify conditions that ensure a monotone relation between decisions' timing and outcomes. These conditions apply to a prevalent class of economic environments. Second, we develop a theory of monotone comparative statics for stopping problems, offering general and unifying qualitative insights into the decision-maker's value and stopping behavior. We apply our results to models of information acquisition, bankruptcy, irreversible investment, and option pricing to explain documented patterns at odds with current theories.

\end{abstract}

	\setcounter{page}{0}
\thispagestyle{empty}
\end{titlepage}
 	
    Optimal stopping plays a crucial role in modeling various economic phenomena. Examples include the pricing of real options,\footnote{E.g., \cite{jacka1991optimal,villeneuve1999exercise,shiryaev1999essentials,chen2007mathematical}.} information acquisition,\footnote{E.g., \cite{wald2004sequential,shiryaev1967two,fudenberg2018speed,che2019optimal,zhong2022optimal}.} R\&D and investment timing,\footnote{E.g., \cite{dixit1993art,dixit1994investment,decamps2006irreversible,hopenhayn2011preemption,bobtcheff2017researcher,ebert2020weighted}.} labor search and negotiations,\footnote{E.g., \cite{jovanovic1979firm,jovanovic1984matching,mccall1990occupational,mcclellan2023dynamic}.} experimentation and bandits,\footnote{E.g., \cite{el1994dynamic,bolton1999strategic,moscarini2001optimal,keller2005strategic,mcclellan2022experimentation}.} the timing of elections, scandals, and the campaigns' behavior in political economy,\footnote{E.g., \cite{keppo2008optimal,gratton2018drop,shahanaghi2021reputation}.} and the choice of capital structure with endogenous bankruptcy,\footnote{E.g., \cite{leland1994corporate,manso2010performance}.} and many more. Yet, the literature has largely focused on stationary decision environments. This focus stems from technical constraints. Abandoning stationarity increases the problem's dimensionality and precludes closed-form solutions. In many applications, however, the decision problem changes with time. Key primitives, such as market price volatility, interest and discount rates, and information arrival and processing rates, often vary as the decision-maker (DM) considers her options. Moreover, even when the information flow is steady, learning speed slows as information accumulates. DMs also often face deadlines or fading opportunities.

    In this paper, we tackle some of the technical challenges associated with nonstationary optimal stopping problems to obtain qualitative insights into the DM's behavior. Our contribution is twofold. (i) We characterize the relationship between decision timing and outcomes in general nonstationary stopping problems. (ii) We develop a theory of monotone comparative statics for nonstationary stopping problems. Finally, we revisit several canonical economic models, showing how our results extend well-known results but also shed light on documented decision patterns at odds with the classical stationary insights.
    \begin{mycenter}[5pt]
        \textit{Stopping and Comparative Statics: General Results}
    \end{mycenter}
    We consider a general nonstationary optimal stopping problem where a DM must choose among alternatives whose appeal evolves according to a diffusion process. Unlike previous literature, our formulation allows the problem primitives to depend on both the current state and calendar time.

% Our first main contribution establishes a link between the value function's time dependence to the evolution of the optimal continuation region (the set of states where the DM opts not to stop) over time. Specifically, we prove that the continuation region (strictly) shrinks over time when the value function (strictly) decreases with time and (strictly) expands when the value function (strictly) increases with time. Proving this (strict) dynamic is particularly complex: the DM's option value depends on the sign of the value function's cross derivative (in time and state), which, unlike the first derivatives, it is usually difficult to ascertain from the problem's primitives. However, we can exploit a classic result from the theory of partial differential equations, the Hopf boundary lemma, to sign the cross derivatives of the value function exactly where needed: at the boundaries of the continuation region. 
The first question in nonstationary stopping problems is how the DM's behavior changes over time. Our first set of results relates the decisions' timing and outcomes by characterizing the evolution of the continuation region (the set of states where the DM opts not to stop). Specifically, we identify conditions under which the DM's value for waiting (strictly) increases or (strictly) decreases over time, implying that the continuation region (strictly) shrinks or (strictly) expands over time (Theorem \ref{theorem:localmonotonestopping} and Corollary \ref{corollary:globalmonotonestopping}). While signing the time-derivative of the value function is sufficient to prove the weak monotonicity of the continuation region, proving \emph{strict} monotonicity is harder. The DM's option value depends on the value function's cross derivative (in time and state), which, unlike the first derivatives, is difficult to derive from the primitives. Fortunately, we can exploit the connection between the value function and the Hamilton-Jacobi-Bellman (HJB) equation (which we establish formally) to prove strict monotonicity. In particular, we use a classic result in PDE theory, the Hopf boundary lemma, to sign the cross derivatives of the value function exactly where we need it: at the boundaries of the continuation region. %Our results thus relate the decisions' timing and outcomes.
% prove that the continuation region (strictly) shrinks over time when the value function (strictly) decreases with time and (strictly) expands when the value function (strictly) increases with time. Proving the weak monotonicity is straightforward.
% This result allows us to link a decisions timing and outcomes in settings 
%     In this setting, we identify conditions that guarantee that the problem's time dependence is monotone (hereafter, \textit{monotone environments}) \textbf{Comment on monotone environments?}. We prove that the continuation region (the set of states at which stopping is suboptimal) strictly shrinks over time in monotone-decreasing environments and expands in monotone-increasing environments. The weak monotonicity of the continuation region is an easy consequence of the monotonicty of the value function. On the other hand, proving the strict monotonicty is more complex. The DM's option value at the boundary depends on the value function's cross derivative (in time and state), which, unlike the first derivatives of the value function, is difficult to derive from the primitives. However, we can exploit a classic result from the theory of partial differential equations, the Hopf boundary lemma, to sign the cross derivatives of the value function at the boundaries of the continuation region and prove the strict monotonicity. Our first set of results thus provide tools to link a decision's timing and outcome (Theorem \ref{theorem:localmonotonestopping} and Corollary \ref{corollary:globalmonotonestopping}).

    The second natural question is that of comparative statics: How do the value functions, continuation regions, and optimal stopping times compare across decision problems? Our second set of results develops a theory of monotone comparative statics for general stopping problems and describes the impact of changes in the problem's primitives on the DM's behavior. Theorem \ref{theorem:compstatsstoppingtimes} shows that the DM waits longer when she is more patient (as already shown in \citet{quah2013discounting}) and when her waiting cost is lower. 
    %\textcolor{red}{This insight generalizes the well-known option value principle: When the value function is convex, the ``option value of waiting'' leads the agent to wait beyond the myopic optimum.}
% Theorem \ref{theorem:compstatsstoppingtimes}    confirms \cite{quah2013discounting}'s result that a more patient DM waits longer and shows the same is true for a DM whose cost of waiting is smaller. 
    Theorem \ref{theorem:compstatsvalues} establishes that the continuation region expands with an increase (decrease) in the drift when the DM's payoff is nondecreasing (nonincreasing) in the state variable. Similarly, when the primitives are convex, the continuation region expands (shrinks) with an increase (decrease) in volatility. So, this result generalizes the well-known option value principle, which states that when the value function is convex, the ``option value of waiting'' leads the agent to wait beyond the myopic optimum. Our proofs build, again, on the connection between the value function and the HJB equation. One \emph{methodological} contribution is to show that ``comparison principles'' (i.e., theorems that rank the solutions of differential equations pointwise) are well-suited to obtain comparative statics for stopping problems.

    \begin{mycenter}[5pt]
    \textit{Applications to Specific Models}
    \end{mycenter}
\textbf{1)} In classical information acquisition problems, a DM evaluates two alternatives while learning about a parameter relevant to her choice. Following \citet{wald2004sequential}, most articles assume the parameter is binary, and learning is stationary. As a result, the DM optimally stops when her beliefs hit a constant boundary. So, decisions' timing and accuracy are uncorrelated. However, recent studies in psychology and neuroscience document positive, negative, or even non-monotonic correlations depending on the context.\footnote{E.g., \cite{drugowitsch2012cost,goldhammer2014time,goldhammer2015more, bolsinova2017modelling,molenaar2018semi,chen2018curvilinear,fudenberg2020testing}.} While these articles assume constant boundaries and focus on parameter estimation, we take a normative approach. We show that the documented time-accuracy dependencies can stem from DMs' rational responses to different information acquisition settings. Our results establish that decision accuracy and timing are positively (negatively) correlated when the discount rate decreases (increases), information cost decreases (increases), or learning speed increases (decreases) over time. Notably, negative correlation also arises when the DM holds any non-binary prior. Additionally, our comparative statics indicate that a DM always acts more slowly when she is more patient or information costs are lower.

\textbf{2)} In many information acquisition problems, the DM faces abrupt events. For instance, information may arrive gradually and abruptly, and the DM may face a (stochastic) deadline or risk missing an opportunity if not seized in time (e.g., a competitor may preempt her). Yet, these settings remain largely unexplored due to their technical challenges.\footnote{The closest reference is from neuroscience: \citet{frazier2007sequential} studies a discrete-time sequential sampling problem with stochastic deadlines, where the arrival rate increases over time.} Our results reveal that slower decisions are less (more) accurate when the pressure from a (stochastic) deadline increases (decreases) or when the arrival rate of abrupt information decreases (increases) over time. %\footnote{Here, we assumed the DM must act at the deadline. Generally, when the deadline is fixed or has an increasing arrival rate, decision timing and accuracy are negatively (positively) correlated if the deadline payoff is sufficiently low (high).} 
Our approach is novel. It bypasses the need to solve complex integro-differential HJB equations associated with jump-diffusion models by focusing on the periods between jumps.
% From a more applied perspective, this result enables us to address, for example, the speed-accuracy tradeoff in decision problems where a DM can miss an opportunity if she does not seize it in time (e.g., a competitor may preempt the DM by hiring the worker or buying the asset the DM was considering) or when learning processes may change abruptly at specific points in time (e.g., a company's earning release, the publication of a scandal, or revelation of a merger). Similarly, our results can help us understand the impact of increasing news coverage on an asset's selling price and price fluctuations induced by the shadow of a crisis.
% Most models then relied on the existence of a closed form solution to derive insights. The setup, therefore, needs to be simple enough.

%the optimal cut-off may no longer be stationary. For example

    \textbf{3)} Recent models of corporate bankruptcy, starting with Leland (1994), view the timing of bankruptcy as endogenously chosen. However, to derive insights, most articles have relied on closed-form solutions. This significantly constrained the models' setups. Recently, \cite{manso2010performance} argued that more general environments are needed to capture features consistent with empirical evidence. Our results indicate that the optimal stopping is still a simple cut-off strategy even in more complex nonstationary environments. We also show that, irrespective of time stationarity, an increase in cash flows or the firm's growth rate delays bankruptcy, while an increase in the discount rate hastens it. Finally, the level of profitability triggering default increases over time if the firm's growth rate slows over time.

    % \textbf{4)} In irreversible investment problems, a DM chooses when (if at all) to make a costly investment whose profitability typically follows a stationary geometric Brownian motion  \citep{dixit1993art}. Profitability and investment timing are, therefore, uncorrelated. However, when profitability follows a more general process, our results establish that investment timing and profitability are typically correlated. For instance, later investments are less profitable when the drift or volatility decrease over time (e.g., as competition increases or as the market stabilizes after a crisis) and when the investment cost decreases over time (e.g., during expansionary phases of monetary policy). Moreover, we show that, conditional on investing, profitability is higher when the drift or volatility are higher, the investment cost is higher, and the discount rate is lower.

    \textbf{4)} In irreversible investment problems, a DM chooses if and when to make a costly investment whose profitability follows a geometric Brownian motion (e.g., \cite{dixit1993art}). A classical prediction is that profitability and investment timing are uncorrelated. However, this is specific to stationary environments. Our results establish that later investments are less profitable when the drift or volatility decreases over time (e.g., as competition increases or the market stabilizes) and when the investment cost decreases over time (e.g., during expansionary phases of monetary policy). Moreover, we show that conditional on investing, profitability is higher when the drift or volatility is higher, the investment cost is higher, and the discount rate is lower.
    
    \textbf{5)} In finance, the price of an American put option is given by the value function of a time-stationary optimal stopping problem, where a DM chooses when to exercise the option. The standard model assumes the underlying asset price follows a geometric Brownian motion \citep{shiryaev1999essentials}, and predicts that (i) options are exercised when the asset price hits a constant lower bound, and (ii) options are exercised at higher asset prices when their strike price is higher, the market interest rate is higher, or the asset price volatility is lower. However, these stationary models have been criticized empirically for the constant volatility assumption \citep{bakshi1997empirical, bates2022empirical}, and the lack of correlation between execution times and asset prices at exercise \citep{carpenter1998exercise}. Our results confirm (ii), while addressing the empirical criticism. They predict a positive (negative) correlation between execution times and asset prices at exercise when interest rates or volatility increase (decrease) over time.

    \begin{mycenter}[5pt]
    \textit{Related literature}
    \end{mycenter}

Stopping problems have a long history in economics, finance, and mathematics, with a literature too vast to fairly summarize. \cite{dixit1993art} provides an excellent introduction, while \cite{peskir2006optimal} and \cite{shiryaev2007optimal} offer more technical treatments. Most articles in economics and finance focus on stationary environments.\footnote{Even in \citet{moscarini2001optimal}, \citet{che2019optimal}, and \citet{zhong2022optimal}, the endogenous optimal learning policy is stationary since the environment is.} Two notable exceptions are \cite{georgiadisharris2024} and \cite{fudenberg2018speed}, which, like us, emphasize the importance of time. While we study \emph{general} problems to obtain qualitative insights across settings, they focus on specific applications. The former derives the optimal information acquisition policy when a DM has to act at some unknown time. The latter studies an information acquisition problem where a DM holds a Gaussian prior over the value of taking one action over the other. The authors show that later decisions are less accurate because (i) learning slows down as information accumulates, and (ii) the DM expects a smaller payoff difference, and thus a lower benefit from accuracy, as time progresses. Section \ref{subsec:learningnonbinary} complements their analysis. We show this negative correlation persists even in a more conservative model (that shuts down (ii)) and extends to \textbf{any} {non-binary} prior. We also show that when the problem's primitives are nonstationary, decision accuracy may, in fact, increase over time, rationalizing other decision patterns documented in psychology \citep{ratcliff2016diffusion} that are at odds with most current theories.%common patterns in both perceptual and cognitive testing.

% Instead, we offer a systematic treatment of \emph{general} nonstationary stopping problems to obtain qualitative insights across many settings. In particular, 

Moreover, our theory of monotone comparative statics is, to our knowledge, novel. 
The only other paper deriving comparative statics for stopping problems, \cite{quah2013discounting}, focuses on the role of patience. Our paper complements their findings by providing comparative statics for other features of the environment.

Meanwhile, the mathematical literature typically has a different focus and rarely study comparative statics for stopping problems. Most papers are concerned with the properties and regularity of the value function \citep{bergman1996general,ekstrom2004convexity,durandard2023existence}, or the regularity of the boundaries \citep{kotlow1973free,de2015note,de2019lipschitz,de2022stopping,audrito2023regularity}. Still, some papers have studied the shape of the continuation region for specific stopping problems, often in the context of American put options \citep{jacka1991optimal,sd1992finite,villeneuve1999exercise,chen2007mathematical,milazzo2023monotonicity}, or as a first step in proving the regularity of the boundaries \citep{friedman1975parabolic,friedman2008partial,de2022stopping}. They usually do not, however, establish, as we do, the \emph{strict} monotonicity of the continuation region.\footnote{\cite{friedman1975parabolic}, \cite{chen2007mathematical}, and \cite{de2022stopping} are exceptions.} So, our results complement the mathematical literature by (i) identifying a general class of problems in which the stopping region is easy to depict and (ii) providing comparative statics. 

    \begin{mycenter}[5pt]
    \textit{Organization of the paper}
    \end{mycenter}
Section \ref{sec:setting} introduces our environment and the general nonstationary stopping problem we study. Section \ref{sec:analysis} presents our two sets of results: the evolution of the DM's behavior in a given problem (Section \ref{subsec:monotoneenvironments}) and our monotone comparative statics theory (Section \ref{subsec:CSEE}). In Section \ref{sec:applications}, we apply our results to several canonical economic problems. Section \ref{sec:extensions} gives two extensions. All the proofs are in the Appendix.

 \section{Setting}\label{sec:setting}
%For simplicity, we assume that (i) the endpoints $\underline{x}$ and $\bar{x}$ are either never attained or absorbing and that, (ii) if $r=0$, $\mathcal{X}$ is bounded. 
% FOOTNOTE? To this end, let $\left( \Omega, \left\{\mathcal{F}_t\right\}_{t\geq 0}, \mathcal{F}, \mathbb{P}\right)$ be a filtered probability space whose filtration $\left\{\mathcal{F}_t\right\}_{t\geq0}$ satisfies the usual conditions.\footnote{See for example \cite{protter2005stochastic}.} 
We study an optimal stopping problem in continuous time defined on the filtered probability space $\left( \Omega, \left\{\mathcal{F}_t\right\}_{t\geq 0}, \mathcal{F}, \mathbb{P}\right)$ which filtration satisfies the usual conditions (see, e.g., \cite{protter2005stochastic}). A risk-neutral decision maker (DM), with time horizon $T\in (0,\infty]$, sequentially chooses whether or not to stop a one-dimensional nonstationary diffusion process $X = \left\{ X_t \right\}_{t\geq 0}$ defined on the open (possibly unbounded) interval $\mathcal{X} = \left( \underline{x}, \bar{x} \right) \subset \mathbb{R}$. As standard, we denote by $\partial \mathcal{X}$ its boundary and by $\Bar{\mathcal{X}}$ its closure. %Finally, define $\mathcal{Y} _T\equiv [0,T)\times \mathcal{X}$.
 %; we denote by $g(X_t)$ the maximum between the two. %by $\mathcal{T}$ the set of $\mathcal{F}$-stopping times taking values in $[0, T]$,
%CCCCCCCCCCC To simplify the exposition, we assume that ${g^i}: \bar{\mathcal{X}} \to \mathbb{R}$ belongs to $\mathcal{C}^{2, \alpha} (\bar{\mathcal{X}})$, the space of twice continuously differentiable functions on $\bar{\mathcal{X}}$ with $\alpha$-H\"{o}lder continuous derivatives, and that $g^a - g^b$ has at most one zero at $x^c \in \mathcal{X}$.
% obtains state-dependent final payoff of $g(X_{t})$, $g$ can be represented as the maximum of two smooth functions: $g = g^1\vee g^2$, with
% 		\begin{itemize}
% 			\item ${g^i}: \bar{\mathcal{X}} \to \mathbb{R}$, $i=1,2$, belongs to $\mathcal{C}^{2, \alpha}(\bar{\mathcal{X}})$.
% 			\item $g^1 - g^2$ has at most one zero at $x^c \in \mathcal{X}$.
% 		\end{itemize}
% , or let the process continue, obtaining a flow payoff $f(t,X_{t})$, and the possibility of a future final payoff $g(X_{t+s})$. 
% At every time $t$ and state $X_t$, the DM chooses whether or not to stop the diffusion process. If she stops, she obtains the final payoff of $g(X_{t})$, or let the process continue, obtaining a flow payoff $f(t,X_{t})$, and the possibility of a future final payoff $g(X_{t+s})$. 
The diffusion process evolves according to
\begin{align}\label{eq:diffusion} 
    X_t = X_0 + \int_{0}^{t} \mu(s, X_s) ds + \int_{0}^{t} \sigma(s, X_s) \, \mathrm{d}B_s, \, \quad \mathbb{P}\text{-a.s.},
\end{align}
where $B$ is the standard one-dimensional Brownian motion. The drift $\mu: [0,T) \times \mathcal{X} \to \mathbb{R}$ and the standard deviation $\sigma : [0,T) \times \mathcal{X} \to \mathbb{R}_{++}$ are allowed to vary with time $t$ and the current state of the process $X_t$. % \footnote{The literature typically assumes uniform ellipticity: $\exists \lambda>0$ such that $\forall (t,x)$, $\sigma(t,x) > \lambda$. We relax this to strict ellipticity, requiring only $\sigma(t,x) >0$, to accommodate, for example, cases where $x$ is a function of the DM's posterior beliefs and learning slows to $0$ as time approaches $\infty$ or as $x$ nears the domain endpoints (as in Brownian learning).} 
% \footnote{The literature typically assumes uniform ellipticity: $\exists \lambda>0$ such that $\forall (t,x)$, $\sigma(t,x) > \lambda$. We relax this assumption, requiring only $\sigma(t,x) >0$ (strict ellipticity) to accommodate, e.g., the possibility that learning slows to $0$ as time goes to $\infty$ or when the belief approaches the endpoints of the domain as is the case for Brownian learning.} 
    Both $\mu$ and $\sigma$ are \textit{twice continuously differentiable}. Assume both endpoints, $\underline{x}$ and $\bar{x}$, are unattainable.\footnote{This is for simplicity. A previous version of this paper also considered absorbing boundaries.}

    Until the process is stopped, the DM obtains a flow payoff of $f(t,X_{t})$, with $f:[0,T) \times \mathcal{X} \to \mathbb{R}$. At any time $t$, the DM may stop the diffusion process and select one of two alternatives, $a$ and $b$, obtaining expected payoffs of $g^a(X_t)$ and $g^b(X_t)$, respectively.\footnote{Our analysis easily extends to cases with $n>2$ alternatives. Moreover, we discuss how to accommodate nonstationary reward functions (i.e., time-dependent $g$) in Remark \ref{remark:nonstationarystoppingpayoff} and Section \ref{subsec:timedependentg}.} The DM discounts her payoff at the (possibly stochastic) rate $r(t,x)$, with $r: [0,T) \times \mathcal{X} \to \mathbb{R}_+$. Both $f$ and $r$ are \textit{twice continuously differentiable}.

    Formally, for all $t \in [0, T)$, denote by $\mathcal{T}(t)$, the set of stopping time taking values in $[t, T]$. At every time and state pair $(t,x)\in [0,T)\times \mathcal{X}$, the DM solves 
	\begin{align}\label{eq:valuefunction}
		V(t, x)  = \,&  \underset{\tau\in \mathcal T(t)}{\sup } \, \mathbb{E}_{(t,x)}\left[ \int_{t}^{\tau} e^{-\int_{t}^{s}r(u,X_u)\, \mathrm{d}u} f(s, X_s)ds + e^{-\int_{t}^\tau r(s,X_s)ds} g\left(X_{\tau}\right) \right], \tag{$\mathcal{V}$} 
        \\ & \text{ subject to: }  X_{t+s}= x + \int_{t}^{t+s} \mu(k, X_k) dk + \int_{t}^{t+s} \sigma(k, X_k) \, \mathrm{d}B_k, \notag
	\end{align}
	where $g(x)=\max\{g^a (x), g^b (x) \}$, and $\mathbb{E}_{(t,x)}$ is the expectation operator associated with the process $X$ starting at $(t,x)$. %defined by the above equation % and $V(t, x)$ denotes the \emph{value function} associated with our problem. 
    We also assume the problem is well-posed:
    \begin{align}
        \mathbb{E}_{(t,x)} \left[ \underset{ t\leq s \leq T}{\sup} \, \left( \int_{t}^{s} e^{-\int_{t}^{s}r(u,X_u)\, \mathrm{d}u} f(s, X_s)ds + e^{-\int_{t}^s r(s,X_s)ds} g\left(X_{s}\right) \right) \right] < \infty. \label{eq:uniformintegrability} 
    \end{align}    
    This inequality holds, e.g., when $f$ and $g$ are bounded and $r(t,x) \geq \underline{r} >0$ for all $(t,x)$. It guarantees that the problem's \emph{value function} $V(t, x)$ is locally bounded.
    
    To simplify the exposition, we assume ${g^i}: \mathcal{X} \to \mathbb{R}_{+}$, $i \in {a,b}$, is non-negative, twice continuously differentiable, with $\alpha$-H\"{o}lder continuous derivatives. We also assume that $g^a - g^b$ has at most one zero, at $x^c \in \mathcal{X}$, ensuring that the DM prefers one alternative when $x \leq x^c$ and the other when $x \geq x^c$. Notably, we allow $g^a$ and $g^b$ to have different derivatives at $x^c$ (e.g., $g^a_t < 0$ and $g^b_t > 0$), as the stopping reward in many applications exhibits this feature. Thus, $g$ may have a convex kink at $x^c$.

    In this setting, the stochastic differential equation \eqref{eq:diffusion} has a unique strong solution, and the process $X$ is Markovian in $(t,x)$ (see  Chapter 21 in \cite{kallenberg2006foundations}). Following the classic theory for Markovian stopping problems, we define the optimal stopping region $\mathcal{S}$ and its complement, the continuation region $\mathcal{C}$, as
% \vspace{-1em}
%     \begin{center}
%         \resizebox{40em}{!}{
% \begin{minipage}{\textwidth}
\begin{align*}
\mathcal{S} = \{ (t,x) \in [0,T)\times \mathcal{X} : V(t,x) = g(x) \} \text{ and } \mathcal{C} = \{ (t,x) \in [0,T)\times \mathcal{X} : V(t,x) > g(x) \}.
\end{align*}
% \end{minipage}
% }
    % \end{center}
	Corollary 2.9 in \cite{peskir2006optimal} then implies that the smallest optimal stopping time is $	\tau_{\mathcal{S}} = \inf \left\{ t'\geq 0 \, : \, \left(t+t', X^{(t,x)}_{t'}\right) \in \mathcal{S} \right\}$. 

% Our model is general. So, for concreteness, it may help focusing on the following running example
% NO NEED FOR THIS INTRO For concreteness, we illustrate our general model through the lenses of a standard information acquisition problem.
% with a simple running example. 
    \begin{example}[\textbf{Information Acquisition}]\label{ex:informationacquisition1}
        %Consider a classical information acquisition problem where 
    A DM must choose between two alternative actions, $d \in \{-1, 1\}$, with payoffs dependent on an unknown parameter $\theta \in \{-1, 1\}$. At any time $t$, the DM can either take action $d$ or observe a signal: 
    \setlength{\abovedisplayskip}{4pt}
    \setlength{\belowdisplayskip}{4pt} 
    	 \begin{align*}
    		Z_t =\int_0^t i(t) \,\theta \, \mathrm{d} u+ \int_0^t \zeta(t) \, \mathrm{d} B_u,
    	\end{align*} 
    where $i(t)$ and $\zeta(t)$ are the signal intensity and noise at time $t$. Decision's accuracy and urgency are governed by a gain function, a flow cost $c(t)\geq 0$, and a constant discount rate $r> 0$. Following \cite{shiryaev1967two}, this information acquisition problem reduces to finding the decision time $\tau$ that maximizes the DM's expected payoff:
   \begin{align} \label{bu0}
    		V^{re}(t, X_t)  = \,& \underset{\tau}{\sup } \, \mathbb{E}\left[e^{-r\tau} \left(a X_{\tau} \vee b(1-X_{\tau})\right) - \int_0^{\tau} e^{-rt}c(t) \, \mathrm{d}t \right] 
    	\tag{$\mathcal{V}^{re}$} 
        \\ & 
    		\text{subject to: }\,\, X_{t+s} = X_t + \int_{t}^s \sigma(u,X_u) \, \mathrm{d}B_u, \notag
    \end{align}
    where $X_0$ is the DM's prior that $\theta=1$, $X_t$ is her posterior belief at time $t$, $a X_{\tau} \vee b(1-X_{\tau})$ is the expected payoff of optimally choosing $d$ at $t=\tau$, and $\sigma(t,x):=\theta\frac{2 \, i(t) }{\zeta(t)}x(1-x)$ measures the DM's learning speed and follows from a standard filtering argument \citep{bain2008fundamentals}. Notably, problem \eqref{bu0} is a special instance of \eqref{eq:valuefunction}, with $\mu(t,X_t)=0$, $g(X_\tau)=a X_{\tau} \vee b(1-X_{\tau})$, $f(t,X_t)=c(t)$, $r(t,X_t)=r$, and $T=\infty$.

    When $i$, $\zeta$, and $c$ are independent of time, this is the classical information acquisition problem proposed by \citet{wald2004sequential} and \citet{shiryaev1967two} and widely analyzed in economics, physics, and psychology. By introducing time nonstationarities, our framework allows us to capture, for example, situations where the DM obtains better signals over time (e.g., as her experience increases) or faces increasing information costs (e.g., as she gets tired or new information becomes harder to find).

    % Note that this is the classical information acquisition problem, proposed by \citet{wald2004sequential} and \citet{shiryaev1967two} and largely analyzed in the economics, physics, and psychology literature, with one key difference. While the literature assumes $i(t)$, $\zeta(t)$, and $c(t)$ to be constant in $t$, our framework allows for time nonstationarities. This allows us to capture, for example, situations where the DM obtains better signals over time (e.g., as her experience increases) or faces increasing information costs (e.g., as she gets tired or new information becomes more difficult to find).

% Note that this is the classical information acquisition problem proposed by \citet{wald2004sequential} and \citet{shiryaev1967two} and widely analyzed in economics, physics, and psychology literature, with one key difference. While the literature typically assumes $i(t)$, $\zeta(t)$, and $c(t)$ are constant over time, our framework allows for time nonstationarities. This enables us to capture situations where the DM obtains better signals over time (e.g., as her experience increases) or faces increasing information costs (e.g., as she gets tired or new information becomes harder to find).

    \end{example}

	\section{Analysis and results}\label{sec:analysis}

\paragraph{} As a first technical step, we establish that the value function associated with  \eqref{eq:valuefunction} is smooth and coincides with the unique solution of the following Hamilton-Jacobi-Bellman equation \hypertarget{link:HJB}{(HJB)}:%\footnote{ see Appendix \ref{subsec:toolbox} and Lemma \ref{lemma:HJB} for details and discussion}

    \vspace{-1em}
    \begin{adjustwidth}{-0in}{-0.25in}
        \begin{align*}\label{eq:HJB}
          \!\!\!\!\!\!\!\begin{cases}
                \max \!\left\{ g(x) \!-\!v(t,x),  v_t(t,x) \!+\! \mathcal{L}^{(t,x)}v(t,x) \!-\!r(t,x)v(t,x) \!+\! f(t,x) \right\} = 0 \text{\small{ a.e. in }} \medmath{[0,T)\times \mathcal{X}} \\
                v(T,x) = g(x) \text{ on } \mathcal{X}, %\tag{HJB}
            \end{cases} 
	   \end{align*} \normalsize
    \end{adjustwidth}
     \vspace{0.5em}
     where $\mathcal{L}^{(t,x)}\phi(t,x) = \frac{1}{2} \sigma^2(t,x) \phi_{xx}(t,x) + \mu(t,x) \phi_{x}(t,x),$ is the infinitesimal generator associated with $X$, and, for a function $\phi$, $\phi_t$, $\phi_x$ and $\phi_{xx}$ denote its derivatives with respect to $t$, $x$, and its second derivative with respect to $x$. The formal result, Lemma \ref{lemma:HJB}, is in Appendix \ref{subsec:toolbox}.\footnote{Lemma \ref{lemma:HJB} accommodates stopping payoffs $g(x)$ with convex kinks (e.g., $g(x)$ is the maximum of linear functions), which is critical for applications where the DM makes a decision upon stopping.}\textsuperscript{,}\footnote{Hereafter, the derivatives of the value functions are standard derivatives. They exist in the continuation region by Lemma \ref{lemma:HJB}.} While technical, this result is key to analyzing nonstationary environments where the value function cannot be solved in closed form.

	\subsection{Monotone Environments}\label{subsec:monotoneenvironments}

    % Endowed with this technical result, in the rest of this section, we derive comparative statics results in a broad class of nonstationary environments where the time dependence is locally ``monotone".\footnote{In what follows, the derivatives of the value functions are standard derivatives. They exist in the continuation region as a consequence of Lemma \ref{lemma:HJB}.} 
    % In this Section, we use that the value function is the unique $L^p$-solution of the Hamilton-Jacobi-Bellman equation to derive comparative statics in a wide class of nonstationary environments where time dependence is locally ``monotone''. 
    % can now study comparative statics 
    % are now ready to state and prove our main result
    %As a first step, we formally define locally monotone environments.\footnote{In what follows, the derivatives of the value functions are standard derivatives. They exist in the continuation region as a consequence of Lemma \ref{lemma:HJB}.} 
    % In these environments, Theorem \ref{theorem:localmonotonestopping} and \ref{theorem:localmonotonestopping} show that the boundaries of the continuation regions are locally monotone.
 
    % They provide a set of conditions, under which Theorem \ref{theorem:localmonotonestopping} and \ref{theorem:localmonotonestopping} shows that the boundaries of the continuation regions are locally monotone.

% Our main results This section focuses on environments where time dependence is `\textit{monotone}'.	
% As a second step of the analysis we define what it means for the time dependence to 
    We first study the evolution of the DM's behavior over time in nonstationary problems. To do so, we focus on environments where time dependence is `\textit{monotone}'.	
    \begin{definition}\label{definition:monotoneenvironments}
		The optimal stopping problem \eqref{eq:valuefunction} is locally \textbf{monotone increasing} on $(\underline{t}, \bar{t})$ if $ V_t(t,x)\geq 0 $ on $(\underline{t}, \bar{t})$ for all $x\in \mathcal{X}$. It is locally \textbf{strictly monotone increasing} if, for all $(t,x) \in \mathcal{C}$ such that $t\in (\underline{t}, \bar{t})$, $V_t(t,x) >0$ and 
    \setlength{\abovedisplayskip}{3pt}
    \setlength{\belowdisplayskip}{3pt}
    \begin{equation}
            \label{eq:IOV}
        \sigma_t(t,x)V_{xx}(t,x)+\mu_t(t,x) V_x(t,x)-r_t(t,x) V(t,x) +f_t(t,x) \geq 0. \tag{IOV}
    \end{equation}
    \setlength{\abovedisplayskip}{5pt}
    \setlength{\belowdisplayskip}{5pt}
	Similarly, \eqref{eq:valuefunction} is locally \textbf{monotone decreasing} on $(\underline{t}, \bar{t})$ if $ V_t(t,x)\leq 0 $ on $(\underline{t}, \bar{t})$ for all $x\in \mathcal{X}$. It is locally \textbf{strictly monotone decreasing} if, for all $(t,x) \in \mathcal{C}$ with $t\in (\underline{t}, \bar{t})$, $V_t(t,x) <0$ and
	\begin{align}\label{eq:DOV}
	   \sigma_t(t,x) V_{xx}(t,x) + \mu_t(t,x) V_x(t,x)-r_t(t,x) V(t,x) + f_t(t,x) \leq 0. \tag{DOV}
	\end{align}	
    % Finally, the optimal stopping problem \eqref{eq:valuefunction} is \textbf{(strictly) monotone} if it is globally (strictly) monotone (i.e., $(\underline{t}, \bar{t}) = (0, T)$).
    Finally, \eqref{eq:valuefunction} is \textbf{(strictly) monotone} if it is locally (strictly) monotone on $(0, T)$.
	\end{definition}
    Intuitively, an optimal stopping problem is monotone if, given the state $x$, the expected payoff of the DM is monotone in time. It is strictly monotone if also the option value of waiting (in \eqref{eq:IOV} and \eqref{eq:DOV}) evolves monotonically over time.\footnote{Since the stopping payoff $g(x)$ is time-independent, then $V_t(t,x)=0$ for all $(t,x)\notin \mathcal{C}$.}

    Our first main result shows that the continuation region shrinks over time when the problem is monotone decreasing and expands over time when the problem is monotone increasing. Formally, we define the strict set order as follows.
  
    % The proof follows immediately from the local monotonicity of the value function.
    % In the following we show how the monotonicity of the environment maps into ''monotonicity" of the stopping region.

    % \begin{definition}\label{def:strictsetorder}[Strict set order]
    % 	The continuation region is increasing over time if, for all $\bar{t}>\underline{t}$, the $t$-sections of the continuation region, $\mathcal{C}^t \coloneqq \left\{ (t,x) \in \mathcal{C} \, : \, x \in \mathcal{X} \right\}$, are such that $\mathcal{C}^{^{\underline{t}}\subset \mathcal{C}^{^{\bar{t}}$; strictly so if also $\partial \mathcal{C}^{^{\underline{t}} \cap \mathcal{X} \subset Int(\mathcal{C}^{^{\bar{t}})$.\\
    %     The continuation region is decreasing over time if, for all $\bar{t}>\underline{t}$, $\mathcal{C}^{^t \coloneqq \left\{ (t,x) \in \mathcal{C} \, : \, x \in \mathcal{X} \right\}$, are such that $\mathcal{C}^{^{\bar{t}}\subset \mathcal{C}^{^{\underline{t}}$; strictly so if also $\partial \mathcal{C}^{^{\bar{t}} \cap \mathcal{X} \subset Int(\mathcal{C}^{^{\underline{t}})$.\footnote{One way to understand the strict set order is as follows: If the continuation region is strictly monotone in the sense of definition \ref{def:strictsetorder}, then its boundary can be written as the union of the graphs of strictly monotone functions.}
    % 	\end{definition}
    \begin{definition}\label{def:strictsetorder}\textbf{(Strict set order)}
    	The continuation region $\mathcal{C}$ is increasing (decreasing) over time if its $t$-sections, $\mathcal{C}^{^t} = \left\{ (t,x) \in \mathcal{C} \, : \, x \in \mathcal{X} \right\}$, are such that, for all $\bar{t}>\underline{t}$, $\mathcal{C}^{^{\underline{t}}}\subset \mathcal{C}^{^{\bar{t}}}$ ( $\mathcal{C}^{^{\underline{t}}}\supset \mathcal{C}^{^{\bar{t}}}$); strictly so if also $\partial \mathcal{C}^{^{\underline{t}}} \cap \mathcal{X} \subset \mathrm{Int}(\mathcal{C}^{^{\bar{t}}})$ ($\partial \mathcal{C}^{^{\bar{t}}} \cap \mathcal{X} \subset \mathrm{Int}(\mathcal{C}^{^{\underline{t}}})$).
	\end{definition}
    % \textcolor{red}{Explain the meaning of strictly decreasing/increasing}
    % \textcolor{blue}{not sure how much we want to emphasize the fact that we are calling strictly decreasing and strictly increasing a continuation region when we never stop.... }
    % \textcolor{purple}{Note that, if we focus on 
    % In settings where the DM always stops when the state is sufficiently close to the boundary of the state space $\underscore{c}, \bar{x}$  (e.g., in essentially all information acquisition problems), the above definition is completely standard. However, when distinguishing between strictly or weakly increasing/decreasing we ignore the boundaries that are imposed on the continuation region by the state space.    
    % % The above definition is quite standard, except for how we treat settings where ; the only convention we use is that not sure how much we want to emphasize the fact that we are calling strictly decreasing a continuation region when we never stop.... 
    % }
    % This definition allows us to characterize (strict) monotonicity of the continuation region when the stopping boundaries are interior. 
   The strict set order is the monotonicity notion we will use hereafter. It states that the continuation region is monotone if its boundaries are monotone. It is strictly monotone if its boundaries are strictly monotone when inside the domain.
	% \begin{remark}\label{remark:unionofstrictlymonotonefunctions}
	% 	If the continuation region is strictly monotone in the sense of definition \ref{def:strictsetorder}, its boundary can be written as the union of the graphs of strictly monotone functions.
    %        % \textcolor{red}{comment on strict order: only for the interior of the domain. Vacuously true if $\partial C_t \cap \mathcal{X} = \emptyset$}
	% \end{remark}

	\begin{theorem}\label{theorem:localmonotonestopping} 
 	\textcolor{white}{.}

    \vspace{-0.5em}
	   \begin{enumerate}[topsep=3.5pt,itemsep=3pt]
	       % \item If the optimal stopping problem is  \textbf{monotone decreasing} on $\left( \underline{t}, \bar{t} \right)$, the optimal continuation region is nonincreasing on $\left( \underline{t}, \bar{t}  \right)$. 
			
            \item If the optimal stopping problem is locally \textbf{(strictly) monotone decreasing} on $(\underline{t}, \bar{t})$, the optimal continuation region is \textbf{(strictly) decreasing} on $(\underline{t}, \bar{t})$.
    		% \item If the optimal stopping problem is \textbf{monotone increasing} on $(\underline{t}, \bar{t} )$, the optimal continuation region is nondecreasing on $(\underline{t}, \bar{t} )$. 
    
    		\item If the optimal stopping problem is locally \textbf{(strictly) monotone increasing} on $(\underline{t}, \bar{t})$, the optimal continuation region is \textbf{(strictly) increasing} on $(\underline{t}, \bar{t})$.
    	\end{enumerate}
    \end{theorem}
    Theorem \ref{theorem:localmonotonestopping} relates the behavior of the continuation region to the environment's local monotonicity. Its usefulness lies in (i) the relative ease of verifying the environment's monotonicity and (ii) the prevalence of monotone problems in the literature. We illustrate both features in Section \ref{sec:applications}. In particular, we show that the monotonicty of the value function and conditions \eqref{eq:IOV} and \eqref{eq:DOV} are often easily verified from the primitives. We also provide sufficient conditions on the primitives that hold in many environment of interests in Sections \ref{subsec:Visconvex} and \ref{subsec:arithmeticandgeometricBM}.
    
    Weak monotonicity follows immediately from the monotonicity of the value function, which guarantees that the option value of waiting, $V(t,x)-g(x)$, is (locally) monotone. Strict monotonicity is harder to prove. Intuitively, the DM stops when the benefits of waiting equal the costs: At the stopping boundary, the DM is indifferent between making a decision and waiting an extra instant. In a monotone environment, the benefit of making an immediate decision at a fixed state level $x$ is constant. At the same time, the cost of waiting an extra instant (as measured by \eqref{eq:IOV} or \eqref{eq:DOV}) is strictly monotone. Therefore, upon waiting, the DM is no longer indifferent, and the level $x$ is either in the interior of the continuation region (when \eqref{eq:IOV} holds) or the interior of the stopping region (when \eqref{eq:DOV} holds). Formally, we leverage the Hopf boundary lemma to prove (by contradiction) that the boundary cannot be flat.

% \textcolor{red}{    While Theorem \ref{theorem:localmonotonestopping} characterizes the local dynamics of the continuation region, it provides limited insight into its overall shape (unless combined with a (standard) single-crossing assumption). For instance, $\mathcal{C}$ might be disjoint, making interpretations more challenging. However, under a (standard) single-crossing assumption, $\mathcal{C}$ is well-behaved. 
% }
While Theorem \ref{theorem:localmonotonestopping} characterizes the evolution of the continuation region, it provides limited insight into its overall shape (e.g., $\mathcal{C}$ may be connected or disjoint). However, a standard single-crossing condition guarantees that $\mathcal{C}$ has convex sections.

  % \textcolor{gray}{  (OLDER) While Theorem \ref{theorem:localmonotonestopping} characterizes the local dynamics of the continuation region, it offers little insight into its shape (e.g., $\mathcal{C}$ may be connected or disjoint). As a result, interpreting the results can be difficult. 
  %   However, under the following single-crossing assumption, we can show that an upper and lower boundary delimits $\mathcal{C}$.}
    \begin{condition}[SC]\label{condition:singlecrossing}
        For all $t \in [0,T)$, the mapping $x \to f(t,x) + \left(\mathcal{L}^{(t,x)} -r(t,x) \right) g(x)$ is nondecreasing on $\left(\underline{x}, x^c\right]$ and nonincreasing on $\left[x^c, \bar{x}\right)$.
    \end{condition}
	Intuitively, Condition C\ref{condition:singlecrossing} ensures that the gain from waiting one extra instant decreases as we move away from the indifference point $x^c$. As a result, the following (standard) proposition shows that the $t$-sections of the continuation region $\mathcal{C}$ are convex, and, hence, that $\mathcal{C}$ lays between two boundaries.
 %\footnote{To see this, formally, observe that, when $g$ is smooth, for all $(t,x) \in [0,T)\times \Bar{\mathcal{X}}$,
	% \begin{align*}
	% V(t,x) - g(x) = \underset{\tau}{\sup}\, \mathbb{E}_{(t,x)} \left[ \int_0^{\tau} e^{-\int_{0}^{s}r(u,X_u)\, \mathrm{d}u} \left(f(s,X_s) + \left(\mathcal{L}^{(s,X_s)} -r(s,x) \right) g(X_s)\right) ds \right].
	% \end{align*}}
    \begin{proposition}\label{prop:convextsectionsstopping}
		Suppose C\ref{condition:singlecrossing} holds. Then there exist functions $\underline{b}:[0,T) \to[\underline{x}\,,x_c]$ and $ \bar{b}:[0,T) \to[x_c\, ,\bar{x}]$ such that 
		% \begin{align*}
		$\mathcal{C} = \left\{ (t,x) \in [0,T] \times \mathcal{X} \, : \, x  \in \left(\underline{b}(t), \bar{b}(t)\right) \right\}.$
		% \end{align*}
	\end{proposition}
	% Proposition \ref{prop:convextsectionsstopping} is standard when $g$ is smooth and $f(t,x) + \left(\mathcal{L}^{(t,x)} -r(t,x) \right)g$ is monotone \citep{villeneuve2007threshold}. However, the usual argument does not easily adapt to our case. We provide a new proof based on the HJB equation in Appendix \ref{app:proofsmainresults}. 
 
    % In monotone problems, Proposition \ref{prop:convextsectionsstopping} facilitates the interpretation of Theorem \ref{theorem:localmonotonestopping}: The continuation region is delimited by monotone upper and lower bounds. As time passes, the DM stops in more extreme states when the problem is monotone increasing, and in more central states when the problem is monotone decreasing. 

   In monotone problems satisfying Condition \ref{condition:singlecrossing}, Theorem \ref{theorem:localmonotonestopping}'s predictions can be refined: The continuation region is delimited by monotone upper and lower bounds. As time passes, the DM stops in more extreme states when the problem is increasing and in more central states when the problem is decreasing.
    \begin{corollary}\label{corollary:globalmonotonestopping}
Suppose \eqref{eq:valuefunction} is \textbf{globally monotone} and satisfies C\ref{condition:singlecrossing}. If \eqref{eq:valuefunction} is:  
    	\begin{enumerate}[topsep=3.5pt,itemsep=3pt]
            \item \textbf{decreasing}, then                 $\mathcal{C}\!=\! \left\{ \left(t, x\right) \in [0,T] \!\times \!\mathcal{X}  : \,  x  \in \left( \underline{b}(t), \bar{b}(t) \right) \right\}$, with $\underline{b}:[0,T) \to[\underline{x}\,,x_c]$ c\`{a}gl\`{a}d {nondecreasing} and $ \bar{b}:[0,T) \to[x_c\, ,\bar{x}]$ c\`{a}dl\`{a}g {nonincreasing}.\footnote{In this case, $\bar{b}$ and $\underline{b}$ are also continuous. See Online Appendix \ref{app:online}.}
            \setlength{\abovedisplayskip}{4pt}
            \setlength{\belowdisplayskip}{4pt} 
            % \begin{align*}
            %     \mathcal{C} \coloneqq \left\{ \left(t, x\right) \in \mathbb{R}_+ \times \mathcal{X} \, : \,  x  \in \left( \underline{b}(t), \bar{b}(t) \right) \right\}.
            % \end{align*}
            % If the problem is \textbf{strictly monotone decreasing} then $\bar{b}$ and $\underline{b}$ are strictly monotone in $\mathcal{X}$.
            % : if $\bar{b}(\underline{t}) < \bar{x}$, then $\bar{b}(t)$ is \textbf{strictly decreasing} on $[\underline{t}, T)$; and if $\underline{b}(\underline{t}) > \underline{x}$, then $\underline{b}(t)$ is \textbf{strictly increasing} on $[\underline{t}, T)$.
              % \vspace{-27pt}

            \item \textbf{increasing}, then $\mathcal{C}\!=\! \left\{ \left(t, x\right) \in [0,T] \!\times \!\mathcal{X}  : \,  x  \in \left( \underline{b}(t), \bar{b}(t) \right) \right\}$, with $\underline{b}:[0,T)\to [\underline{x},x_c]$ c\`{a}dl\`{a}g {nonincreasing} and $ \bar{b}:[0,T) \to[x_c\, ,\bar{x}]$ c\`{a}gl\`{a}d  {nondecreasing}.
      %       \setlength{\abovedisplayskip}{4pt}
      %       \setlength{\belowdisplayskip}{4pt} 
    		% \begin{align*}
    		% 	\mathcal{C} \coloneqq \left\{ \left(t, x\right) \in \mathbb{R}_+ \times \mathcal{X} \, : \,  x  \in \left( \underline{b}(t), \bar{b}(t) \right) \right\}.
    		% \end{align*}
    		% When the problem is \textbf{strictly monotone increasing}: if $\bar{b}(\bar{t}) < \bar{x}$, then $ \bar{b}(t)$ is \textbf{strictly increasing} on $[0, \bar{t})$; and  if $\underline{b}(\bar{t}) > \underline{x}$, then $ \underline{b}(t)$ is \textbf{strictly decreasing} on $[0, \bar{t})$.
		\end{enumerate}
  Moreover, if \eqref{eq:valuefunction} is \textbf{strictly monotone}, then $\bar{b}$ and $\underline{b}$ are strictly monotone in $\mathcal{X}$.\footnote{$\bar{b}$ and $\underline{b}$ may be flat at the boundary (and equal to $\bar{x}$ and $\underline{x}$) if the DM prefers never stopping.} 
  % at $t$ (i.e., $\bar{b}(t)=\bar{x}$ and $\underline{b}(t)=\underline{x}$), then $\bar{b}$ and $\underline{b}$ would be flat (and equal to $\bar{x}$ and $\underline{x}$) in $t$}
  % \footnote{If the DM strictly prefers never stopping at $t$ (i.e., $\bar{b}(t)=\bar{x}$ and $\underline{b}(t)=\underline{x}$), then $\bar{b}$ and $\underline{b}$ would be flat (and equal to $\bar{x}$ and $\underline{x}$) in $t$}

  % to wait even at $\bar{x}$ ($\underline{x}$) at time $t$, then $\bar{x}$ ($\underline{x}$) would stay flat a}
	\end{corollary}

	This corollary will be the crucial result for most of our applications. 
	
	% Theorem \ref{theorem:localmonotonestopping}, Proposition \ref{prop:convextsectionsstopping}, and Corollary \ref{corollary:globalmonotonestopping} are proved for a time independent stopping reward $g$. In Section \ref{subsec:timedependentg}, we discuss how to extend these results for nonstationary reward functions.

    \begin{examplecont}{\ref{ex:informationacquisition1}}[\textbf{Information Acquisition continued}]\label{ex:informationacquisition2}
        
% In our running example, we show how, our monotone comparative statics can be use to characterize the relation between decision timing and accuracy.  For any choice $d\in \mathcal{D}$, the accuracy of $d$ at time $t$ is the probability $P(d,t)\in [0,1]$ that $d$ is the correct choice conditional on DM stopping and choosing $d$ at time $t$.
% \begin{definition}
% Accuracy is said to be increasing (decreasing) if $P$ is monotone increasing (decreasing) in $t$ for all $d\in \mathcal{D}$ and strictly so for at least one $d\in \mathcal{D}$; Accuracy is said to be constant if $P$ is constant in $t$ for all $d\in \mathcal{D}$.
% \end{definition} 

% \footnote{ 
% %Continuity immediately comes from Lemma \ref{lemma:continuousvaluefunction}. 
%  For A\ref{assumption:singlecrossing}, it is sufficient to realize that the stopping payoff $g$ is the maximum of two linear function in this case.} 
% For A\ref{assumption:Viscontinuous}, see Section \ref{subsec:Visconvex} and Lemma \ref{lemma:Viscontinuous}.}  
% Moreover, since $r>0$, the DM would always stop in the interior of $\mathcal{X}=[0,1]$. 
    In Example \ref{ex:informationacquisition1}, $\mathcal{X}=(0,1)$ is the set of attainable posteriors, $g$ is the maximum of two linear functions, and $r>0$. Thus, C\ref{condition:singlecrossing} holds, and Corollary \ref{corollary:globalmonotonestopping} characterizes $\mathcal{C}$ as a function of the problem's primitives. 
% the accuracy of $d=1$ at the stopping time $\tau$ is $X_\tau$, and the accuracy of $d=-1$ is $1-X_\tau$
% there is an obvious mapping between stopping boundaries and decision accuracy in this setting. %: if $\bar{b}(t)$ is closer to $1$ and $\underline{b}(t)$ is closer to $0$ then decisions that occur at time $t$ are more accurate. 
% Since $\theta\in\{-1,1\}$ and $X_t$ is the posterior that $\theta=1$, there is a direct mapping between 
% this characterization has direct implications for decision accuracy. 
    Since $\theta\in\{-1,1\}$, $X_\tau\in \{\bar{b}(\tau),\underline{b}(\tau)\}$ is the posterior probability that $\theta=1$ upon stopping, Corollary \ref{corollary:globalmonotonestopping} has direct implications for decision accuracy. Formally, the accuracy of $d\in \{-1,1\}$ is the probability $P(d,t)\in [0,1]$ that $d=\theta$ given the DM stops and chooses $d$ at time $t$. 
% \begin{definition}
    Accuracy is said to be increasing (decreasing) if $P(d,t)$ is increasing (decreasing) in $t$ for all $d\in \mathcal{D}$ and strictly so for at least one $d\in \mathcal{D}$. %; Accuracy is said to be constant if $P$ is constant in $t$ for all $d\in \mathcal{D}$.
% \end{definition} 

    \textbf{Classical benchmark.} In the setting \`{a} la Wald, the flow cost $c$, the signal intensity $i$, and the noise $\zeta$ are time-stationary.\footnote{This implies that the learning speed $\theta\frac{2i}{\zeta}x(1-x)$ is also independent of time.} Thus, the associated problem $\mathcal{V}^{re}$ is both monotone increasing and decreasing, with $V^{re}_t=0$. Thus, Corollary \ref{corollary:globalmonotonestopping} implies that $\mathcal{C}$ is bounded by two constant boundaries, $\overline{b}(t)=\overline{\pi}$ and $\underline{b}(t)=\underline{\pi}$, delivering the classical result that accuracy is constant over time (see Figure 1).\footnote{For a textbook treatment, see \cite{peskir2006optimal}, Chapter VI.}.  

% in $\mathcal{X}$. When $X_t$ exceeds $\overline{b}(t)=\overline{\pi}\in(0,1)$, the DM takes action $d=1$, and when $X_t$ falls below a lower threshold $\underline{b}(t)\underline{\pi} \in(0,1)$, she takes action $d=-1$ 
  
 % Since the DM's problem at time $t$ is uniquely pinned down by $X_t$, its solution consists of two time-constant boundaries in the posteriors' space. When $X_t$ exceeds an upper threshold $\overline{\pi}\in(0,1)$, the DM takes action $d=1$, and when $X_t$ falls below a lower threshold $\underline{\pi} \in(0,1)$, she takes action $d=-1$ (see Figure 1, left).\footnote{For a textbook treatment, see \cite{peskir2006optimal}, Chapter VI.} 

\textbf{Results.} Suppose $i$, $\zeta$, or $c$ are not stationary. If $c$ is increasing (decreasing) or the learning speed $\sigma(t,x)=\theta\frac{2i}{\zeta}x(1-x)$ is decreasing (increasing) in $t$, and at least one strictly so, then $V_t^{re}<0$ ($V_t^{re}>0$) and, by Theorem \ref{theorem:localmonotonestopping} and Corollary \ref{corollary:globalmonotonestopping}, $\overline{b}(t)$ and $\underline{b}(t)$ move closer (farther) over time (see Figure 1). %$\mathcal{C}$ shrinks (expands) over time. 
Finally, since  $P(1,t)=\overline{b}(t)$ and $P(-1,t)=1-\underline{b}(t)$, we can prove the following.%\footnote{Proposition \ref{prop:IAconvexproblemboundaries} is a special case of Proposition \ref{prop:convexproblemboundaries}, whose proof is in Appendix \ref{app:Visconvex}.} 

% Thus, Theorem \ref{theorem:localmonotonestopping} and Corollary \ref{corollary:globalmonotonestopping} imply that $\mathcal{C}$ is bounded by an upper bound $\overline{b}(t)$ which decreases (increases) in $t$ if $c(t)$ increases (decreases) and $\sigma(t,x) \equiv \frac{i(t)}{\zeta(t)} x(1-x)$ decreases (increases) over time; and a lower bound $\underline{b}(t)$ that evolves in the opposite direction (see Figure 1). Finally, exploiting the fact that  $P(1,t)=\overline{b}(t)$ and $P(-1,t)=1-\underline{b}(t)$, we prove the following.

    \begin{proposition}\label{prop:IAconvexproblemboundaries}
		The nonstationary information acquisition problem \eqref{bu0} displays  
   \setlength{\abovedisplayskip}{4pt}
\setlength{\belowdisplayskip}{4pt} 
		\begin{itemize}[topsep=3.5pt,itemsep=3pt]
			\item Decreasing accuracy if $t \to \frac{i(t)}{\zeta(t)}$ and $t \to -c(t)$ are nonincreasing, with at least one strictly decreasing; and 
   
			\item Increasing accuracy if  $t \to \frac{i(t)}{\zeta(t)}$ and $t \to -c(t)$ are nondecreasing, with at least one strictly increasing, and $T =\infty$.\footnote{If $T<\infty$, the result would also hold, provided that the payoff of the DM at the boundary is large enough.} 
		\end{itemize}
	\end{proposition}

    \noindent Proposition \ref{prop:IAconvexproblemboundaries} establishes that later decisions are less accurate when the signal-to-noise ratio $\frac{i(t)}{\zeta(t)}$ decreases or the information cost ${c(t)}$ increases over time (e.g., when the DM gets tired or when most relevant news is discovered earlier). Conversely, later decisions are more accurate when $\frac{i(t)}{\zeta(t)}$ increases or ${c(t)}$ decreases over time (e.g., when the DM gains experience in collecting and evaluating information).
    % This finding bridges the gap between theory and empirics, showing how the documented correlations between decision timing and accuracy may result from decision makers' optimal responses to non-stationarities in learning environments.

    % For instance, when the flow learning cost $f(t,x)$ is constant, Proposition \ref{prop:convexproblemboundaries} establishes that decision timing and accuracy are negatively correlated if the DM's learning speed $\sigma(t,x)$ decrease over time (e.g., as the DM becomes more tired or the residual uncertainty decreases). Conversely, it establishes a positive correlation between decision timing and accuracy when the learning speed $\sigma(t,x)$ increases over time (e.g., as the DM gains more experience evaluating the signal she receives).
    % }

  % \textbf{Relevance:} The independence of decision accuracy and time has been a core predictions of classical theoretical frameworks. Yet, it has also been largely challenged by empirical studies in neuroscience and psychology (\citet{bolsinova2017modelling,molenaar2018semi,chen2018curvilinear,goldhammer2014time,goldhammer2015more,fudenberg2020testing}), which found a positive, negative or even curvilinear correlation between accuracy and response time in different situations. Our analysis partially bridges the gap between theory and empirics. We show that these patterns may result from DMs' optimal responses to specific nonstationary features of the learning environments.
  
  \vspace{-1.5em}
\begin{figure}[H]\label{NSR}
\flushleft
\begin{tikzpicture}
    \node at (-5.2,0) {\includegraphics[scale=0.52]{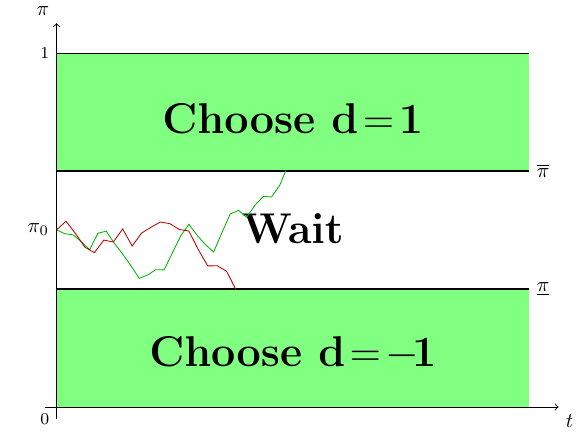}};
    \node at (0,0) {\includegraphics[scale=0.52]{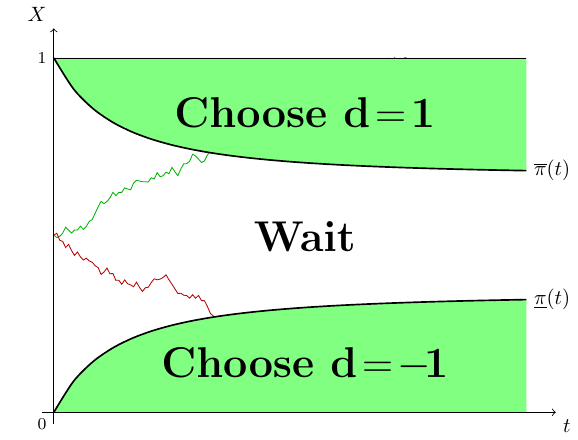}};
    \node at (5.2,0) {\includegraphics[scale=0.52]{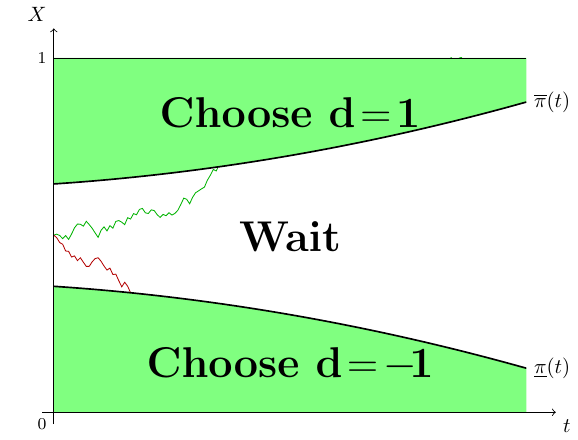}};
\end{tikzpicture}
\caption{ \textit {\textbf{Left.} When $c$ and $\sigma(t,x)= \frac{i(t)}{\zeta(t)}x(1-x)$ are constant in $t$ (the classical framework): $\mathcal{C}^{^t}$ is constant over time, implying constant decision accuracy. \textbf{Center.} When $c(t)$ increases and $\sigma(t,x)$  decreases in $t$: $\mathcal{C}$ shrinks over time, implying decreasing decision accuracy. \textbf{Right.} When $c(t)$ decreases and $\sigma(t)$  increases in $t$: $\mathcal{C}$ expands over time, implying increasing decision accuracy}}
\end{figure}
    
    \end{examplecont}

    \subsection{Monotone Comparative Statics} \label{subsec:CSEE}
 
    In this section, we compare decision outcomes across different economic environments. Let $V^{^{{\sigma}, \mu, f,r, g}}$, $\mathcal{C}^{^{{\sigma}, \mu, f,r, g}}$, and $\tau^{\mathsmaller{{\sigma}, \mu, f,r, g}}$ be the value function, the continuation region, and the stopping time associated with the optimal stopping problem in (\ref{eq:valuefunction}), parameterized by the \emph{functions} $\sigma$, $\mu$, $f$, $r$, and $g$. 
    % First, we show the DM stops later when she is pointwise more patient (lower $r$) and when her flow payoffs are pointwise higher (higher $f$).   
    First, we focus on settings where the DM is pointwise more patient (lower $r$) or faces pointwise higher flow payoffs ($f$).               
    % \begin{theorem}
    %  \label{corollary:compstatsstoppingtimes} \textcolor{white}{.}
    %         % Let $\tau^{\mathsmaller{{\sigma}, \mu, f,r, g}}$ be the smallest optimal stopping time in the stopping problem with parameters $(\sigma, \mu, f,r, g)$.

    %         \begin{enumerate}
    %             \item Suppose $\bar{r} \geq \underline{r}$. Then $V^{^{{\sigma}, \mu, f,\bar{r}, g}} \leq V^{^{{\sigma}, \mu, f, \underline{r}, g}}$, $\mathcal{C}^{^{{\sigma}, \mu, f,\bar{r}, g}} \subset \mathcal{C}^{^{{\sigma}, \mu, f, \underline{r}, g}}$, and $\tau^{\mathsmaller{{\sigma}, \mu, f,\bar{r}, g}} \leq \tau^{\mathsmaller{{\sigma}, \mu, f, \underline{r}, g}}$ $\mathbb{P}$-a.s..
    
    %             \item Suppose $\bar{f} \geq \underline{f}$. Then $V^{^{{\sigma}, \mu, \bar{f}, r, g}} \geq V^{^{\sigma, \mu, \underline{f},r, g}}$,  $\mathcal{C}^{^{\sigma, \mu, \underline{f},r, g}} \subset \mathcal{C}^{^{{\sigma}, \mu, \bar{f}, r, g}}$, and $\tau^{\mathsmaller{{\sigma}, \mu, \bar{f}, r, g}} \geq \tau^{\mathsmaller{\sigma, \mu, \underline{f},r, g}}$. $\mathbb{P}$-a.s..
    %         \end{enumerate}
        
    % \end{theorem}
    \begin{theorem}\label{theorem:compstatsstoppingtimes}  
        Suppose $\bar{r} \geq \underline{r}$ and $\bar{f} \geq \underline{f}$. Then $V^{^{{\sigma}, \mu, \bar{f}, \underline{r}, g}} \geq V^{^{\sigma, \mu, \underline{f},\bar{r}, g}}$ and $\mathcal{C}^{^{\sigma, \mu, \underline{f},\bar{r}, g}} \subset \mathcal{C}^{^{{\sigma}, \mu, \bar{f}, \underline{r}, g}}$. Moreover, $\tau^{\mathsmaller{{\sigma}, \mu, \bar{f}, \underline{r}, g}} \geq \tau^{\mathsmaller{\sigma, \mu, \underline{f},\bar{r}, g}}$ $\mathbb{P}$-a.s.. 
    \end{theorem}
    This result complements \citet{quah2013discounting}. It shows, as they do, that a lower stochastic discount rate results in a higher value function, a smaller continuation region, and a longer stopping time even in time nonstationary environments. It also shows a similar dynamic arises when the flow payoffs are higher. Intuitively, when flow payoffs are higher, waiting is more attractive. So, the DM only settles when stopping provides a higher benefit (i.e., the continuation region is broader). Formally, the proof of Theorem \ref{theorem:compstatsvalues} derives the results from the analysis of the HJB equation and a comparison principle.

Next, we study the impact of a pointwise increase in the drift or variance of the diffusion process. We show that (i) a higher $\mu$ benefits (harms) the DM and expands (shrinks) the continuation region if the value function is increasing (decreasing) in $x$, and (ii) a higher $\sigma$ benefits (harms) the DM and expands (shrinks) the continuation region if the value function is convex (concave) in the state $x$.\footnote{Corollary \ref{corollary:strictcompstats} in Appendix \ref{app:proofCSEE} offers a strict version of the comparative statics of Theorems \ref{theorem:compstatsstoppingtimes} and \ref{theorem:compstatsvalues}.}

%on continuation regions and value functions.
 \begin{theorem}\label{theorem:compstatsvalues}
\textcolor{white}{.}

  \vspace{-0.5em}      % Assume that $\bar{\sigma} \geq \underline{\sigma}$, $\bar{r} \geq \underline{r}$, $\bar{\mu} \neq \underline{\mu}$, and $\bar{f} \geq \underline{f}$ (all pointwise). Then: 
      \begin{enumerate} 
             % \item Suppose $\bar{r} \geq \underline{r}$. Then $V^{^{{\sigma}, \mu, f,\bar{r}, g}} \leq V^{^{{\sigma}, \mu, f, \underline{r}, g}}$ and $\mathcal{C}^{^{{\sigma}, \mu, f,\bar{r}, g}} \subset \mathcal{C}^{^{{\sigma}, \mu, f, \underline{r}, g}}$. 

             %  \item Suppose $\bar{f} \geq \underline{f}$. Then $V^{^{{\sigma}, \mu, \bar{f}, r, g}} \geq V^{^{\sigma, \mu, \underline{f},r, g}}$ and $\mathcal{C}^{^{\sigma, \mu, \underline{f},r, g}} \subset \mathcal{C}^{^{{\sigma}, \mu, \bar{f}, r, g}}$.

                \item Let $\bar{\sigma} \geq \underline{\sigma}$. If $\exists \sigma\in \{\underline{\sigma},\bar{\sigma}\}$ such that $V^{^{{\sigma}, \mu, f,r, g}}$ is convex (concave) in $x$, then $V^{^{\bar{\sigma}, \mu, f,r, g}} \!\!\geq\! V^{^{\underline{\sigma}, \mu, f, r, g}}\!\!$ and $\mathcal{C}^{^{\underline{\sigma}, \mu, f,r, g}} \!\! \subset \! \mathcal{C}^{^{\bar{\sigma}, \mu, f,r, g}}$ ($V^{^{\bar{\sigma}, \mu, f,r, g}} \!\! \leq \!V^{^{\underline{\sigma}, \mu, f, r, g}}\!\!$ and $\mathcal{C}^{^{\bar{\sigma}, \mu, f,r, g}} \!\! \subset \! \mathcal{C}^{^{\underline{\sigma}, \mu, f,r, g}}$\!).

            % \item  Suppose $\bar{\mu} \neq \underline{\mu}$. If either $\bar{\mu} \,  \text{sign}\left(V^{^{{\sigma}, \underline{\mu}, f,r, g}}_x\right)$ is greater than $\underline{\mu} \, \text{sign}\left(V^{^{{\sigma}, \underline{\mu}, f,r, g}}_x\right)$ or $\bar{\mu} \,  \text{sign}\left(V^{^{{\sigma}, \bar{\mu}, f,r, g}}_x\right)$ is greater than $\underline{\mu} \, \text{sign}\left(V^{^{{\sigma}, \bar{\mu}, f,r, g}}_x\right)$, then $V^{^{{\sigma}, \bar{\mu}, f,r, g}} \geq V^{^{{\sigma}, \underline{\mu}, f,r, g}}$ and $\mathcal{C}^{^{{\sigma}, \underline{\mu}, f,r, g}} \subset \mathcal{C}^{^{{\sigma}, \bar{\mu}, f,r, g}}$.
% \item  Let $\bar{\mu} \neq \underline{\mu}$. 
% If $\exists \mu\in \{\underline{\mu},\bar{\mu}\}$ such that $\bar{\mu} \,  \text{sign}\left(V^{^{{\sigma}, {\mu}, f,r, g}}_x\right)\geq \underline{\mu} \, \text{sign}\left(V^{^{{\sigma}, {\mu}, f,r, g}}_x\right)$, then $V^{^{{\sigma}, \bar{\mu}, f,r, g}} \geq V^{^{{\sigma}, \underline{\mu}, f,r, g}}$ and $\mathcal{C}^{^{{\sigma}, \underline{\mu}, f,r, g}} \subset \mathcal{C}^{^{{\sigma}, \bar{\mu}, f,r, g}}$.
            \item Let $\bar{\mu} \geq \underline{\mu}$. If $\exists \mu\in \{\underline{\mu},\bar{\mu}\}$ such that $V^{^{{\sigma}, {\mu}, f,r, g}}$ increases (decreases) in $x$, then $V^{^{{\sigma}, \bar{\mu}, f,r, g}} \!\!\geq \!V^{^{{\sigma}, \underline{\mu}, f,r, g}}\!\!$ and $\mathcal{C}^{^{{\sigma}, \underline{\mu}, f,r, g}}\!\! \subset \! \mathcal{C}^{^{{\sigma}, \bar{\mu}, f,r, g}}$ ($V^{^{{\sigma}, \bar{\mu}, f,r, g}} \!\!\leq \! V^{^{{\sigma}, \underline{\mu}, f,r, g}}\!\!$ and $\mathcal{C}^{^{{\sigma}, \bar{\mu}, f,r, g}} \!\!\subset \! \mathcal{C}^{^{{\sigma}, \underline{\mu}, f,r, g}} \!$).
        \end{enumerate}
    \end{theorem}
    As the proof of Theorem \ref{theorem:compstatsstoppingtimes}, the proof of Theorem \ref{theorem:compstatsvalues} derives the results from the HJB equation and a comparison principle. Intuitively, when $V$ is convex, the decision-maker is risk-loving. An increase in volatility, therefore, benefits the decision-maker as it increases the DM's option value. The DM is more willing to wait. As an illustration, it may be helpful to consider the case when the state process represents the DM's belief. Here, the logic is simple. An increase in volatility indicates that the DM receives more information per unit of time. So, the benefits of waiting one more instant also increase. Similarly, the DM favors a higher state when the value function increases. Naturally, as a higher drift induces higher values of the state, the DM prefers the process whose drift is the largest.  

    One consequence of Theorem \ref{theorem:compstatsvalues} is that the \emph{option value principle} (i.e., the fact that the DM waits beyond the myopic optimum to stop when the value function is convex) extends to general nonstationary environments.\footnote{See, e.g., Section 5.1 in \cite{strulovici2015smoothness}.} 
    Specifically, Theorem \ref{theorem:compstatsvalues} implies that the continuation region is larger for $\sigma, \mu >0$ than for $\sigma= \mu =0$. Since the DM always stops at the myopic optimum when $\sigma =\mu=0$, we conclude that the ``option value'' leads the DM to wait beyond the myopic optimum.

    % Formally, suppose that the value function is convex (and nondecreasing).  Instead, Theorem \ref{theorem:compstatsvalues} guarantees that the continuation region is larger when $\sigma, \mu >0$. So, the ``option value'' leads the DM to wait beyond the myopic optimum.
    
    % Formally, suppose that the value function is convex (and nondecreasing). If $\sigma =\mu=0$, the DM optimally stops at the myopic optimum. Instead, Theorem \ref{theorem:compstatsvalues} guarantees that the continuation region is larger when $\sigma, \mu >0$. So, the ``option value'' leads the DM to wait beyond the myopic optimum.
    
    Contrary to Theorem \ref{theorem:compstatsstoppingtimes}, Theorem \ref{theorem:compstatsvalues} does not deliver predictions on the optimal stopping time. That is because changes in $\mu$ and $\sigma$ generally have an ambiguous effect on stopping times. While a larger continuation region implies, ceteris paribus, longer stopping times, $\sigma$ and $\mu$ affect the speed at which the state process may reach such boundaries. So, it is challenging to obtain clear-cut conclusions on stopping times in general environments.\footnote{For example, \cite{goncalves2024speed} shows that the expected value of the optimal stopping times is single-peaked for the stationary Wald problem.} However, Theorem \ref{theorem:compstatsvalues} may still yield prediction on the stopping times under more structure. In Example \ref{ex:bankruptcies1} below, we show that an increase in $\mu$ unambiguously increases (almost surely) the optimal stopping time.

    \begin{remark}[\textbf{Nonstationary Stopping Payoff}]\label{remark:nonstationarystoppingpayoff}
        The proofs of Theorems \ref{theorem:compstatsstoppingtimes} and \ref{theorem:compstatsvalues} do not use the stationarity of $g$. So, both Theorem \ref{theorem:compstatsstoppingtimes} and \ref{theorem:compstatsvalues} remain valid, as are, when the stopping payoff also depends on time, i.e., $g:[0,T]\times \mathcal{X} \to \mathbb{R}$, where $g$ is the maximum of functions in $\mathcal{C}^{1,2}\left([0,T]\times \mathcal{X}\right)$.
    \end{remark}
    
    % Thus, a higher variance in the diffusion process benefits (harms) the DM and expands (shrinks) the continuation region if the value function is convex (concave) in the state $x$. Similarly, a higher drift in the diffusion process benefits (harms) the DM and expands (shrinks) the continuation region if the value function is increasing (decreasing) in $x$. 

% Thus, the continuation region would be smaller in response to a higher $$ hen the variance of the diffusion proce

    \begin{examplecont}{\ref{ex:informationacquisition2}}[\textbf{Information Acquisition continued}]\label{ex:informationacquisition3}
        % \textcolor{red}{Weird: Longer decision time means that the DM observes strictly more information in Blackwell sense: how can their decision be less accurate? Need to distinguish accurate decsion from better decision?}
        In Example \ref{ex:informationacquisition1}, Theorem \ref{theorem:compstatsstoppingtimes} implies that,  regardless of the information process, a (weakly) more patient DM facing (weakly) lower information costs will stop later and make more accurate decisions at any stopping time. In principle, longer decision times and a broader continuation region might have opposite effects on overall expected accuracy. A broader $\mathcal{C}$ implies that the DM acts only on more extreme posteriors at any given time, but later decisions are (ceteris paribus) less accurate in strictly monotone decreasing environments (by Proposition \ref{prop:IAconvexproblemboundaries}). 
        However, we can show that the first effect always dominates: a more patient DM with lower information costs is always more accurate. %Indeed  Since the continuation region is larger and % (e.g., when $\frac{i(t)}{\zeta(t)}$ and $-c(t)$ (strictly) decrease with $t$). 

        Similarly, since $V^{re}$ is convex in $x$, Theorem \ref{theorem:compstatsvalues} implies that, conditional on the stopping time $\tau$, the DM makes more accurate choices when her learning speed $\sigma(x,t)=\frac{i(t)}{\zeta(t)} x(1-x)$ is higher. However, the impact of $\sigma$ on stopping times is ambiguous. Intuitively, a higher $\sigma$ has two countervailing effects: (i) it increases the value of waiting, expanding the $\mathcal{C}$, but (ii) it also reduces the time at which the belief process exits any given $\mathcal{C}$. 
        % Whether $\tau$ increases or decreases depends on which of the two effects dominates. 
        In general, the impact of increased $\sigma$ on $\tau$ is non-monotonic: $\tau$ converges to $0$ both when $\sigma$ goes to $\infty$ and when $\sigma$ goes to $0$. In the first case, the DM stops immediately because she learns the relevant parameter $\theta$ instantly. In the latter, she stops immediately because she knows she will never learn anything about $\theta$. Unfortunately, a more complete characterization of stopping times is unfeasible in our general framework. In a simpler time-stationary setting, \citet{goncalves2024speed} establishes that the relation between $\frac{i(t)}{\zeta(t)}$ and the expected $\tau$ is inverse U-shaped.
    \end{examplecont}

    \begin{example}[\textbf{Strategic Default}]\label{ex:bankruptcies1}
        Recent models of corporate bankruptcy, since \citet{leland1994corporate}, view the timing of bankruptcy as endogenously chosen by equityholders based on the firm's performance $x$, which follows a diffusion process. Following \cite{quah2013discounting}, we assume equityholders receive an increasing payout rate $\delta(x)$, but must pay a (weakly) decreasing coupon rate $c(x)$ to debtholders. Upon default, equityholders suffer a penalty $g(x)$, decreasing in $x$. The equityholders' problem is then
        \begin{align*}
            V(x) = \underset{\tau}{\sup }\, \mathbb{E}_x\left[ \int_0^{\tau} e^{-rt} \left(\delta(X_t) - c(X_t) \right)\mathrm{d}t + e^{-r\tau} g(X_{\tau}) \right].
        \end{align*}
        In \cite{leland1994corporate}, $\delta(x) = \delta x$, $c(x) =c$, $g(x) =0$, and $X$ follows a stationary geometric Brownian motion with constant drift $\mu$ and volatility $\sigma^2$. Thus, equityholders default if and only if $X_t \leq \kappa/(\kappa+1) (1-\mu/r) c/\delta$, where $\kappa = \left( \mu - \sigma^2/2 + \sqrt{\mu - \sigma^2/2 + 2r\sigma^2}\right)/\sigma^2$ and comparative statics can be read directly from the formula. Equityholders default for higher $x$ when $r$ and $c$ are higher, or $\delta$ and $\mu$ are lower. 
        
        While providing a good first approximation, this stationary model cannot capture empirically relevant patterns like mean reversion of the cash-flow process \citep{sarkar2003trade} or the negative relation between cash-flow volatility and level \citep{froot1993risk}.\footnote{This point is already made in \citet{manso2010performance}.}
        % However, as argued by \cite{manso2010performance}, more general processes are needed to capture empirically relevant patterns like mean reversion of the cash-flow process \citep{sarkar2003trade} or the negative relation between cash-flow volatility and level \citep{froot1993risk}. 
        Our results show that cutoff strategies remain optimal in nonstationary environments and provide comparative statics for $\delta,$ $ c$, and $\mu$ when $X$ is a general nonstationary diffusion (in addition to those for $r$ obtained by \cite{quah2013discounting}).\footnote{When $g$ is negative, the problem does not fit our assumptions. However, all our results extend if we consider $V-g$ instead.} If $\delta(t, \cdot ) - c(t, \cdot)$ is nondecreasing in $x$, Proposition \ref{prop:convextsectionsstopping} implies that $\tau^* = \inf\left\{ t\geq 0 \, : \, X_t \leq \underline{b}(t) \right\}$ and V is nondecreasing in $x$.
        \begin{proposition}\label{prop:Vmonotonictywrtx}
            If the flow and stopping payoffs ($f$ and $g$) are nondecreasing (nonincreasing) in $x$, then the value function $V$ is nondecreasing (nonincreasing) in $x$.
        \end{proposition}
        By Theorems \ref{theorem:compstatsstoppingtimes} and \ref{theorem:compstatsvalues}, $\underline{b}(\cdot)$ decreases with $\delta$ and $\mu$ and increases with $c$. Consequently, $\tau$ increases with $\delta$ and $\mu$ and decreases with $c$. When the benefit of operating ($\delta - c$) is larger, the opportunity cost of stopping is larger, and equityholders wait longer before defaulting. Similarly, when the firm's potential ($\mu$) is larger, the option value is larger, and equityholders are more willing to stick with the firm even in troubled times (lower $x$). Our results confirm the robustness of this intuition. The comparative statics holds for any nondecreasing $\delta(t, \cdot) - c(t, \cdot)$, nondecreasing penalty $g(\cdot)$, and any diffusion process.
    \end{example}

    Finally, although Theorems \ref{theorem:compstatsstoppingtimes} and \ref{theorem:compstatsvalues} take the stopping reward $g$ to be fixed, we can still obtain comparative statics with respect to the stopping payoff $g$. Both Theorems apply to $V - g$ as long as $g(x)$ is regular enough (e.g., $g\in \mathcal{C}^{2,4}([0,T]\times \mathcal{X})$), since
    \begin{align*}
        V(t,x) - g(x) & = \underset{\tau \in \mathcal{T}(t)}{\sup } \, \mathbb{E}_{(t,x)}\bigg[ \int_t^{\tau} e^{-r(s-t)} \bigg( f(s,X_s) + \mu(s, X_s) g_x(X_s) \\
        & \qquad + \frac{\sigma^2}{2}(s, X_s) g_{xx}(X_s) -r(s,X_s) g(X_s) \bigg)ds\bigg],
    \end{align*}
    as a result of It\^{o}'s lemma and the optional sampling theorem. The example below illustrates this observation.

    \begin{examplecont}{\ref{ex:bankruptcies1}}[\textbf{Strategic Default continued}]\label{ex:bankruptcies2}
        $V(t,x) - g(x)$ solves the stopping problem when the stopping payoff is zero and the flow payoff is $\delta(t,x) - c(t, x) + \mu(t, x) g_x(x) + \frac{\sigma^2}{2}(t, x) g_{xx}(x) -r g(x)$. If $g(x) = G \leq 0$ for all $x$ (i.e., the equityholders pay a fixed penalty for defaulting), the optimal default threshold $\underline{b}(t)$ is increasing in $G$. Therefore, $\tau^*$ is decreasing with $G$: The equityholders default earlier when the penalty is smaller and wait longer when the penalty is more severe.
    \end{examplecont}

\section{Applications and Examples}\label{sec:applications}
Theorems \ref{theorem:localmonotonestopping} and \ref{theorem:compstatsvalues}, and Corollary \ref{corollary:globalmonotonestopping} make assumptions about the value function, an endogenous object. Nonetheless, they are valuable tools because, in many stopping problems, one can verify from first principles that the value function is (strictly) monotone and satisfies \eqref{eq:IOV} or \eqref{eq:DOV}. This section shows these assumptions are met in two prominent classes of problems in economics and finance. As a result, our analysis delivers novel predictions in several applications touching upon information acquisition, irreversible investment, and option princing.

    % In so doing, we also develop tools to check whether an optimal stopping problem is locally (strictly) monotone. These same methods can be similarly employed to verify whether the value function satisfies the assumptions of Theorems \ref{theorem:localmonotonestopping} and \ref{theorem:compstatsstoppingtimes} in other environments. 
	
 % \paragraph{} To illustrate the power of our main results, we show how they deliver novel predictions in various applications.	
 % In Section \ref{subsec:learningnonbinary}, we focus on information acquisition problems where, unlike the classical models, the relevant state is not binary. We show that when the state is drawn from any non-binary distribution, (i) the problem is time-nonstationary, and (ii) decision accuracy always decreases over time. This extends the results of the seminal paper by \cite{fudenberg2018speed}, which established these findings for normally distributed states and linear payoffs. In Section \ref{subsec:deadline}, we consider information acquisition problems where abrupt changes may occur at stochastic times. For example, we examine the possibility that the DM perfectly learns the state or is forced to take action at a stochastic deadline.

    \subsection{Martingale Convex Stopping Problems }\label{subsec:Visconvex}
    The key challenge in applying our main results is verifying \eqref{eq:IOV} (or \eqref{eq:DOV}). However, this task is simpler in \textit{martingale convex stopping problems}.
    
    % if: (i) the the value function is convex in $ x $, (ii) the absence of drift, and (iii) the constant discount rate.

    %Under standard assumptions on the primitives, establishing the strict monotonicity of the environment then boils down to signing $V_t(t,x)$. %(Corollary \ref{corollary:convexstoppingproblemmonotone} below).
    % First, we formally define convex environments as follows.
    
	\begin{definition}\label{def:convexenvironment} 
		We say that problem \eqref{eq:valuefunction} is martingale convex (MC-problem) if 
        \begin{itemize}[topsep=3.5pt,itemsep=3pt]
            \item[(1)] $f$ and $g$ are convex in $x$, with $g$ strictly convex at $x^c$,
            \item[(2)] $X$ is a pure diffusion process, i.e., $\mu(t,x) =0$,
            \item[(3)] the discount rate is constant $r(t,x) =r \geq  0$ and, if $r=0$, there exists $\epsilon >0$ and $\bar{t} \in [0,\infty)$ such that $f(t,x)<-\epsilon$ for all $t\geq \bar{t}$ and $x \in \mathcal{X}$.\footnote{If $T<\infty$, condition (ii) holds vacuously.}
        \end{itemize}
    \end{definition}
\noindent Note that most information acquisition problems, including our running example, are convex. When convex, problem \eqref{eq:valuefunction} simplifies to 
    \begin{align*}
         V(t,x) =\,\,\, & \underset{\tau}{\sup } \, \mathbb{E}_{(t,x)}\left[ \int_{t}^{\tau} e^{-rs} f(s, X_s)ds + e^{-r\tau} g\left(X_{\tau}\right) \right], \tag{$\mathcal{V}$} 
    \\
	& \text{subject to } \,\,\,\,\,\,\,\,	X_{t+s} = X_t + \int_{t}^s \sigma(u,X_u) \, \mathrm{d}B_u.
	\end{align*}

    \begin{proposition}\label{prop:Visconvex}
        In an MC-problem, the value function $V$ is convex in $x$. % for all $t \in [0,T)$. 
	\end{proposition}
    \noindent The proof of Proposition \ref{prop:Visconvex} uses a discrete-time approximation argument and a key result on the propagation of convexity by \cite{bergman1996general}. The convexity of $V$ greatly simplifies the analysis. Verifying the environment's (strict) monotonicity boils down to signing $V_t$: \eqref{eq:IOV} holds if $V_t>0$, $\sigma_t\geq 0$, and $f_t\geq 0$; \eqref{eq:DOV} holds if the inequalities are reversed. Moreover, the sign of $V_t(t,x)$ matches those of $\sigma_t(t,x)$ and $f_t(t,x)$ (when the two agree). So, an MC-problem is strictly increasing if $\sigma_t\geq 0$ and $f_t\geq 0$, with at least one strict inequality (and decreasing if the inequalities are reversed). 
    %So, establishing the strict monotonicity of the environment boils down to signing $V_t(t,x)$. 
    % Moreover, since we can show that $V_t(t,x)>0$ if $\sigma(t,x)\geq 0$ and $f(t,x)\geq 0$, with one inequality being strict (and $V_t(t,x)<0$ if the inequalities are reversed), we can conclude that an MC-problem is strictly monotone increasing if $\sigma_t(t,x)\geq 0$ and $f_t(t,x)\geq 0$, with one inequality being strict (and decreasing if the inequalities are reversed).
Combining this logic with Corollary \ref{corollary:globalmonotonestopping}, we establish the following proposition (which proof is in Appendix \ref{app:Visconvex}).
\begin{proposition}\label{prop:convexproblemboundaries}
     %and that stopping boundaries are in the interior of $\mathcal{X}$.
    The optimal continuation region of an MC-problem such that C\ref{condition:singlecrossing} holds is $\mathcal{C} = \left\{ \left(t, x\right) \in \mathbb{R}_+ \times \mathcal{X} \, : \,  x \in \left( \underline{b}(t), \bar{b}(t) \right) \right\}$. %Moreover, if  $\bar{b}(t),\underline{b}(t)\in \mathrm{int}(\mathcal{X}) $
		% \begin{itemize}[topsep=3.5pt,itemsep=3pt]
		% 	\item $t\to \bar{b}(t)$ is \textbf{(strictly) decreasing} and continuous and $t\to \underline{b}(t) \in \mathcal{X}$ is \textbf{increasing} and continuous on $[0,\infty)$ if $t \to \sigma(t,x)$ and $t \to f(t,x)$ are nonincreasing (with at least one being strictly decreasing); and 
		% 	\item $t \to \bar{b}(t) \in \mathcal{X}$ is \textbf{(strictly) increasing} and $t\to \underline{b}(t)\in \mathcal{X}$ is \textbf{(strictly) decreasing} on $[0,\infty)$ if $t \to \sigma(t,x)$ and $t \to f(t,x)$ are nondecreasing (with at least one being strictly increasing), and $T =\infty$.\footnote{If $T<\infty$, the result would also hold, provided that the payoff of the DM at the boundary is large enough.}
		% \end{itemize}
  		\begin{itemize}[topsep=3.5pt,itemsep=3pt]
			\item If $\sigma_t\leq 0$ and $f_t\leq 0$,  then$t\to \bar{b}(t)$ is \textbf{nonincreasing} and $t\to \underline{b}(t)$ is \textbf{nondecreasing} on $[0,\infty)$.    
  			\item If $\sigma_t\geq 0$ and $f_t\geq 0$, and $T =\infty$, then $t \to \bar{b}(t) $ is \textbf{nondecreasing} and $t\to \underline{b}(t)$ is \textbf{nonincreasing} on $[0,\infty)$.\footnote{If $T<\infty$, the result holds when the DM's payoff at the boundary is large enough.} 
		\end{itemize}
    Moreover if we also have that, for all $(t,x)\in T\times \mathcal{X}$ either $\sigma_t\neq0$ or $f_t\neq0$, then $\bar{b}$ and $\underline{b}$ are strictly monotone on the interior of the codomain.\footnote{Note that Proposition \ref{prop:convexproblemboundaries} does not rule out cases where the DM never stops.}
  % \begin{itemize}[topsep=3.5pt,itemsep=3pt]
		% 	\item $\bar{b}(t)$ is \textbf{(strictly) decreasing} and continuous and ${b}(t) \in \mathcal{X}$ is \textbf{(strictly) increasing} and continuous on $[0,\infty)$ if $\sigma_t(t,x)\geq 0$ and $f_t(t,x)\geq 0$ (with at least one strict inequality); and 
		% 	\item  $ \bar{b}(t)$ is \textbf{(strictly) increasing} and $\underline{b}(t)\in \mathcal{X}$ is \textbf{(strictly) decreasing} on $[0,\infty)$ if $\sigma_t(t,x)\geq 0$ and $f_t(t,x)\geq 0$ (with at least one strict inequality), and $T =\infty$.\footnote{If $T<\infty$, the result would also hold, provided that the payoff of the DM at the boundary is large enough.}
		% \end{itemize}
	\end{proposition}
	% The proof of Proposition \ref{prop:convexproblemmonotonicity} is in Appendix \ref{app:Visconvex}.\footnote{Note that Proposition \ref{prop:convexproblemboundaries} does not rule out cases where the DM never stops.} %Without loss, we can assume $\Bar{b}$ ($\underline{b}$) is strictly decreasing (increasing) in $t$ if $\Bar{b}(t)\in \mathbb{R}\setminus \mathcal X$.   } 
\noindent Thus, the continuation region shrinks over time when the flow payoff $f(t,x)$ decreases in $t$ and the state dynamics slow down, i.e., $ \sigma(t,x) $ decreases in $ t $; strictly so if $ f $ or $\sigma$ strictly decreases in $ t $ and the DM stops for some $x\in \mathcal{X} $.

     Finally, note that Theorems \ref{theorem:compstatsstoppingtimes} and \ref{theorem:compstatsvalues} apply to MC-problems to deliver comparative statics with respect to the features of the environment.

% \subsection{Information Acquisition Problems}
\subsubsection{Learning with Non-Binary Priors}\label{subsec:learningnonbinary}
% Our first application focuses on the speed-accuracy tradeoff in information acquisition problems whose only departure from the classical framework \`a la Wald is that the DM may learn about a relevant parameter $\theta$ that is non binary.

% We consider same setting as in Section \ref{sec:RE} with two differences: (i) we assume that the signal process and flow costs are constant, as in the classical framework \`a la Wald ($i(t)=i,\zeta(t)=\zeta$, and $c(t)=c$ for all $t$); and (ii) unlike the classical framework, the prior the relevant parameter $\theta$ is allowed to be non binary.  
% Our first application analyzes the speed-accuracy tradeoff emerging in standard information acquisition problems when the parameter to be learned, $\theta$, is non-binary. Recall (e.g., from Section \ref{sec:RE2}) that in the classical framework—with by binary $\theta$ and constant flow cost $c$, discount rate $r$, signal intensity $i$, and noise $\zeta$— accuracy is constant over time. However, this result stems from the time stationarity of the learning process, a knife-edge feature of the binary-prior setting. With non-binary priors, the learning speed, and thus the information acquisition problem, becomes time-nonstationary.
Our first application analyzes the speed-accuracy tradeoff emerging in standard information acquisition problems when the parameter to be learned, $\theta$, is non-binary. Recall (e.g., example \ref{ex:informationacquisition2}) that accuracy is constant over time in the classical setting—characterized by binary $\theta$ and constant $c$, $r$, $i$, and $\zeta$. However, this result stems from the time stationarity of the learning process, a knife-edge feature of the binary-prior setting. With non-binary priors, the learning speed, and thus the information acquisition problem, becomes time-nonstationary.

In a seminal paper, \cite{fudenberg2018speed} abandon the binary prior assumption. They consider a DM choosing between two alternatives $\{l,r\}$ whose (unknown) payoff difference, $\theta=\left( \theta^l- \theta^r \right)\sim \mathcal N (\theta_0,\sigma_0)$, is learned through a Brownian signal with drift $\left( \theta^l- \theta^r \right)$. In this Gaussian setting, the authors establish a negative correlation between decisions' timing and accuracy. However, accuracy (i.e., the probability of making the correct choice) is equally affected by all errors, irrespective of their size $\left|\theta^l - \theta^r\right|$. Thus, two forces contribute to decreasing accuracy in their setting: (i) a \textit{learning effect}, where the DM updates her beliefs less as she accumulates more information, and (ii) a \textit{relevance effect}, where the DM expects a smaller payoff difference $\left|\theta^l - \theta^r\right|$, and thus a lower benefit from making the correct choice, as time progresses.
%since accuracy is defined as the probability of making the correct decision (choosing $r$ when $\theta^r \geq \theta^l$ and $l$ otherwise), capturing solely the frequency of errors by the DM and ignoring their magnitude. Every error affects accuracy equally, regardless of its size $\left|\theta^l - \theta^r\right|$.  Consequently, two forces contribute to their decreasing accuracy result: (i) a \textit{learning effect}, i.e., the fact that DM updates her beliefs less as she accumulates more and more information, and (ii) a \textit{relevance effect}, i.e., the fact that, as time progresses, the DM expects a smaller payoff difference $\left|\theta^l - \theta^r\right|$ and, thus, benefits less from making the correct choice.
Both forces appear essential in their analysis.

% To isolate the learning effect from the relevance effect, we consider an alternative model where the payoff difference, $\theta=\left| \theta^l- \theta^r \right|$ is constant and known, effectively shutting down the relevance effect. This more conservative model allows us to demonstrate that the negative correlation between decision timing and accuracy persists in this setting, and, moreover, holds for \textbf{any} non-binary prior. This finding complements the results established in \cite{fudenberg2018speed}.

% INTRO? To focus on the learning effect, we consider an alternative model that neutralizes the relevance effect by assuming that the payoff difference, $\left| \theta^l- \theta^r \right|$, is constant and known while keeping the learning process non-binary. We show that the negative correlation between decision timing and accuracy persists even in this more conservative model and extends to \textbf{any} non-binary prior (not only Gaussian). This finding complements the results established in \cite{fudenberg2018speed}.

To isolate the learning effect, we adopt an alternative model that retains a non-binary learning process (e.g.,  $\theta=\left( \theta^l- \theta^r \right)\sim \mathcal N (\theta_0,\sigma_0)$) but neutralizes the relevance effect by assuming the payoff difference only depends on the sign of $\theta$, i.e., whether $l$ is preferable to $r$ and vice versa. This captures, for example, settings where the DM's rewards/punishments for making the correct/incorrect decision are constant.
 By showing that the negative correlation between decision timing and accuracy persists even in this more conservative model and extends to \textbf{any} non-binary prior (not only Gaussian), we complement the results of \cite{fudenberg2018speed}.

\textbf{Setting.} Our setting parallels the classical (time-stationary) information acquisition model of Example \ref{ex:informationacquisition1}, where flow cost $c$, discount rate $r$, signal intensity $i$, and noise $\zeta$ are constant, and the value of choosing $d \in \{-1,1\}$ hinges on the sign of $\theta$. The only difference is that $\theta$ is a real-valued random variable with distribution $F \in \Delta(\mathbb{R})$ that may not be binary. 
 %To avoid triviality, we assume that $0<F(0) <1$. 
        The DM learns about $\theta$ by observing the Brownian signal 
 %    \textcolor{purple}{Our certain difference drift-diffusion model parallel the classical Wald hypothesis testing problem with one crucial difference: while the value of choosing an alternative is either $0$ or $1$ depending on the sign of the true state $\theta$, we do not assume that the unknown parameter $\theta$ is binary. Instead, in our model, $\theta$ is a real-valued random variable distributed according to $F \in \Delta(\mathbb{R})$. The DM observes a noisy Brownian signal }
		$Z_t = \theta \,t+ \int_0^t \zeta \, \mathrm{d} B_u,$
	and chooses $\tau$ and $d $ to maximize her expected  gain:\setlength{\abovedisplayskip}{4pt}
\setlength{\belowdisplayskip}{4pt} 
	\begin{align*}
		\mathbb{E} \left[ e^{-r\tau}\left( a \mathbbm{1}_{\{ d=1, \theta \geq 0\}} + b \mathbbm{1}_{\{ d=-1, \theta <0\}} \right)- \int_{0}^{\tau} e^{-rt}c \, \mathrm{d}t \right],
	\end{align*}
	where $a,b\geq0$, $c>0$, and $r\geq 0$. From the same filtering argument as in \cite{ekstrom2015bayesian}, the problem admits the following equivalent formulation:
	\begin{align}\label{eq:nonbinarylearning}
		V^{nb}(t,x) = \underset{\tau\geq t}{\sup } \, \mathbb{E}_{(t,x)}\left[ e^{-r(\tau-t)} \left( aX_{\tau} \vee b(1-X_{\tau}) \right) - \int_{t}^{\tau}e^{-r(s-t)} c  \, \mathrm{d}s\right], \tag{$\mathcal{V}^{nb}$}
	\end{align}
	  \setlength{\abovedisplayskip}{0pt}
\setlength{\belowdisplayskip}{4pt} 
	\begin{align*}
		\text{subject to } \,\,\,\,\,\,\,\, X_{t+s} = X_t + \int_{t}^{s} \sigma\left( u, X_u\right) \, \mathrm{d}B_u,
	\end{align*}
	where $X_t$ is the DM’s time $t$ posterior that $\theta\geq 0$, and $\sigma(t,x) \in (0,1)$ is the volatility of her posteriors, i.e., her learning speed. 

Since the problem is convex, $ r > 0 $, and $ g(x) $ is the maximum of two linear functions, then C\ref{condition:singlecrossing} holds, and the DM always stops in $\mathcal{X} = (0,1)$. Thus, we can leverage Proposition \ref{prop:convexproblemboundaries} to prove that when $\sigma$ is strictly decreasing, the continuation region is bounded by an upper limit $\overline{\pi}(t)$ that decreases and a lower limit $\underline{\pi}(t)$ that increases over time (see Fig.\ref{NSR}). 
% $V^{nb}$ is continuous and differentiable, and A\ref{assumption:singlecrossing} holds, we can easily show that the DM always stops in the interior of $\mathcal{X}=[0,1]$ and that within the continuation region, $V_t^{nb}(t,x)<0$ if $\sigma_t(x,t)<0$. Thus, we can leverage Theorem \ref{theorem:localmonotonestopping} and Corollary \ref{corollary:globalmonotonestopping} to prove that when $\sigma$ is strictly decreasing, the continuation region is bounded by an upper limit $\overline{\pi}(t)$ that decreases and a lower limit $\underline{\pi}$ that increases over time (see Figure \ref{NSR}.1).
This is particularly relevant since,  following \cite{ekstrom2015bayesian}, we can establish that DM's learning speed $\sigma$ is strictly decreasing in $t$ whenever the prior on $\theta$ is not binary, i.e., $\left| \mathrm{supp}(F)\right| >2$ (see Lemma \ref{lemma:propertiesofsigma} in the appendix). Thus, exploiting the direct link between stopping boundaries and decision accuracy, 
$P(1,t)=\overline{\pi}(t)$ and $P(-1,t)=1-\underline{\pi}(t)$ , we prove the following.\footnote{Proposition \ref{proposition:nonbinarylearning} is a special case of Proposition \ref{prop:convexproblemboundaries}, which proof is in Appendix \ref{app:Visconvex}.}

%  Moreover, following \cite{ekstrom2015bayesian}, we can establish that the DM's learning speed $\sigma$ strictly decreased in $t$ whenever the prior on $\theta$ is not binary, i.e., whenever $F$ contains more than 2 elements (see Lemma \ref{lemma:propertiesofsigma} in the appendix).
%   \setlength{\abovedisplayskip}{4pt}
% \setlength{\belowdisplayskip}{4pt} 

 % \footnote{Lemma \ref{lemma:propertiesofsigma} follows immediately from the proof of Proposition 3.8 and Corollary 3.10 in \cite{ekstrom2015bayesian}.}
 % \begin{lemma}\label{lemma:propertiesofsigma}
	%    % (i) $\sigma \in \mathcal{C}^{1,2,\alpha}$. (ii) For all $x \in (0,1)$, $t\to \sigma(t,x)$ is strictly decreasing if the support of $F$ contains more than 2 elements.
 %      The learning speed $\sigma$ is in  $\mathcal{C}^{1,2,\alpha}$. Moreover, for all $x \in (0,1)$, $t\to \sigma(t,x)$ is strictly decreasing if the support of $F$ contains more than 2 elements.
	% \end{lemma}
	% Lemma \ref{lemma:propertiesofsigma} follows immediately from the proof of Proposition 3.8 and Corollary 3.10 in \cite{ekstrom2015bayesian}. 
 % This allows us to prove that accuracy decreases over time in the \textit{certain difference} drift-diffusion model presented below.
 
	% Thus, using Lemma \ref{lemma:propertiesofsigma} (which also implies that $\sigma$ is bounded above by a constant), Proposition \ref{prop:convexproblemmonotonicity} allows us to prove that accuracy decreases over time. % in the \textit{certain difference} drift-diffusion model.
 
	\begin{proposition}\label{proposition:nonbinarylearning}
		 The information acquisition problem \eqref{eq:nonbinarylearning} displays decreasing accuracy whenever $F$ is not binary.
	\end{proposition}

% 	\textcolor{red}{Comment on always decreasing accuracy for non-binary distributions.}
	%Note that Proposition \ref{proposition:nonbinarylearning} is not tied to one particular prior over $\theta$. Accuracy is decreasing in the decision time independently of the DM's prior $F$ as long as it is not binary. In this sense, the predictions obtained in the classical Wald model (with binary prior) are not robust.
    Proposition \ref{proposition:nonbinarylearning} is not tied to one particular prior. It proves that decisions' timing and accuracy are strictly negatively correlated for any non-binary prior $F$. In this sense, our proposition shows that the classical prediction in Wald-like models that decision accuracy is constant over time, which arises when the prior is binary, is severely not robust.

 \vspace{-1.5em}
\begin{figure}[H]
\flushleft
\begin{center}
\begin{tikzpicture}
    \node at (7,0) {\includegraphics[scale=0.6]{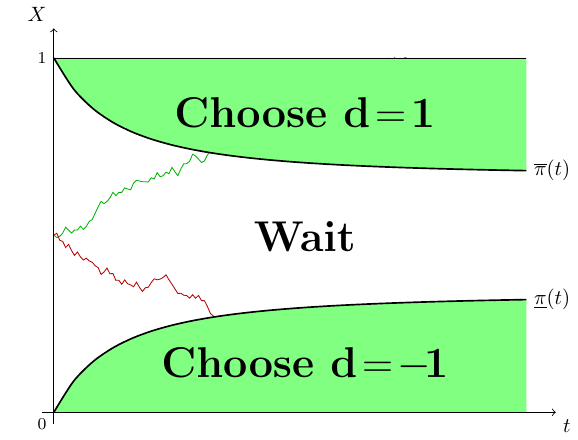}};
    \node at (0,0) {\includegraphics[scale=0.6]{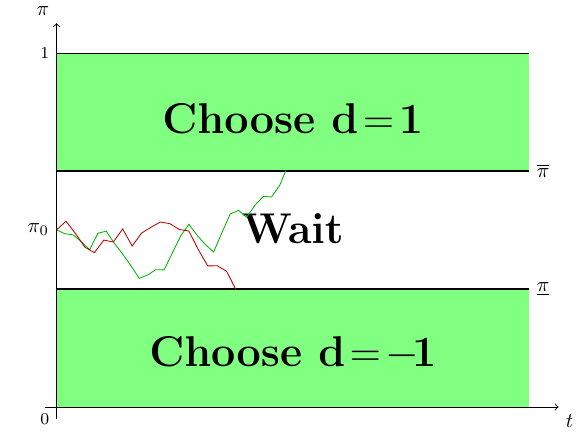}};
\end{tikzpicture}
\end{center}
\caption{ \textit {On the left, the classical information acquisition problem with binary prior: the continuation region in the belief space is constant over time, implying constant decision accuracy. On the right, the same problem, but with any non-binary prior: continuation region shrinks over time, implying decreasing decision accuracy.}}
\label{NSR}
\end{figure}

    \subsubsection{Gradual and Abrupt Learning, and Stochastic Deadlines}\label{subsec:deadeventsline}

    % \textcolor{red}{Transform to talk about Poisson arrivals and show that we can accommodate intergro-differential equations: Take the actual value function in the integral.}
    
% This section analyzes information acquisition problems that are subject to abrupt changes at stochastic times. These include instances where information arrives both gradually and abruptly, as well as scenarios where the DM risks being forced to make a decision or losing her chance. %For simplicity, we assume that the abrupt information is resolutive (this assumption can be relaxed; see Appendix for extensions).
   This section analyzes the effect of abrupt changes in information acquisition problems. These include instances where information arrives both gradually and abruptly, as well as scenarii where the DM may be forced to decide or lose the opportunity to at a stochastic time.

\textbf{Setting:} the DM's problem mirrors the classical time-stationary framework of Example \ref{ex:informationacquisition1} with constant $r$, $i$, $\zeta$, and (for simplicity) $c=0$. The only difference is that at any moment the DM opts to wait, the process may exogenously terminate, leaving the DM with a payoff that may depend on her posterior. This termination time, denoted by $\delta$, is governed by a time-nonstationary arrival rate, $\alpha(t, X_t)$, which itself may depend on the DM’s current posterior $X_t$ about the relevant state $\mu \in \{-1, 1\}$. We define $\gamma: [0,1] \to \mathbb{R}$ as the function mapping the DM's posterior at $\delta$ to her termination payoff $\gamma(X_{\delta})$.  
% Notably, $\gamma(X_{\delta})$ can also capture, in a reduced-form fashion, the expected value for the DM of a "new" decision problem triggered by $\delta$ (or, in the case of gradual and abrupt learning, the value of the same problem after abrupt information is revealed) 
% In this paper, we focus on Markov stochastic arrival times, where terimi arises at a time-nonstationary arrival rate $a(t, X_t)$ that may also depend on the current state $X_t$. 
% is abruptly revealed at a time-nonstationary arrival rate $a(t, X_t)$ that may also depend on the current posterior $X_t$.\footnote{This reflects, e.g., the possibility that the DM may adjust the intensity of her search for new (resolutive) information based on her current belief, $X_t$. } Since waiting is costly, the DM immediately takes the most favorable action $d=\mu$ when $\mu$ is revealed. The DM's problem then consists of finding decision time $\tau$ and rule $d \in \{-1,1\}$ to maximize the expected total gain. 
The DM's problem thus consists of finding a decision time $\tau$ and a rule $d \in {-1,1}$ to maximize the expected total gain: 
% \setlength{\abovedisplayskip}{4pt}
% \setlength{\belowdisplayskip}{4pt} 
 % \begin{align*}
	% 	\mathbb{E} \left[\mathbbm{1}_{\{ \delta  \geq \tau \}} e^{-r\tau}\left(a \mathbbm{1}_{\{ d=\mu =1\}} +  b \mathbbm{1}_{\{ d=\mu =-1\}} \right)  + \mathbbm{1}_{\{ \delta <\tau \}} e^{-r\delta}\left(a \mathbbm{1}_{\{\mu =1\}} +  b \mathbbm{1}_{\{  \mu =-1\}} \right) - \int_{0}^{\tau \wedge \delta} e^{-rt}c \mathrm{d}t\right]. 
	% \end{align*}
  % \begin{align*}
	$\mathbb{E} \left[\mathbbm{1}_{\{ \delta  \geq \tau \}} e^{-r\tau}\left(a \mathbbm{1}_{\{ d=\mu =1\}} +  b \mathbbm{1}_{\{ d=\mu =-1\}} \right)  + \mathbbm{1}_{\{ \delta <\tau \}} e^{-r\delta}\gamma(X_\delta) %- \int_{0}^{\tau \wedge \delta} e^{-rt}c \mathrm{d}t
  \right]$, where $a,b,r\geq0$. We can rewrite this as an optimal stopping problem:
    \vspace{0.3em}
  \begin{align}\label{eq:valuefunctiondeadline}
		V^{\delta}(t,x) = \underset{\tau\in \mathcal{T}}{\sup }\, \mathbb{E} \left[\mathbbm{1}_{\left\{ \tau \leq \delta \right\}} e^{-r\tau}g(X_\tau)  
  + \mathbbm{1}_{\left\{ \tau > \delta \right\}}e^{-r\delta}\gamma( X_{\delta}) %- \int_{0}^{\tau \wedge \delta} e^{-rt}c \mathrm{d}t
  \right], \tag{$\mathcal{V}^{\delta}$}
	\end{align}
 	\setlength{\abovedisplayskip}{0pt}
 \begin{align*}
		\text{subject to } \,\,\,\,\,\,\,\, X_{t+s} = X_t + \int_{t}^{s} \sigma\left(  X_u\right) \, \mathrm{d}B_u,
	\end{align*} 
	where $X_t$ is the DM’s time $t$ posterior that $\mu=1$, $\sigma(x) = \frac{i}{\zeta} x(1-x)$ is her learning speed, and $g(X_\tau)$ is her payoff from optimally selecting $d\in\{-1,1\}$ at the stopping time: $g(x):= \left( a x \vee b(1-x) \right)$ for all $x\in [0,1]$. %\footnote{    We also assume that the volatility of the process is smooth (i.e.,  $\sigma \in \mathcal{C}^{2,\alpha}$).}
    As we show in Appendix \ref{app:randomdeadlines}, we can reformulate this stopping problem to fit our framework. Corollary \ref{corollary:globalmonotonestopping} then yields the following.
    \begin{proposition}\label{prop:randomdeadlinesboundaries1}
		In the information acquisition problem \eqref{eq:valuefunctiondeadline}, the optimal continuation region is $\mathcal{C} = \left\{ \left(t, x\right) \in \mathbb{R}_+ \times \mathcal{X} \, : \,  x \in \left( \underline{b}(t), \bar{b}(t) \right) \right\}$. Moreover,
        \begin{itemize}
			\item $t\to \bar{b}(t)$ is \textbf{(strictly) decreasing} and continuous and $t\to \underline{b}(t)$ is \textbf{(strictly) increasing} and continuous on $\mathcal{X}$
            \begin{itemize}
                \item if $\alpha(t,x)$ is (strictly) decreasing in $t$ and $\gamma(x) \geq V^{\delta}(t,x)$ for all $t$; or %a \vee b$; or %\geq \max_{x\in [0,1]} g(x)$ ; or
                \item if $\alpha(t,x)$ is (strictly) increasing in $t$ and $\gamma(x) \leq V^{\delta}(t,x)$ for all $t$.
        \end{itemize}

   % $a(t,x)$ is (strictly) decreasing in $t$ and $\gamma(x) \geq V^{\delta}(t,x)$ for all $x\in \mathcal{X}, t\in T$, or if $a(t,x)$ is (strictly) increasing in $t$ and $\gamma(x) \leq V^{\delta}(t,x)$ for all $x\in \mathcal{X}, t\in T$; and
			\item $t\to \bar{b}(t)$ is \textbf{(strictly) increasing} and $t\to \underline{b}(t)$ is \textbf{(strictly) decreasing} on $\mathcal{X}$ 
      \begin{itemize}
              \item if $\alpha(t,x)$ is (strictly) increasing in $t$ and $\gamma(x) \geq V^{\delta}(t,x)$ for all $t$; or %a \vee b$; or %\geq \max_{x\in [0,1]} g(x)$ ; or
            \item if $\alpha(t,x)$ is (strictly) decreasing in $t$ and $\gamma(x) \leq V^{\delta}(t,x)$ for all $t$.
   \end{itemize}
   % if $a(t,x)$ is (strictly) increasing in $t$ and $\gamma(x) \geq V^{\delta}(t,x)$ for all $x\in \mathcal{X}, t\in T$, or if $a(t,x)$ is (strictly) decreasing in $t$ and $\gamma(x) \leq V^{\delta}(t,x)$ for all $x\in \mathcal{X}, t\in T$. 
		\end{itemize}     
% {  both when the deadline forces the DM to choose (\textit{forced decision}) or when it take away the DM's opportunity to choose and delivers zero payoff (\textit{forgone opportunity}).}

	\end{proposition}
    The proof of Proposition \ref{prop:randomdeadlinesboundaries1} is in Appendix \ref{app:randomdeadlines}.  This result shows that the shape of the continuation region, and thus the speed-accuracy tradeoff, hinges on whether $\delta$ is good news ($\gamma(x) \geq V^{\delta}(t,x)$) or bad news ($\gamma(x) \leq V^{\delta}(t,x)$) in settings where $a(t,x)$ is monotone in time. Although the statement imposes conditions directly on the value function to maintain generality, in many examples (including those below), the comparison between $\gamma$ and $V^{\delta}$ can be made based on the primitives, without computing $V^{\delta}(t,x)$ explicitly.

\begin{example}[\textbf{Forced Decision and Forgone Opportunity}]
    Suppose $\delta$ represents a stochastic deadline when the DM loses the opportunity to act or has to act immediately. For instance, the DM could be considering a particular investment and be preempted by a rival. In both cases, $\gamma (x)\leq g(x)\leq \inf_t V^\delta (t,x)$. Thus,  Proposition \ref{prop:randomdeadlinesboundaries1} implies that earlier decisions are more accurate if and only if the time pressure decreases over time (i.e., $\alpha(t,x)$ decreases in $t$).
\end{example}

\begin{example}[Gradual and Abrupt Learning]
    Suppose that $\delta$ represents the stochastic arrival of some abrupt information. In this setting, the DM learns about $\theta$ both gradually, through the usual Brownian signal $Z$, and abruptly, at a nonstationary Poisson arrival rate $\alpha(t, X_t)$ that may also depend on the current posterior $X_t$ (reflecting, e.g., that the DM adjusts the intensity of her search for new information based on her current belief). If the abrupt signal perfectly reveals $\theta$, the DM always makes the correct decision at $\delta$. Thus, $\gamma(x)=x(a \mathbbm{1}_{\{X_\delta =1\}}) +  (1-x) (b \mathbbm{1}_{\{ X_\delta =-1\}})\geq \sup_{t} V^{\delta}(t,x)$. Proposition \ref{prop:randomdeadlinesboundaries1} then implies that, apart from decisions prompted by $\delta$, which are always correct, earlier decisions are more accurate if and only if the abrupt signal's likelihood increases over time (i.e., $\alpha(t,x)$ increases in $t$). A similar logic applies when the abrupt information is not resolutive. In this case, $\gamma(x)$ represents the expected value of the DM's problem when her posterior jumps from $x$ to $X'$, with $\mathbbm{E}[X'|x]=x$. Intuitively, if the DM does not stop at $(t,x)$, it is because acquiring information is still valuable. Thus, in the continuation region, we still have $\gamma(x) \geq V^\delta(t,x)$.
\end{example} 
% %A2
% the only reason for the DM to wait in this problem is acquiring information thus 

% as long as, in expectation, the DM benefits from having more information 

% . In that case, %$\gamma(x)=\mathbbm{E}[V(t,s)|x]$
% . All we need to show is that the value function is convex and thus abrupt information would alwa 

    \subsection{Stopping of Arithmetic and Geometric Brownian Motions}\label{subsec:arithmeticandgeometricBM} 

    Arithmetic and geometric Brownian motion have been essential in the study of financial markets since Louis Bachelier formalized the concept of Brownian motion to model prices. They are central, for example, to the celebrated Black-Scholes model and modern option pricing theory \citep{black1973pricing, merton1973theory}, and both have been used to model investment opportunities. %\footnote{The proof of the results in this Section are in Appendix \ref{app:aBMandgBM}}.

    % As already noted at the beginning of Section \ref{subsec:Visconvex},  Theorems \ref{theorem:localmonotonestopping} and \ref{theorem:compstatsvalues}, and Corollary \ref{corollary:globalmonotonestopping} make assumptions on the value function: an endogenous object. 
% This section applies our results to a second popular class of problems: the optimal stopping of (non-stationary) arithmetic and geometric Brownian motion. These processes have been pivotal in analyzing financial markets since Louis Bachelier formalized Brownian motion to model prices. They are fundamental, for instance, to the renowned Black-Scholes model and modern option pricing theory \citep{black1973pricing, merton1973theory}.
    % The proof of the results in this Section are in Appendix \ref{app:aBMandgBM}.

    % The main challenge is, once again, to verify \eqref{eq:IOV} (or \eqref{eq:DOV}). 
% When the state evolves according to a (non-stationary) arithmetic or geometric Brownian motion, the conditions for \eqref{eq:IOV} or \eqref{eq:DOV} can be readily deduced from the primitives of the model. 
Formally, in this section, we consider the following special case of problem \eqref{eq:valuefunction}:
\begin{align*}\label{eq:valuefunctionAG}
     V^{\mathsmaller{BM}}(t,x) = \underset{\tau}{\sup } \, \mathbb{E}_{(t,x)}\left[ \int_{t}^{\tau} e^{-\int_t^s r(u)\, \mathrm{d}u} f(s, X_s)ds + e^{-\int_t^{\tau} r(u)\, \mathrm{d}u} g\left(X_{\tau}\right) \right], \tag{$\mathcal{V}^{\mathsmaller{BM}}$} 
\end{align*}
subject to 
\begin{align}\label{eq:aBM}
	X_{t+s} = X_t + \int_t^s \mu(u)\, \mathrm{d}u +  \int_{t}^s \sigma(u) \, \mathrm{d}B_u, \tag{aBM}
\end{align}
or
\begin{align*}\label{eq:gBM}
    X_{t+s} = X_t + \int_t^s X_u \mu(u)\, \mathrm{d}u +  \int_{t}^s X_u \sigma(u) \, \mathrm{d}B_u, \tag{gBM}
\end{align*}
where $\mu$, $\sigma$, and $r$ depend on time only. To ensure the problem is well-posed (i.e., \eqref{eq:uniformintegrability} holds), we assume $f$ and $g$ grow at most linearly, $\mu$ and $\sigma$ are bounded, $r$ is positive and uniformly bounded away from zero, and, in the case of geometric Brownian motion \eqref{eq:aBM}, there exists $\epsilon > 0$ such that $\mu(t) < r(t) - \epsilon$ for all $t$. %\textcolor{blue}{Finally we assume that $f_x$ and $g_x$ have the same sign.}

    % where the drift and variance of the arithmetic and geometric Brownian motions, as well as the discount rate $r$, depend only on time. To ensure that the problem is well-posed (i.e., \eqref{eq:uniformintegrability} holds), we also assume that $f$ and $g$ grow at most linearly, $\mu$ and $\sigma$ are bounded, $r$ is positive and uniformly bounded away from zero, and, in the case of geometric Brownian motion, there exists $\epsilon > 0$ such that $\mu(t) < r(t) - \epsilon$ for all $t$.

% , and the assumptions underlying Theorem \ref{theorem:localmonotonestopping}
In these settings, the problem's monotonicity is readily deduced from its primitives. Indeed, (i) by Proposition \ref{prop:Vmonotonictywrtx}, $V^{\mathsmaller{BM}}_x$ has the same sign as $g_x$ {and $f_x$ (when these two coincide)}, and (ii) $V^{\mathsmaller{BM}}$ is convex in $x$ when $f$ and $g$ are convex in $x$.\footnote{See Proposition \ref{prop:aBMandgBMconvexity} in Appendix \ref{app:aBMandgBM}.} 
  % \begin{proposition}\label{prop:aBMandgBMconvexity}
  %       If $f$ and $g$ are convex in $x$, then the value function associated with \eqref{eq:valuefunctionAG} is convex in $x$.
  %   \end{proposition}
% both arithmetic and geometric Brownian motions preserve convexity.
%     \begin{proposition}\label{prop:aBMandgBMconvexity}
%         If $f$ and $g$ are convex in $x$, then $V^{\mathsmaller{BM}}(t,x)$ is convex in $x$.
%     \end{proposition}
    % Moreover, Proposition \ref{prop:Vmonotonictywrtx} guarantees that the sign of $V_x$ is constant and equal to that of $g_x$. 
    So, when $g$ and $f$ are (weakly) convex in $x$, \eqref{eq:IOV} holds if $f_t(t,x), -r_t(t), \mu_t(t)g_x(x), \sigma_t(t) \geq 0$, and \eqref{eq:DOV} holds if $f_t(t,x), -r_t(t), \mu_t(t)g_x(x), \sigma_t(t) \leq 0$. Corollary \ref{corollary:globalmonotonestopping} then yields:
    \begin{corollary}\label{corollary:alignedproblemmonotonicity} 
        Let $T = \infty$. Suppose that C\ref{condition:singlecrossing} holds, and that $g$ and $f$ are (weakly) convex in $x$ and both nondecreasing or both nonincreasing, and. The optimal continuation region for \eqref{eq:valuefunctionAG} is $ \mathcal{C} = \left\{ \left(t, x\right) \in \mathbb{R}_+ \times \mathcal{X} \, : \,  x \in \left( \underline{b}(t), \bar{b}(t) \right) \right\}$. Moreover, 
		\begin{itemize}[topsep=3.5pt,itemsep=3pt]
			\item If $f_t(t,x), -r_t(t), \mu_t(t) g_x(x), \sigma_t(t) \geq 0$ (with at least one inequality strict), then $t\to \bar{b}(t)$ is \textbf{(strictly) decreasing} and $t\to \underline{b}(t)$ is \textbf{(strictly) increasing}. 
			\item If $f_t(t,x), -r_t(t), \mu_t(t)g_x(x), \sigma_t(t) \leq 0$ (with at least one inequality strict), $t\to \bar{b}(t) \in \mathcal{X}$ is \textbf{(strictly) increasing} and $t\to \underline{b}(t)$ is \textbf{(strictly) decreasing}. 
		\end{itemize}
    \end{corollary}

    \subsubsection{Application: Irreversible Investment}\label{subsec:irreversibleinvestment}
    % A canonical problem in economics has been that of the optimal timing of investment. Most classical models \citep{dixit1993art,dixit1994investment,ebert2020weighted} posits that the investor receives a payoff $X - I$ upon investing in a project, where the value of the project $X$ evolves according to a geometric Brownian motion with drift $\mu$ and variance $\sigma^2$, and $I$ is the investment cost. The investor thus solves the stopping problem:

One canonical economic problem fitting the above framework is that of irreversible investment \citep{dixit1993art,dixit1994investment,ebert2020weighted}. An investor chooses when (if at all) to invest in a project whose value $X$ evolves according to a geometric Brownian motion as in \eqref{eq:gBM}. The investor's payoff from investing is simply $X - I$, where $I\geq 0$ is the investment cost. The investor then solves the optimal stopping problem (which is well-posed as long as $\mu < r$):
    \begin{align*}
        \underset{\tau}{\sup} \, & \mathbb{E}\left[ e^{-r\tau}\left(X_{\tau} -I\right) \right] \,\, & \text{subject to: } X_{t+s} = X_t + \int_t^s X_u \mu(u)\, \mathrm{d}u +  \int_{t}^s X_u \sigma(u) \, \mathrm{d}B_u.
    \end{align*}
      % $\tau^* = \inf \left\{ t\geq 0 \, :\, X_{t} \geq b^* \right\}$. 
      
    \textbf{Classical benchmark.} Following \citet{dixit1993art}, the literature has focused on settings where $\mu$, $\sigma$, and $r$ are time-independent, showing that the optimal policy consists of a simple hitting time: the DM invests when the profitability $X$ hits $b^* = I /(1 -1/\kappa) $ for $\kappa = \left( \sqrt{\left(\mu -\frac{1}{2}\sigma^2\right)^2 + 2\sigma^2 r} -\left(\mu -\frac{1}{2}\sigma^2\right)\right)/{\sigma^2}$ (see, e.g., \cite{dixit1994investment}). Thus, in the classical time-stationary case, (i) profitability and timing of investment are uncorrelated, and (ii) conditional on investment, profitability must be higher when the investment cost is higher, the drift and variance of the geometric Brownian motion are higher, and the discount rate is lower.

    % Moreover, profitability and the timing of investment are uncorrelated.

    \textbf{Results.} %Our results allow us to generalize these insights to time-nonstationary (monotone) environments. %In particular, if $I(t)$, $\mu(t)$, and $\sigma(t)$ are decreasing in $t$,
    We study whether insights (i) and (ii) extend to general nonstationary (monotone) environments. While the optimal policy is still a simple hitting time $\tau^* = \inf \left\{ t \geq 0 \, :\, X_{t} \geq b(t) \right\}$, the profitability level $b(t)$ that triggers the investment is no longer constant. Applying Corollary \ref{corollary:alignedproblemmonotonicity} to $V^{\mathsmaller{BM}}(t,x) - (x - I(t))$ shows that $b(t)$ decreases (increases) in $t$ when $I(t)$, $\mu(t)$ and $\sigma(t)$ decrease (increases) in $t$.\footnote{$x - I(t)$ needs not be nonnegative. However, the investor can always obtain a payoff of $0$ by never stopping. So, the problem is equivalent to one with stopping payoff $(x-I(t))^+$, which fits our assumptions.} As a result, investment timing and profitability will be correlated, violating (i). For example, later investment will be less profitable when the investment cost $I(t)$ decreases in $t$ (e.g., during expansionary phases of monetary policy), when the profitability variance $\sigma(t)$ decreases in $t$ (e.g., as the market gradually stabilizes in the aftermath of a crisis), when the profitability drift $\mu(t)$ decreases (e.g., as competition increases).
    
    % Conversely, if $I(t)$, $\mu(t)$ and $\sigma(t)$ are increasing, the investment profitability threshold $b(t)$ is increasing over time. 
    On the other hand, Theorems \ref{theorem:compstatsstoppingtimes} and \ref{theorem:compstatsvalues} establish that the comparative statics in (ii) are robust and extend to nonstationary environments. 
    \begin{proposition}\label{prop:irreversibleinvestmentcompstats}
        $b(t)$ is increasing (pointwise) in $\mu$, $\sigma$, $-r$ and $I$. Moreover, the time to investment $\tau^*$ is decreasing in $r$ and increasing in $I$.
    \end{proposition}
    Proposition \ref{prop:irreversibleinvestmentcompstats} ensures that the monotonicity of the optimal investment threshold holds for general investment cost and nonstationary profitability processes, confirming our economic intuition. When the discount rate decreases, the investor is more willing to wait for future opportunities. Moreover, when the drift or volatility of the value of the project increases, the likelihood of better future opportunities increases. So, the wedge between the total and present investment value, i.e., the option value, increases. Hence, the investor waits for better opportunities to invest.

    \subsubsection{Application: Pricing and American Options}\label{subsec:americanput}

   % Modern finance theory \citep{jacka1991optimal,shiryaev1999essentials,peskir2006optimal} posits that the arbitrage-free price of the American put option with strike price $K$ and an infinite horizon (perpetual option) is given by
% \,\, & \text{subject to: } X_{t+s} = X_t + \int_t^s X_u \mu \, \mathrm{d}u +  \int_{t}^s X_u \sigma \, \mathrm{d}B_u,
 In finance, the arbitrage-free price of an American put option with strike price $K$ and an infinite horizon (perpetual option) is typically modeled as the value function of a time-stationary optimal stopping problem:
\begin{align}\label{Vao}
    V^{\mathsmaller{AO}}(t,x)=&\underset{\tau}{\sup} \, \mathbb{E}\left[ e^{-r\tau} \left(K - X_{\tau}\right)^+ \right] \tag{$\mathcal{V}^{\mathsmaller{AO}}$}\\ & \text{subject to: } X_{t+s} = X_t + \int_t^s X_u r \, \mathrm{d}u +  \int_{t}^s X_u \sigma \, \mathrm{d}B_u, \notag
\end{align}
where $r(t)$ is the interest rate, $X_t$ is the spot price of the underlying asset, and $\sigma(t)$ is the spot price volatility \citep{jacka1991optimal,shiryaev1999essentials,peskir2006optimal}. 
% where the market price $X$ of the underlying asset follows a stationary geometric Brownian motion with constant drift $r$ and volatility $\sigma$ \citep{jacka1991optimal,shiryaev1999essentials,peskir2006optimal}. 

\textbf{Classical benchmark.} Following \citet{jacka1991optimal}, the literature has focused on settings where $\sigma$ and $r$ are time-independent. In this stationary setting, the value function has a closed-form solution, and the DM exercises the option when $X_t \leq b^* = \frac{K}{1+ \frac{\sigma^2}{2r}}$ (see, e.g., Chapter VII in \citet{peskir2006optimal}).  Thus, options are exercised at higher spot prices $X_t$, when their strike price $K$ is higher, the interest rate is $r$ higher, or the spot price volatility $\sigma$ is lower. However, one may wonder whether these insights are robust, especially as the stationary formulation has been criticized empirically: Execution times and spot prices at exercise tend to be correlated \citep{carpenter1998exercise}, and the constant volatility assumption appears to lack explanatory power \citep{bakshi1997empirical,bates2022empirical}.
    
    % We then consider the American put option pricing problem when the interest rate $r$ and the volatility $\sigma$ are time-nonstationary. Proposition \ref{prop:americanput} confirms that relaxing the constant volatility assumption (and the constant interest rate assumption) can explain different correlation patterns between the exercise time and price
    % does not change the qualitative insights and
       \textbf{Results.} Proposition \ref{prop:americanput} confirms that relaxing the constant volatility assumption (and/or the constant interest rate assumption) when studying American put options can explain the documented correlation patterns between the execution times and spot prices at exercise, without affecting the classical comparative statics. This results is an immediate consequence of Corollary \ref{corollary:alignedproblemmonotonicity}, and Theorems \ref{theorem:compstatsstoppingtimes} and \ref{theorem:compstatsvalues}).  
    \begin{proposition}\label{prop:americanput}
        In \eqref{Vao}, the optimal stopping time is $\tau = \inf\left\{t \geq 0 \, : \, X_t \leq {b}(t)\right\}$. If $-r(t)$ and $\sigma(t)$ are increasing (decreasing) in $t$, then $b(t)$ is decreasing (increasing) in $t$. Finally, both $V^{\mathsmaller{BM}}(t,x)$ and $b(t)$ are decreasing (pointwise) in $r$ and $\sigma$.
    \end{proposition}
    While confirming the robustness of (ii), our results reveal a positive (negative) correlation between execution times and spot prices at exercise when interest rates or volatility increase (decrease) over time.
% It demonstrates that when the interest rate or the volatility increases, the price of the American Put Option goes down, and options are exerted at higher spot prices.

    \section{Extensions}\label{sec:extensions}

    \subsection{Time-dependent stopping reward}\label{subsec:timedependentg}

    In Remark \ref{remark:nonstationarystoppingpayoff} in Section \ref{subsec:CSEE}, we observed that Theorem \ref{theorem:compstatsstoppingtimes} and \ref{theorem:compstatsvalues} extend to nonstationary stopping payoffs. This turns out to be the case of Theorem \ref{theorem:localmonotonestopping} too, under one additional assumption.
    \begin{assumption}\label{assumption:nonstationaryg}
        \begin{itemize}[topsep=3.5pt,itemsep=3pt]
            \item[(i)] $g = g^1\vee g^2$, with ${g^i}: [0,T]\times \mathcal{X}[0,T) \times \mathcal{X} \to \mathbb{R}$, $i=1,2$, belongs to $\mathcal{C}^{1,2, \alpha}([0,T]\times \mathcal{X})$.
		    \item[(ii)] For all $t \in [0,T)$, $x \to g^1(t,x) - g^2(t,x)$ is strictly monotone and crosses zero at $x^c(t) \in \mathcal{X}$ and $x \to f(x,t) + \left( \partial_t +\mathcal{L}^{(t,x)} -r \right)g$ is nondecreasing on $(\underline{x},x^c(t)]$ and nonincreasing on on $[x^c(t), \bar{x})$. 
            \item[(iii)] When the stopping problem is strictly decreasing, $g^1$ is supermodular in $(t,x)$ and $g^2$ is submodular in $(t,x)$; and, when the stopping problem is strictly increasing, $g^1$ is submodular in $(t,x)$ and $g^2$ is supermodular in $(t,x)$.
		\end{itemize}
    \end{assumption}
    When this assumption is satisfied, the proof and result of Theorem \ref{theorem:localmonotonestopping} extend, as are, to nonstationary $g$. The proof of Theorem \ref{theorem:localmonotonestoppingnonstationaryg} is therefore omitted.
    \begin{theorem}\label{theorem:localmonotonestoppingnonstationaryg} 
	   Suppose Assumption \ref{assumption:nonstationaryg} holds. Then		
	   \begin{enumerate}[topsep=3.5pt,itemsep=3pt]
            \item If $V-g$ is nonincreasing in $t$ (and \eqref{eq:DOV} holds) on $\left( \underline{t}, \bar{t} \right)$, the optimal continuation region is \textbf{(strictly) decreasing} on $(\underline{t}, \bar{t})$.
    		% \item If the optimal stopping problem is \textbf{monotone increasing} on $(\underline{t}, \bar{t} )$, the optimal continuation region is nondecreasing on $(\underline{t}, \bar{t} )$. 
    
    		\item If $V-g$ is nondecreasing in $t$ (and \eqref{eq:IOV} holds) on $\left( \underline{t}, \bar{t} \right)$, the optimal continuation region is \textbf{(strictly) decreasing} on $(\underline{t}, \bar{t})$.
    	\end{enumerate}
    \end{theorem}
    
    \subsection{Optimal stopping of a controlled diffusion}\label{subsec:control}
    
    In the main section of the paper, we derived our comparative statics result for pure stopping problem. Our techniques extend to mixed stopping and control problems:
    \begin{align}\label{eq:valuefunctioncontrol}
        V^c(t,x) = \underset{A = \{A_s\}_{s\geq t} \in \mathcal{A}(t), \tau \in \mathcal{T}(t)}{\sup } & \, \mathbb{E}^A_{(t,x)}\bigg[ \int_t^\tau e^{-\int_t^s r(A_u, u, X_u)\, \mathrm{d}u }f(A_s, s,X_s) ds \notag \\
        & \qquad \qquad +  e^{-\int_t^{\tau} r(A_s, s, X_s)ds } g(X_{\tau})\bigg],
    \end{align}
    \setlength{\abovedisplayskip}{0pt}
    \begin{align}\label{eq:controlleddiffusion}
        \text{subject to } X_{t+s} = X_t + \int_t^{t+s} \mu(A_{u}, u, X_u) \, \mathrm{d}u +  \int_t^{t+s} \sigma(A_{u}, u, X_u) \, \mathrm{d}B_u 
    \end{align}
    \setlength{\abovedisplayskip}{4pt}
\setlength{\belowdisplayskip}{4pt} 
    As above, $\mathcal{T}(t)$ is the set of stopping time greater than $t$. $\mathcal{A}(t)$ is the set of admissible control, which we will take to be all controls $A$ taking value in some compact set $\mathbb{A}$ for which the controlled stochastic differential equation \eqref{eq:controlleddiffusion} has a weak solution. 
    
    To prove the pendent of Theorem \ref{theorem:localmonotonestopping} in that setting, we first need to define monotone environments for control and stopping problems. 
    \begin{definitionp}{\ref*{definition:monotoneenvironments}$'$}\label{definition:monotoneenvironmentscontrol}
        The optimal stopping and control problem \eqref{eq:valuefunction} is locally \textbf{monotone increasing} on $(\underline{t}, \bar{t})$ if $ V_t(t,x)\geq 0 $ on $(\underline{t}, \bar{t})$ for all $x\in \mathcal{X}$. It is locally \textbf{strictly monotone increasing} if, for all $(t,x) \in \mathcal{C}$ such that $t\in (\underline{t}, \bar{t})$ and all $a\in \mathbb{A}$, $V_t(t,x) >0$ and
		\begin{equation}\label{eq:controlledIOV}
			\sigma_t(a,t,x)V_{xx}(t,x)+\mu_t(a,t,x) V_x(t,x)-r_t(a,t,x) V(t,x) +f_t(a, t,x) \geq 0.\tag{cIOV}
        \end{equation}
        The optimal stopping and control problem \eqref{eq:valuefunction} is locally \textbf{monotone decreasing} on $(\underline{t}, \bar{t})$ if $ V_t(t,x)\geq 0 $ on $(\underline{t}, \bar{t})$ for all $x\in \mathcal{X}$. It is locally \textbf{strictly monotone decreasing} if, for all $(t,x) \in \mathcal{C}$ with $t\in (\underline{t}, \bar{t})$ and all $a\in \mathbb{A}$, $V_t(t,x) <0$ and
		\begin{align}\label{eq:controlledDOV}
		    \sigma_t(a, t,x) V_{xx}(t,x) + \mu_t(a, t,x) V_x(t,x)-r_t(a, t,x) V(t,x) + f_t(a, t,x) \leq 0. \tag{cDOV}
		\end{align}	
    \end{definitionp}
    Moreover, we also make the following additional assumption. It is not needed, but it simplifies the analysis: It guarantees the existence of a well-behaved optimal control, thus forgoing the need to work with approximately optimal control.
    \begin{assumption}\label{assumption:Lipschizselection}
        There exists a measurable selection $a:[0,T)\times \mathcal{X}\to A$ that maximizes the left hand side of \eqref{eq:HJBc} such that $\sigma(a(t,x),t,x)$ is Lipschitz continuous.
    \end{assumption}

    \begin{theoremp}{\ref*{theorem:localmonotonestopping}$'$}\label{theorem:localmonotonestoppingcontroldiffusion}
        Suppose that Assumption \ref{assumption:Lipschizselection} holds. Then		
        \begin{enumerate}[topsep=3.5pt,itemsep=3pt]
            \item If the optimal stopping and control problem is \textbf{(strictly) monotone decreasing} on $(\underline{t}, \bar{t})$, the optimal continuation region is \textbf{(strictly) decreasing} on $(\underline{t}, \bar{t})$.
            
    		\item If the optimal stopping and control problem is \textbf{(strictly) monotone increasing} on $(\underline{t}, \bar{t})$, the optimal continuation region is \textbf{(strictly) increasing} on $(\underline{t}, \bar{t})$.
    	\end{enumerate}
    \end{theoremp}
    The proof of Theorem \ref{theorem:localmonotonestoppingcontroldiffusion} is in Appendix \ref{app:proofoftheoremcontrolledstopping}. It follows the same steps as the proof of Theorem \ref{theorem:localmonotonestopping}. However, two complications arise. First, we have to ensure that we can ``take the time derivative'' of the \emph{nonlinear} HJB equation associated to a control problem. Second, the ``derivative equation'' may have irregular coefficients. To deal with these difficulties, we invoke the Envelope Theorem to ``take the derivative'' of the HJB equation. We then rely on a regularity results in \cite{durandard2023existence} to replace Theorem 3.5.10 in \cite{friedman2008partial}. Finally we replace the version of the Hopf's Boundary Lemma for classical solutions we used in the proof of Theorem \ref{theorem:localmonotonestopping} by a version that applies to strong solutions (Lemma 1.39 in \cite{wang2021nonlinear}). 

% \textcolor{red}{VAGUE BUT MAYBE AN INTERESTING HINT TO WHAT CAN BE DONE... STILL I WILL NOT RESIST IF YOU PREFER TAKING THIS OUT} 
% A classical reference for optimal stopping problems of controlled diffusion is \citet{moscarini2001optimal}, which studies a stopping problem \`a la Wald where the DM can select, at every instant of time $t$ its learning speed $\sigma_t$ at an experimentation cost cost $c(\sigma_t,x_t)$ that is increasing in the selected speed.\footnote{Technically, in \citet{moscarini2001optimal}, the DM selects the drift of the signal diffusion problem but, by standard arguments, we can reformulate their problem as optimal stopping control problem whose underlying process is a $[0,1]$-valued martingale on the DM's belief $x$ about the relevant binary parameter and the DM controls the volatility of this process.} 
% In this setting, our framework and comparative statics would allow to address, for example, settings where the cost of experimentation increases or decreases over time. 

    Similarly, we can extend Theorem \ref{theorem:compstatsstoppingtimes} and \ref{theorem:compstatsvalues} to obtain comparative statics for mixed stopping and control problems. As in Section \ref{subsec:CSEE}, define $V^{^{{\sigma}, \mu, f,r, g}}$, $\mathcal{C}^{^{{\sigma}, \mu, f,r, g}}$, and $\tau^{\mathsmaller{{\sigma}, \mu, f,r, g}}$ to be the value function, the continuation region, and the stopping time associated with the optimal control and stopping problem in (\ref{eq:valuefunctioncontrol}), parameterized by the \emph{functions} $\sigma$, $\mu$, $f$, $r$, and $g$. 

    \begin{theoremp}{\ref*{theorem:compstatsstoppingtimes}$'$}\label{theorem:compstatsstoppingtimescontroldiffusion}    
        Suppose A\ref{assumption:Lipschizselection} holds. If $\bar{r} \geq \underline{r}$ and $\bar{f} \geq \underline{f}$ for all $a \in \mathbb{A}$, then $V^{^{{\sigma}, \mu, \bar{f}, \underline{r}, g}} \geq V^{^{\sigma, \mu, \underline{f},\bar{r}, g}}$ and $\mathcal{C}^{^{\sigma, \mu, \underline{f},\bar{r}, g}} \subset \mathcal{C}^{^{{\sigma}, \mu, \bar{f}, \underline{r}, g}}$. 
    \end{theoremp}
    The ranking on stopping times in Theorem \ref{theorem:compstatsstoppingtimes} does not carry to fully general mixed control and stopping problems. A change in the flow payoff or discount factor may affect the optimal policy and, thus, the state process. However, Theorem \ref{theorem:compstatsstoppingtimescontroldiffusion} may still yield comparative statics on stopping times in specific applications.

    \begin{theoremp}{\ref*{theorem:compstatsvalues}$'$}\label{theorem:compstatsvaluescontroldiffusion}
        Suppose A\ref{assumption:Lipschizselection} holds.
        
        \vspace{-0.5em}   
        \begin{enumerate} 
            \item If $\bar{\sigma} \geq \underline{\sigma}$ for all $a \in \mathbb{A}$ and $V^{^{\bar{\sigma}, \mu, f,r, g}}$ or $V^{^{\underline{\sigma}, \mu, f,r, g}}$ is convex (concave) in $x$, then $V^{^{\bar{\sigma}, \mu, f,r, g}} \!\!\geq\! V^{^{\underline{\sigma}, \mu, f, r, g}}\!\!$ and $\mathcal{C}^{^{\underline{\sigma}, \mu, f,r, g}} \!\! \subset \! \mathcal{C}^{^{\bar{\sigma}, \mu, f,r, g}}$ ($V^{^{\bar{\sigma}, \mu, f,r, g}} \!\! \leq \!V^{^{\underline{\sigma}, \mu, f, r, g}}\!\!$ and $\mathcal{C}^{^{\bar{\sigma}, \mu, f,r, g}} \!\! \subset \! \mathcal{C}^{^{\underline{\sigma}, \mu, f,r, g}}$\!). 

           \vspace{-0.4em}
            \item If $\bar{\mu} \geq \underline{\mu}$ for all $a \in \mathbb{A}$ and $V^{^{{\sigma}, \underline{\mu}, f,r, g}}$ or $V^{^{{\sigma}, \bar{\mu}, f,r, g}}$ increases (decreases) in $x$, then $V^{^{{\sigma}, \bar{\mu}, f,r, g}} \!\!\geq \!V^{^{{\sigma}, \underline{\mu}, f,r, g}}\!\!$ and $\mathcal{C}^{^{{\sigma}, \underline{\mu}, f,r, g}}\!\! \subset \! \mathcal{C}^{^{{\sigma}, \bar{\mu}, f,r, g}}$ ($V^{^{{\sigma}, \bar{\mu}, f,r, g}} \!\!\leq \! V^{^{{\sigma}, \underline{\mu}, f,r, g}}\!\!$ and $\mathcal{C}^{^{{\sigma}, \bar{\mu}, f,r, g}} \!\!\subset \! \mathcal{C}^{^{{\sigma}, \underline{\mu}, f,r, g}} \!$).
        \end{enumerate}
    \end{theoremp}
    Theorems \ref{theorem:compstatsstoppingtimescontroldiffusion} and \ref{theorem:compstatsvaluescontroldiffusion} follow directly from the proofs of Theorem \ref{theorem:compstatsstoppingtimes} and \ref{theorem:compstatsvalues} by plugging in the diffusion and payoffs associated with the optimal control (given by Assumption \ref{assumption:Lipschizselection}). Their proofs are, therefore, omitted.

    A fully developed application of Theorems \ref{theorem:localmonotonestoppingcontroldiffusion}, \ref{theorem:compstatsstoppingtimescontroldiffusion}, and \ref{theorem:compstatsvaluescontroldiffusion} is beyond the scope of this paper. Still, we briefly illustrate these results in the setting of \cite{moscarini2001optimal}. In this paper, the authors study a stationary information acquisition problem similar to that of Example \ref{ex:informationacquisition1}, in which, however, the DM also decides how fast to learn and incurs an experimentation cost that increases with the chosen speed. Theorem \ref{theorem:localmonotonestoppingcontroldiffusion} indicates that, for example, when the experimentation cost increases over time, the DM's later decisions are less accurate. Moreover, when the cost function is higher, decisions would also be less accurate by Theorem \ref{theorem:compstatsstoppingtimescontroldiffusion}.

	 \fontsize{12}{14}\selectfont
	 \bibliographystyle{chicago}
\bibliography{References.bib}

	\pagebreak

    \appendix
    \section{Appendix: Main results}\label{app:mainresults}
     	\subsection{Optimal stopping problems: toolbox}\label{subsec:toolbox}
	
    % As a preliminary step towards characterizing the shape of the continuation region, we establish that the value function is the unique $L^p$-solution of the Hamilton-Jacobi-Bellman equation. To this end, we first need to introduce a few definitions.
    
    \textbf{Notations:} $W^{1, 2, p}\left( [0,T)\times \mathcal{X} \right)$, $1\leq p\leq \infty$, is the (Sobolev) space of functions that are (i) twice differentiable a.e. in $\mathcal{X}$, (ii) once differentiable a.e. in time, and (iii) with \textit{weak} derivatives in $L^p([0,T)\times \mathcal{X})$. ${W}^{1,2,p}_{loc}([0,T)\times \mathcal{X})$, is the space of all functions whose restriction to $\mathcal{Y}\subset [0,T)\times\mathcal{X}$ belongs to ${W}^{1,2,p}(\mathcal{Y})$ for every compact subset $\mathcal{Y}\subset [0,T)\times \mathcal{X}$.
    
    \begin{definition}\label{def:strongsolution}
		A function $v \in {W}^{1,2,p}_{loc}([0,T)\times \mathcal{X}) \cap \mathcal{C}^0( [0,T] \times \mathcal{X})$ is an $L^p$-solution of the Hamilton-Jacobi-Bellman equation \hypertarget{link:HJB}{(HJB)} if
    \vspace{-1.5em}
\begin{adjustwidth}{-0in}{-0.25in}
        \begin{align*}\label{eq:HJB}
          \!\!\!\!\!\!\!\begin{cases}
                \max \!\left\{ g(x) \!-\!v(t,x),  v_t(t,x) \!+\! \mathcal{L}^{(t,x)}v(t,x) \!-\!r(t,x)v(t,x) \!+\! f(t,x) \right\} = 0 \text{\small{ a.e. in }} \medmath{[0,T)\times \mathcal{X}} \\
                v(T,x) = g(x) \text{ on } \mathcal{X}. %\tag{HJB}
            \end{cases} 
	   \end{align*} \normalsize
\end{adjustwidth}
	   % where $v_t(t,x), v_x(t,x)$ and $v_{xx}(t,x)$ are the weak derivatives of $v$.
	\end{definition}
    Moreover, let $\mathcal{C}^{1,2, \alpha}\left( \mathcal{C}\right)$ denote the space of functions that are continuously differentiable with respect to time with $\frac{\alpha}{2}$-H\"{o}lder continuous derivatives and twice continuously differentiable with respect to $x$, with $\alpha$-H\"{o}lder continuous derivatives. With these definitions in mind, we prove the following result.
	\begin{lemma}\label{lemma:HJB}
		Let $p \in (3, \infty]$. Suppose the value function is continuous. Then the value function in \eqref{eq:valuefunction} is the unique $L^p$-solution of the Hamilton-Jacobi-Bellman equation \hyperlink{link:HJB}{(HJB)}. Moreover, $V \in \mathcal{C}^{1,2, \alpha}\left( \mathcal{C}\right)$ for some $\alpha>0$.
    \end{lemma}

    \begin{proof}[\textbf{Proof of Lemma \ref{lemma:HJB}}]
		We want to show that $V$ is an $L^p$-solution of \hyperlink{link:HJB}{(HJB)}.

        First, observe that, if $T<\infty$, $V(T,x) = g(x) \text{ on } \mathcal{X}$. Next, let $(\tilde{t},\tilde{x}) \in [0,T)\times \mathcal{X}$. There exists $\epsilon>0$ such that $C_{\epsilon}(\tilde{t},\tilde{x}) = [\tilde{t}, \tilde{t}+D)\times B_D(\tilde{x}) \subset [0,T)\times \mathcal{X}$. Consider the following Hamilton-Jacobi-Bellman equation restricted to $C_{\epsilon}(\tilde{t},\tilde{x})$:
        \begin{align*}
            \begin{cases}
                \max\left\{\left( \partial_t + \mathcal{L}^{(t,x)} -r(t,x) \right)v(t,x) + f(t,x), g(x) - v(t,x)\right\} =0  \text{ if }  (t,x) \in  C_{\epsilon}(\tilde{t},\tilde{x}), \\
                v(t,x) = V(t,x) \text{ if } (t,x) \in \partial C_{\epsilon}(\tilde{t},\tilde{x}).
            \end{cases}
        \end{align*}
        By Lemma \ref{lemma:Viscontinuousintandinx} below, $t \to V(t,\tilde{x}+\epsilon)$ and $t \to V(t,\tilde{x}-\epsilon)$ are continuous. The same lemma guarantees that $x \to V(\tilde{t}+\epsilon,x)$ is continuous in $x$ on $[\tilde{x}-\epsilon, \tilde{x}+\epsilon]$. Therefore, by Theorem 2 in \cite{durandard2023existence}, there exists a unique $L^p$-solution $v$ of the above. The following standard verification argument shows that it coincides with the value function. Let $\tau_{C_{\epsilon}} = \inf \left\{ t\geq \tilde{t} \, : \, (t,X_t) \not \in C_{\epsilon}(\tilde{t},\tilde{x}) \right\}$. By the Snell envelope theorem, for all $(t,x) \in {C}_{\epsilon}(\tilde{t}, \tilde{x})$,
		\begin{align*}
            V(t,x) = \mathbb{E}_{(t,x)} & \Bigg[ \int_{0}^{\tau_{\mathcal{S}}\wedge \tau_{C_{\epsilon}}} e^{-\int_{t}^{s} r(u,X_u)\, \mathrm{d}u } f(s,X_s) ds \\
            & + e^{-\int_{t}^{\tau_{\mathcal{S}} \wedge \tau_{C_{\epsilon}}} r(u,X_u)\, \mathrm{d}u } \left( V\left(\tau_{C_{\epsilon}}, X_{\tau_{C_{\epsilon}}}\right) \mathbbm{1}_{\{ \tau_{\mathcal{S}} \geq \tau_{C_{\epsilon}}\}} + g\left(X_{\tau_{\mathcal{S}}}\right) \mathbbm{1}_{\{ \tau_{\mathcal{S}} < \tau_{C_{\epsilon}}\}}  \right)  \Bigg].
        \end{align*}
        But by Krylov-It\^{o}'s formula (e.g., Theorem 2.10.1 in \cite{krylov2008controlled}), 
        \begin{align*}
            v(t,x) = \mathbb{E}_{(t,x)} & \Bigg[ \int_{0}^{\tau_{\mathcal{S}}\wedge \tau_{C_{\epsilon}}} e^{-\int_{t}^{s} r(u,X_u)\, \mathrm{d}u } f(s,X_s) ds \\
            & + e^{-\int_{t}^{\tau_{\mathcal{S}} \wedge \tau_{C_{\epsilon}}} r(u,X_u)\, \mathrm{d}u } \left( V\left(\tau_{C_{\epsilon}}, X_{\tau_{C_{\epsilon}}}\right) \mathbbm{1}_{\{ \tau_{\mathcal{S}} \geq \tau^D\}} + g\left(X_{\tau_{\mathcal{S}}}\right) \mathbbm{1}_{\{ \tau_{\mathcal{S}} < \tau_{C_{\epsilon}}\}}  \right)  \Bigg].
        \end{align*}
        So, $V(t,x) = v(t,x)$ for all $(t,x) \in {C}_{\epsilon}(\tilde{t}, \tilde{x})$. Since, $(\tilde{t}, \tilde{x})$, this shows that $V$ is a $L^p$-solution of \hypertarget{link:HJB}{(HJB)}. 
        Finally, in $\mathcal{C}$, 
        \begin{align*}
            \left( \partial_t + \mathcal{L}^{(t,x)} - r(t,x) \right)v(t,x) + f(t,x) = 0.
        \end{align*}
        Therefore $V \in \mathcal{C}^{1,2,\alpha}$ by Theorem 3.5.10 in \cite{friedman2008partial} for some $\alpha>0$. 
        
        There only remains to show that $V$ is the unique $L^p$-solution of \hypertarget{link:HJB}{(HJB)}. This follows from the comparison principle (Lemma \ref{lemma:comparisonprinciple}) below.
    \end{proof}

    \begin{lemma}\label{lemma:Viscontinuousintandinx}
        \begin{enumerate}
            \item For all $x \in \mathcal{X}$, $t \to V(t,x)$ is continuous in $[0, T]$. 
            \item For all $t \in [0, T]$, $x \to V(t,x)$ is continuous in $\mathcal{X}$. 
        \end{enumerate}
	\end{lemma}
	
	\begin{proof}[\textbf{Proof}]
        The proofs of 1 and 2 are standard and follow the same steps, hence we omit the proof of 2. 
        
        Let $x \in \mathcal{X}$ and let $t, t' \in [0,T)$ with $t'\geq t$. Let 
		\begin{align*}
		V^M(t,x) = \underset{\tau \in \mathcal{T}(t)}{\sup } \, \mathbb{E}\left[ \int_{t}^{\tau} e^{-r(s-t)}f(s, X_s)\wedge M ds + e^{-r(\tau-t)} g\left(X_{\tau}\right) \wedge M \right];
		\end{align*}
		and observe that $V^M \uparrow V$ pointwise by the monotone convergence theorem. Since $V$ is locally bounded, for all $\epsilon>0$, there exists $M$ such that 
		\begin{align*}
		\left| V(t',x) - V(t,x) \right| & \leq \left| V(t',x) - V^M(t',x) \right| + \left| V^M(t',x) - V^M(t,x) \right| + \left| V^M(t,x) - V(t,x) \right| \\
		& \leq \left| V^M(t',x) - V^M(t,x) \right| + 2 \epsilon.
		\end{align*}
		Thus, it is enough to show that $t \to V(t,x)$ is continuous for $f$ and $g$ bounded above. Repeating the same argument, we also have that it is enough to consider $f$ and $g$ bounded below. So, suppose that $f$ and $g$ are bounded for the remainder of the proof. Hence, $V$ is bounded.

		Then, by Snell's envelope theorem, for all $\bar{t} \geq 0$,
		\begin{align*}
		\left| V(t',x) - V(t,x) \right| & \leq \underset{\tau}{\sup}\,  \mathbb{E} \Bigg[ \int_{0}^{\tau\wedge \bar{t}} e^{-rs}\left| f\left(s + t', X^{t',x}_s\right)  - f\left(s + t, X^{t,x}_s\right) \right| ds \\
		& \quad + e^{-r\tau} \mathbbm{1}_{\{ \tau \leq \bar{t} \}} \left|g\left(X^{t',x}_{\tau}\right) - g\left(X^{t,x}_{\tau}\right) \right| \Bigg] \\
		& \quad +\mathbb{E}_{(t',x)}\left[ e^{-r\bar{t}} \mathbbm{1}_{\{ \tau \geq \bar{t}\}} \left| V\left(\bar{t}, X_{\bar{t}}\right) \right| \right] + \mathbb{E}_{(t,x)}\left[ e^{-r\bar{t}} \mathbbm{1}_{\{ \tau \geq \bar{t}\}} \left| V\left(\bar{t}, X_{\bar{t}}\right) \right| \right],
		\end{align*}
		where the supremum is taken over all stopping times weakly smaller than $T$. By \eqref{eq:uniformintegrability}, we can choose $\bar{t}$ large enough so that
		\begin{align*}
			\mathbb{E}_{(t',x)} &  \left[ e^{-r\bar{t}} \mathbbm{1}_{\{ \tau \geq \bar{t}\}} \left| V\left(\bar{t}, X_{\bar{t}}\right) \right| \right] + \mathbb{E}_{(t,x)}\left[ e^{-r\bar{t}} \mathbbm{1}_{\{ \tau \geq \bar{t}\}} \left| V\left(\bar{t}, X_{\bar{t}}\right) \right| \right] < \epsilon.
		\end{align*} 
		Let $\epsilon>0$. In the remaining of the proof, we then take $\bar{t}$ such that
		\begin{align*}
		\left| V(t',x) - V(t,x) \right| \leq \underset{\tau}{\sup}\,  \mathbb{E}& \Bigg[ \int_{0}^{\tau\wedge \bar{t}} e^{-rs} \left| f\left(s + t', X^{t',x}_s\right)  - f\left(s + t, X^{t,x}_s\right) \right| ds \\
		& \qquad + e^{-r\tau} \mathbbm{1}_{\{ \tau \leq \bar{t} \}} \left|g\left(X^{t',x}_{\tau}\right) - g\left(X^{t,x}_{\tau}\right) \right| \Bigg] + \epsilon.
		\end{align*}
		Since $g$ and $f$ are locally Lipschitz continuous, there exists $C >0$ such that
		\begin{align*}
		\left| V(t',x) - V(t,x) \right| \leq \mathbb{E}& \Bigg[ C \bar{t}\wedge 1 \left( \left| t'-t\right| + \underset{0\leq s \leq \bar{t}}{\sup } \, \left| X^{t',x}_s - X^{t,x}_s \right|\right)  \Bigg] + 2 \epsilon,
		\end{align*}
        where we used that the boundary point of $\mathcal{X}$ are not attainable and that $f$ and $g$ are bounded. By standard estimates (see, e.g., Theorem 2.5.9 in \cite{krylov2008controlled} and H\"{o}lder inequality), 
		\begin{align*}
		\mathbb{E}& \Bigg[ \underset{0\leq s \leq \bar{t}}{\sup } \, \left| X^{t',x}_s - X^{t,x}_s \right| \Bigg] \leq K \left( 1+ \left| x\right|^2 \right) \sqrt{t'-t}
		\end{align*}
		where $K>0$ depends on $\bar{t}$ and the bounds on $\sigma$ and $\mu$ only. Thus, $t \to V(t,x)$ is continuous in $t$ for all $x \in \mathcal{X}$.
	\end{proof}

    \begin{lemma}\label{lemma:comparisonprinciple}
        Suppose that $u,v \in W^{1,2,p}_{loc}\left([0,T)\times \mathcal{X}\right)$ are such that
        \begin{align*}
            \begin{cases}
                \max \!\left\{ g(x) \!-\!v(t,x),  v_t(t,x) \!+\! \mathcal{L}^{(t,x)}v(t,x) \!-\!r(t,x)v(t,x) \!+\! f(t,x) \right\} \leq 0 \text{\small{ a.e. in }} \medmath{[0,T)\times \mathcal{X}} \\
                v(T,x) \geq g(x) \text{ on } \mathcal{X}, %\tag{HJB}
            \end{cases}
        \end{align*}
        and
        \begin{align*}
            \begin{cases}
                \max \!\left\{ g(x) \!-\!u(t,x),  u_t(t,x) \!+\! \mathcal{L}^{(t,x)}u(t,x) \!-\!r(t,x)u(t,x) \!+\! f(t,x) \right\} \geq 0 \text{\small{ a.e. in }} \medmath{[0,T)\times \mathcal{X}} \\
                u(T,x) \leq g(x) \text{ on } \mathcal{X}. %\tag{HJB}
            \end{cases}
        \end{align*}
        Then $u\leq v$.
    \end{lemma}

    \begin{proof}[\textbf{Proof.}]
        Let $u,v$ be as defined above. Consider $(t,x) \in [0,T)\times \mathcal{X}$. Observe that $v \geq g$ on $[0,T)\times \mathcal{X}$. Hence, if $u(t,x) \leq g(x)$, we are done. So, suppose that $u(t,x) > g(x)$ and let $\tau = \inf \left\{ s\geq t \, : \, u(s, X_s^{t,x}) = g(X_s^{t,x})\right\}$. By Krylov-It\^{o}'s Lemma (Theorem 2.10.2 in \cite{krylov2008controlled}), 
        \begin{align*}
            u(t,x) & = \mathbb{E}\bigg[ \int_t^{\tau} e^{-\int_t^s r(u, X_u)\, \mathrm{d}u} \left( r(s,X_s)u(s,X_s) - u_t(s,X_s) - \mathcal{L}^{(s,X_s)}u(s,X_s) \right) ds \\
            & \qquad \qquad \qquad \qquad +e^{-\int_t^\tau r(u, X_u)\, \mathrm{d}u} g(X_{\tau}) \bigg] \\
            & \leq \mathbb{E}\left[ \int_t^{\tau} e^{-\int_t^s r(u, X_u)\, \mathrm{d}u} f(s, X_s) ds +e^{-\int_t^\tau r(u, X_u)\, \mathrm{d}u} g(X_{\tau}) \right] \\
            & \leq \mathbb{E}\left[ \int_t^{\tau} e^{-\int_t^s r(u, X_u)\, \mathrm{d}u} f(s, X_s) ds +e^{-\int_t^\tau r(u, X_u)\, \mathrm{d}u} v(\tau, X_{\tau}) \right] \\
            & \leq \mathbb{E}\bigg[ \int_t^{\tau} e^{-\int_t^s r(u, X_u)\, \mathrm{d}u} \left( r(s,X_s)v(s,X_s) - v_t(s,X_s) - \mathcal{L}^{(s,X_s)}v(s,X_s) \right) ds \\
            & \qquad \qquad \qquad \qquad +e^{-\int_t^\tau r(u, X_u)\, \mathrm{d}u} v(, \tau, X_{\tau}) \bigg] \\
            & \leq v(t,x),
        \end{align*}
        which concludes the proof.
    \end{proof}

	\subsection{Omitted proofs for Section \ref{subsec:monotoneenvironments}}\label{app:proofsmainresults}

\begin{proof}[\textbf{Proof of Theorem \ref{theorem:localmonotonestopping}}]
		The weak monotonicity of the continuation region %(points 1. and 3.) 
  follows immediately from the local monotonicity of the value function. 
  
        Next, we prove strict monotonicity when the optimal stopping problem is locally strictly monotone decreasing on $\left( \underline{t}, \bar{t} \right)$. The argument for a locally strictly monotone increasing optimal stopping problem is identical, hence omitted. %From point 1, since the optimal stopping problem is locally monotone decreasing, so is the continuation region. There remains to show that the boundaries are strictly monotone when they are in $\mathcal{X}$. 
        From 1., we know that the continuation region is monotone decreasing. So, there only remains to show that the boundaries are strictly monotone (when they are in $\mathcal{X}$).
		
		The proof is by contradiction. Suppose that there exists $\underline{t}, \bar{t} \in [0,T)$ and $B \in \mathcal{X}$ such that $(\underline{t}, B)$ and $(\bar{t}, B) \in \partial \mathcal{C}$. In this proof, we focus on the ``upper boundary'' of the continuation region, i.e., we assume there exists $\epsilon>0$ such that for all $(\underline{t},\bar{t}) \times (B-\epsilon, B) \subset \mathcal{C}$. The proof for the ``lower boundary'' is identical.
		
		Consider the rectangular domain $\mathcal{Y} = [\underline{t}, \bar{t}) \times (B-\epsilon,B)$ with parabolic boundary $\partial \mathcal{Y} = \left( [\underline{t},\bar{t}] \times \left( \left\{ B-\epsilon \right\} \cup \left\{ B\right\} \right)   \right) \cup\left( \left\{\bar{t}\right\} \times (B-\epsilon,B) \right) \subset \mathcal{C}$. By Lemma \ref{lemma:HJB}, we know that the value function $V$ is the unique $L^p$-solution of the boundary value problem
		\begin{align*}
    		\begin{cases}
        		v_t(t,x) + \frac{\sigma(t,x)^2}{2} v_{xx}(t,x) + \mu(t,x) v_x(t,x) -r(t,x)v(t,x) + f(t,x) = 0 \text{ if }  (t,x) \in  \mathcal{Y}, \\
        		v(t,x) = V(t,x) \text{ if } (t,x) \in \partial \mathcal{Y}.
    		\end{cases}
		\end{align*}
		By Theorem 2 and Corollary 1 in \cite{durandard2022smoothness}, $V_t \in \mathcal{C}^0\left( \bar{\mathcal{Y}} \right)$. Then, Theorem 3.5.10 in \cite{friedman2008partial} implies that $V_t(t,x) \in \mathcal{C}^{1,2, \alpha}\left( [\underline{t}, \bar{t}) \times \left(B-\epsilon, B\right)\right)$ and solves
		\begin{align*}
		\begin{cases}
		v_{tt}(t,x) + \frac{\sigma(t,x)^2}{2} v_{txx}(t,x) + \mu(t,x) v_{tx}(t,x) -r(t,x)v_t(t,x) \\
		\quad = -\sigma(t,x) {\sigma_t(t,x)} v_{xx}(t,x) - \mu_t(t,x) v_{x}(t,x) + r_t(t,x)v(t,x)  - f_t(t,x) \text{ if }  (t,x) \in  \mathcal{Y}, \\
		v_t(t,x) = V_t(t,x) \text{ if } (t,x) \in \partial \mathcal{Y}.
		\end{cases}
		\end{align*}
		Therefore, on $\mathcal{Y}$,
		\begin{align*}
    		v_{tt}(t,x) + & \frac{\sigma(t,x)^2}{2} v_{txx}(t,x) + \mu(t,x) v_{tx}(t,x) -r(t,x)v_t(t,x)\\
    		& = - \sigma(t,x) {\sigma_t(t,x)} v_{xx}(t,x) - \mu_t(t,x) v_{x}(t,x) + r_t(t,x) v(t,x)  - f_t(t,x) \\
    		& \geq 0.
		\end{align*}
        The inequality follows from (DOV) in Definition \ref{definition:monotoneenvironments}. Then, by Hopf's boundary Lemma (Lemma 1.23 in \cite{wang2021nonlinear}), $v_{tx}(t,B^{-}) >0$ for all $t\in (\underline{t}, \bar{t})$, implying $V_x(t, B^-)$ is strictly increasing for $t\in (\underline{t}, \bar{t})$. However, for all $t\in (\underline{t}, \bar{t})$, $V(t,B) = g(x)$ and $V \in W^{1,2,p}_{loc}\left( [0,T)\times\mathcal{X}\right)$ imply  $V_x(t,B) = g_x(x)$, which is constant on $(\underline{t}, \bar{t})$: a contradiction. 
		
		Thus, the upper boundary of the continuation region cannot be flat. %This concludes the proof.
	\end{proof}
 
	\begin{proof}[\textbf{Proof of proposition \ref{prop:convextsectionsstopping}}]
		By Lemma \ref{lemma:HJB}, the value function is the unique $L^p$-solution of the HJB equation
		\begin{align*}
			\begin{cases}
				\max \left\{ g(x) -v(t,x), \left( \partial_t + \mathcal{L}^{(t,x)} -r(t,x) \right)v(t,x) + f(t,x) \right\}  = 0 \text{ in } [0,T)\times \mathcal{X} \\
				v(T,x) = g(x) \text{ on } \mathcal{X}.
			\end{cases} \tag{HJB}
		\end{align*}
		Let $\bar{b}(t) = \inf \left\{ x\in \left[x^c, \bar{x}\right) \, : \, (t,x) \in \mathcal{S} \right\}$ and define
		\begin{align*}
			\tilde{v}(t,x) = \begin{cases}
				v(t,x) \text{ if } t < b(t)\\
				g(x) \text{ if } t \geq b(t).
			\end{cases}
		\end{align*}
		Then $\tilde{v}(t,x)$ is also a $L^p$ solution of equation \hyperlink{link:HJB}{(HJB)} by Condition C\ref{condition:singlecrossing}. Moreover, by the comparison principle in \cite{durandard2023existence} (Corollary 1), there is a unique $L^p$-solution. Therefore $v(t,x) = \tilde{v}(t,x)$ and $\mathcal{S} \cap \left\{ (t,x) \in \mathbb{R}_+ \times \mathcal{X} \, : \, x \geq x^c \right\} = \left\{ (t,x) \in \mathbb{R}_+ \times \mathcal{X} \, : \, x \geq \bar{b}(t) \right\}$. 
		
		\noindent By the same argument, $\mathcal{S} \cap \left\{ (t,x) \in \mathbb{R}_+ \times \mathcal{X} \, : \, x \leq x^c \right\} = \left\{ (t,x) \in \mathbb{R}_+ \times \mathcal{X} \, : \, x \leq \underline{b}(t) \right\}$, with $\underline{b}(t) = \sup \left\{ x\in \left(\underline{x}, x^c\right] \, : \, (t,x) \in \mathcal{S} \right\}$.
	\end{proof}

 \begin{proof}[\textbf{Proof of Corollary \ref{corollary:globalmonotonestopping}}]
		The existence of boundaries $\underline{b}$ and $\bar{b}$ follows from Proposition \ref{prop:convextsectionsstopping}. The monotonicity follows from Theorem \ref{theorem:localmonotonestopping}. Finally, since $V$ is continuous, $\mathcal{S}$ is closed, and, hence, $\underline{b}$ is upper semicontinuous and $\bar{b}$ is lower semicontinuous. Together with their monotonicity, this ensures they are c\`{a}dl\`{a}g or c\`{a}gl\`{a}d.
	\end{proof}

\subsection{Omitted proofs for Section \ref{subsec:CSEE}}\label{app:proofCSEE}

\begin{proof}[\textbf{Proof of Theorem \ref{theorem:compstatsstoppingtimes}}.]
    Let $\underline{r}(t,x) \leq \bar{r}(t,x)$ and $\underline{f}(t,x) \leq \bar{f}(t,x)$. By Lemma \ref{lemma:HJB}, $V^{\sigma, \mu, \underline{f},\bar{r}, g}$ is the unique $L^p$-solution of \eqref{eq:HJB} with $r(t,x) =\bar{r}(t,x)$ and $f(t,x) = \underline{f}(t,x)$. Therefore, since $V^{\sigma, \mu, \underline{f},\bar{r}, g} \geq 0$, it is a $L^p$-supersolution of
        \begin{align*}
            \begin{cases}
                \max \!\bigg\{ g(x) \!-\!V^{\sigma, \mu, \underline{f},\bar{r}, g}(t,x),  V^{\sigma, \mu, \underline{f},\bar{r}, g}_t(t,x) +  \mu(t,x) V^{\sigma, \mu, \underline{f},\bar{r}, g}_x(t,x) \! + \! \frac{{\sigma}^2(t,x)}{2} V^{\sigma, \mu, \underline{f},\bar{r}, g}_{xx}(t,x) \\
                \qquad \qquad \qquad \qquad \qquad \qquad \qquad \qquad - \underline{r}(t,x)V^{\sigma, \mu, \underline{f},\bar{r}, g}(t,x) \!+\! \bar{f}(t,x) \bigg\} \geq 0 \text{\small{ a.e. in }} \medmath{[0,T)\times \mathcal{X}} \\
                \bar{V}(T,x) \geq g(x) \text{ on } \mathcal{X}, %\tag{HJB}
            \end{cases} 
        \end{align*}
        But, by Lemma \ref{lemma:HJB} again, $V^{\sigma, \mu, \bar{f},\underline{r}, g}$ is the unique $L^p$-solution of \eqref{eq:HJB} with $r(t,x) =\underline{r}(t,x)$ and $f(t,x) = \bar{f}(t,x)$. Thus, by a comparison principle (Lemma \ref{lemma:comparisonprinciple}), $V^{\sigma, \mu, \underline{f},\bar{r}, g} \leq V^{\sigma, \mu, \bar{f},\underline{r}, g}$ in $[0,T] \times \mathcal{X}$.

        Finally, since $g \leq V^{\sigma, \mu, \underline{f},\bar{r}, g}$, $\mathcal{C}^{^{\sigma}, \mu, \underline{f},\bar{r}, g} = \left\{V^{\sigma, \mu, \underline{f},\bar{r}, g} = g\right\} \subset \left\{V^{\sigma, \mu, \bar{f},\underline{r}, g} = g\right\} = \mathcal{C}^{\sigma, \mu, \bar{f},\underline{r}, g}$, which immeditaely implies that $\tau^{\sigma, \mu, \underline{f},\bar{r}, g} \leq \tau^{\sigma, \mu, \bar{f},\underline{r}, g}$.
\end{proof}

    \begin{proof}[\textbf{Proof of Theorem \ref{theorem:compstatsvalues}}.]
        The proofs of cases 1. and 2. both follow the same steps as the proof of Theorem \ref{theorem:compstatsstoppingtimes}. So, to keep the paper concise, we only prove the first statement.
    
        Let $\underline{\sigma}(t,x) \leq \bar{\sigma}(t,x)$. The associated value functions are respectively $\underline{V}$ and $\bar{V}$. Suppose that $\bar{V}$ is convex. The proof for the cases when $\bar{V}$ is concave, when $\underline{V}$ is convex, or when $\underline{V}$ is concave are similar, hence omitted. By Lemma \ref{lemma:HJB}, $\bar{V}$ is the unique $L^p$-solution of \eqref{eq:HJB} with $\sigma =\bar{\sigma}$. Therefore, $\bar{V}$ is a $L^p$-subsolution of
        \begin{align*}
            \begin{cases}
                \max \!\bigg\{ g(x) \!-\!\bar{V}(t,x),  \bar{V}_t(t,x) +  \mu(t,x)\bar{V}_x(t,x) \! + \! \frac{\underline{\sigma}^2(t,x)}{2} \bar{V}_{xx}(t,x) \\
                \qquad \qquad \qquad \qquad \qquad \qquad \qquad \qquad - r(t,x)\bar{V}(t,x) \!+\! f(t,x) \bigg\} \leq 0 \text{\small{ a.e. in }} \medmath{[0,T)\times \mathcal{X}} \\
                \bar{V}(T,x) \geq g(x) \text{ on } \mathcal{X}, %\tag{HJB}
            \end{cases} 
        \end{align*}
        But, by Lemma \ref{lemma:HJB} again, $\underline{V}$ is the unique $L^p$-solution of \eqref{eq:HJB} with $\sigma(t,x) =\underline{\sigma}(t,x)$. Thus, by a comparison principle (Lemma \ref{lemma:comparisonprinciple}), $\bar{V} \geq \underline{V}$ in $[0,T] \times \mathcal{X}$.

        Finally, since $g \leq \underline{V}$, $\mathcal{C}^{\underline{\sigma}, \mu, f,r, g} = \left\{\underline{V} = g\right\} \subset \left\{\bar{V} = g\right\} = \mathcal{C}^{\bar{\sigma}, \mu, f,r, g}$.
    \end{proof}

    \begin{proof}[\textbf{Proof of Proposition \ref{prop:Vmonotonictywrtx}.}]
            Let $t\in [0, T]$ and $\bar{x} \geq x \in \mathcal{X}$. If $t=T$, $V(T,x) = g(x)$ and we are done. So suppose that $t<T$. Let $\tau^{(t,x)}$ be the optimal stopping time associated with the problem $V(t,x)$. By standard results (e.g. Theorem IX.3.7 in \cite{revuz2013continuous}), $X_s^{(t,x)} \geq X_s^{(t,x')}$ for all $s\geq t$, $\mathbb{P}$-a.s.. Then
            \begin{align*}
                V(t,x) & = \mathbb{E}_{(t,x)}\left[\int_t^{\tau^{(t,x)}} e^{-\int_t^s r(u, X_u)\, \mathrm{d}u} f(s, X_s) ds + e^{-\int_t^{\tau^{(t,x)}} r(u, X_u) \, \mathrm{d}u} g(X_{\tau^{(t,x)}}) \right] \\
                & \leq \mathbb{E}_{(t,x')}\left[\int_t^{\tau^{(t,x)}} e^{-\int_t^s r(u, X_u)\, \mathrm{d}u} f(s, X_s) ds + e^{-\int_t^{\tau^{(t,x)}} r(u, X_u) \, \mathrm{d}u} g(X_{\tau^{(t,x)}}) \right] \\
                & \leq \underset{\tau \geq t}{\sup } \mathbb{E}_{(t,x')} \left[\int_t^{\tau} e^{-\int_t^s r(u, X_u)\, \mathrm{d}u} f(s, X_s) ds + e^{-\int_t^{\tau} r(u, X_u) \, \mathrm{d}u} g(X_{\tau}) \right] \\ 
                & = V(t,x').
            \end{align*}
            This concludes the proof.
        \end{proof}

\begin{corollary}\label{corollary:strictcompstats}
       Let $\mathcal{C}^{^{{\sigma}, \mu, f,r, g}}$ be the continuation region for the optimal stopping problem with parameters ${\sigma, \mu, f,r, g}$.  
        \begin{enumerate}
            \item Suppose that the stopping problem with parameter ${\bar{\sigma}, \bar{\mu}, \bar{f}, \bar{r}, g}$ is globally strictly monotone decreasing with $V^{^{\bar{\sigma}, \bar{\mu}, \bar{f}, \bar{r},g}}$ convex in $x$, and there exists $\epsilon >0$ such that $\bar{\sigma}(t+\epsilon,x) \geq \underline{\sigma}(t,x), \bar{\mu}(t+\epsilon,x) \text{sign}\left(V^{^{\bar{\sigma}, \bar{\mu}, \bar{f}, \bar{r},g}}_x\right) \geq \underline{\mu}(t,x) \text{sign}\left(V^{^{\bar{\sigma}, \bar{\mu}, \bar{f}, \bar{r},g}}_x\right), \bar{f}(t+\epsilon,x) \geq \underline{f}(t,x), \bar{r}(t+\epsilon,x) \geq \underline{r}(t,x) $ for all $(t,x) \in [0,T)\times mathcal{X}$. Then $\mathcal{C}^{^{\underline{\sigma}, \underline{\mu}, \underline{f},\underline{r}, g}}$ is a strict subset of $ \mathcal{C}^{^{\bar{\sigma}, \bar{\mu}, \bar{f}, \bar{r}, g}}$.

            \item Suppose that the stopping problem with parameter ${\bar{\sigma}, \bar{\mu}, \bar{f}, \bar{r}, g}$ is globally strictly monotone increasing with $V^{^{\bar{\sigma}, \bar{\mu}, \bar{f}, \bar{r},g}}$ convex in $x$, and there exists $\epsilon >0$ such that $\bar{\sigma}(t-\epsilon,x) \geq \underline{\sigma}(t,x), \bar{\mu}(t-\epsilon,x) \text{sign}\left(V^{^{{\sigma}, \bar{\mu}, f,r, g}}_x\right) \geq \underline{\mu}(t,x) \text{sign}\left(V^{^{{\sigma}, \bar{\mu}, f,r, g}}_x\right), \bar{f}(t-\epsilon,x) \geq \underline{f}(t,x), \bar{r}(t-\epsilon,x) \geq \underline{r}(t,x) $ for all $(t,x) \in [0,T)\times \mathcal{X}$. Then $\mathcal{C}^{^{\underline{\sigma}, \underline{\mu}, \underline{f},\underline{r}, g}}$ is a strict subset of $ \mathcal{C}^{^{\bar{\sigma}, \bar{\mu}, \bar{f}, \bar{r}, g}}$. 
        \end{enumerate}
    \end{corollary}
\begin{proof}
    These strict inclusion results directly follow from combining Theorem \ref{theorem:localmonotonestopping} with Theorems \ref{theorem:compstatsstoppingtimes} and \ref{theorem:compstatsvalues}.
\end{proof}

    \subsection{Omitted proofs for Section \ref{subsec:Visconvex}}\label{app:Visconvex}
 
	\begin{proof}[\textbf{Proof of Proposition \ref{prop:Visconvex}}]
        We first show that $V$ is convex in $x$. For all $n \in \mathcal{N}$, let $\mathcal{T}_n(t)$ be the set of stopping times in $\mathcal{T}(t)$ taking value in
        \begin{align*}
            \mathcal{T}_n(t) = \left\{ s_k \, : \, s_0 = 0 \text{ and } s_{k+1} = s_k + 2^{-n} \text{ for all } k \in \mathbb{N} \right\} \cap [0,T\wedge n].
        \end{align*}
        Observe that, for all $n \in \mathbb{N}$, $\mathcal{T}_n(t) \subset \mathcal{T}_{n+1}(t)$. Define
        \begin{align*}
	       V_n(t,x) = \underset{\tau \in \mathcal{T}_n(t)}{ \sup } \, \mathbb{E}_{(t,x)} \left[ \int_{t}^{\tau} e^{-r(s-t)} f(s,X^{(t,x)}_s)ds + e^{-r(\tau -t)} g\left( X_{\tau} \right)\right].
        \end{align*}
		Then $V_n(T\vee n, x) = g(x)$ is convex in $x$. By Theorem 2 in \cite{bergman1996general}, convexity is preserved for one dimensional diffusion. Therefore, for all $t \in \left( T\wedge n -2^{-n} , T\wedge n \right)$,
		\begin{align*}
		x\to \mathbb{E}_{(t,x)} \left[ e^{-r(T\wedge n-t)} g\left(X_{T\wedge n}\right) \right]
		\end{align*}
		is convex in $x$. Moreover, by Fubini's theorem,
		\begin{align*}
		\mathbb{E}_{(t,x)} & \left[\int_{t}^{T\wedge n} e^{-r\left(s - t\right)} f(s,X_s) ds\right]  = \int_{t}^{T\wedge n} \mathbb{E}_{(t,x)} \left[ e^{-r\left(s - t\right)} f(s,X_s) \right] ds.
		\end{align*}
		Then, using Theorem 2 in \cite{bergman1996general} again, for all $t \in \left( T\wedge n - 2^{-n}, T\wedge n \right)$ and all $s \in \left( t, T\wedge n \right)$,
		\begin{align*}
			x \to \mathbb{E}_{(t,x)} \left[ e^{-r\left(s - t\right)} f(s,X_s) \right]
		\end{align*}
		is convex. Therefore, for all $t \in \left( T\wedge n - 2^{-n}, T\wedge n \right)$,
		\begin{align*}
		x \to \mathbb{E}_{(t,x)} & \left[\int_{t}^{T\wedge n} e^{-r\left(s - t\right)} f\left(s,X^{(t,x)}_s\right) ds\right]
		\end{align*}
		is convex, and thus
		\begin{align*}
		V_n(t,x) = \mathbb{E}_{(t,x)} \left[ e^{-r(T\wedge n-t)} g\left(X^{(t,x)}_{T\wedge n} \right) \right] + \mathbb{E}_{(t,x)}\left[\int_{t}^{T\wedge n} e^{-r\left(s - t\right)} f\left(s, X_{s} \right) ds\right]
		\end{align*}
		is also convex in $x$ for all $t \in \left( T\wedge n -2^{-n}, T\wedge n \right)$. At time $T\wedge n-2^{-n}$, the value function is given by the dynamic programming equation:
		\begin{align*}
		% V_n& \left( T\wedge n-2^{-n}, x\right) \\
		% & = \max \left\{ g(x), \mathbb{E}_{(T\wedge n -2^{-n}, x)}\left[ e^{-r 2^{-n} } V_n\left(T\wedge n, X_{T\wedge n}\right) + \int_{T\wedge n -2^{-n}}^{T\wedge n} e^{-r\left(s - \left(T\wedge n -2^{-n} \right)\right)} f(s, X_s) ds \right] \right\},
  V_n& \left( T\wedge n-2^{-n}, x\right) \\
		& \text{\scalebox{0.9}{$= \max \left\{ g(x), \mathbb{E}_{(T\wedge n -2^{-n}, x)}\left[ e^{-r 2^{-n} } V_n\left(T\wedge n, X_{T\wedge n}\right) + \int_{T\wedge n -2^{-n}}^{T\wedge n} e^{-r\left(s - \left(T\wedge n -2^{-n} \right)\right)} f(s, X_s) ds \right] \right\},$}}
		\end{align*}
		which is convex in $x$ as the maximum of two convex functions. Proceeding recursively shows that $x \to V_n(t,x)$ is convex (in $x$) for all $t \in [0,T\wedge n]$.
		
		To conclude the proof, we show that $V_n$ converges pointwise to $V$ as $n \to \infty$, and, hence, $V(t,x)$ is convex in $x$. To see this, let $M >0$ and consider the alternative stopping problems 
		\begin{align*}
		V^M(t,x) = \underset{\tau \in \mathcal{T}(t)}{\sup } \, \mathbb{E}\left[ \int_{t}^{\tau} e^{-r(s-t)}f(s, X_s) ds + e^{-r(\tau-t)} g\left(X_{\tau}\right) \wedge M \right]; \text{ and} \\
		% \end{align*}
		% and 
		% \begin{align*}
		V^M_n(t,x) = \underset{\tau \in \mathcal{T}_n(t)}{\sup } \, \mathbb{E}\left[ \int_{t}^{\tau} e^{-r(s-t)}f(s, X_s) ds + e^{-r(\tau-t)} g\left(X_{\tau}\right) \wedge M \right];
		\end{align*}
		By the monotone convergence theorem, $V^M_n(t,x) \uparrow V_n(t,x)$ and $V^M(t,x) \uparrow V(t,x)$ as $M\to \infty$. Consider then $\tau \in \mathcal{T}(t)$ and let $\tau_n = \inf\left\{ \tilde{\tau} \in \mathcal{T}_n(t) \, : \, \tilde{\tau} \geq \tau \, \,  \mathbb{P}\text{-a.s.}  \right\}$. Then $\tau_n \in \mathcal{T}_n(t)$ and $\tau_n \to \tau$ $\mathbb{P}$-a.s. as $n \to \infty$. By the dominated convergence theorem, 
		\begin{align*}
          \text{\scalebox{0.8}{$
        		\Bigg| \mathbb{E}_{(t,x)} \left[ \int_{t}^{\tau} e^{-r(s-t)}f(s, X_s) ds + e^{-r(\tau-t)} g\left(X_{\tau}\right) \wedge M \right] 
          - \mathbb{E}_{(t,x)}\left[ \int_{t}^{\tau_n} e^{-r(s-t)}f(s, X_s) ds + e^{-r(\tau_n-t)} g\left(X_{\tau_n}\right) \wedge M \right]\Bigg| 
          $}}\\
         \text{\scalebox{0.9}{$ \leq \mathbb{E}_{(t,x)}\left[ \left| \int_{\tau}^{\tau_n} e^{-r(s-t)}f(s, X_s) ds + e^{-r(\tau-t)} g\left(X_{\tau}\right) \wedge M - e^{-r(\tau_n-t)} g\left(X_{\tau_n}\right) \wedge M \right|\right] \to 0
         $}}
		\end{align*}
		as $n\to \infty$ (since either $e^{-r\tau}g(X_{\tau}) \wedge M$ is continuous and uniformly integrable and $e^{-rs}f(s, X_s)$ is integrable if $r >0$, or $g(\cdot) \wedge M$ is bounded and continuous, $f$ is locally bounded, and we can focus on $\tau$ such that $\mathbb{E}\left[ \tau \right]<K$ for some $K>0$ by Lemma \ref{lemma:finitetau} below if $r =0$). Therefore,
		\begin{align*}
		\underset{n \to \infty}{\liminf } \, V^M_n(t,x) \geq V^M(t,x),
		\end{align*}
		and, since $V^M_n(t,x) \leq V^M(t,x)$ for all $(t,x) \in [0,T)\times \mathcal{X}$ and all $n \in \mathbb{N}$, it follows that $V^M_n$ converges pointwise to $V^M$ as $n \to \infty$. Moreover, since $\mathcal{T}_n(t) \subset \mathcal{T}_{n+1}(t)$, $V^M_n \uparrow V^M$. Thus, we can interchange the order of the limits (using that $V(t,x)$ is locally bounded). Thus, $V_n$ converges pointwise to $V$ as $n \to \infty$.
	\end{proof}
	
\begin{proof}[\textbf{Proof of Proposition \ref{prop:convexproblemboundaries}}]
As a first step,  note that (by direct inspection of the definitions) an MC-problem is {strictly monotone}{decreasing} if, for all $(t,x) \in \mathcal{C}$, $V_t(t,x) <0$, $\sigma_t(t,x) \leq 0$, and $f_t(t,x) \leq 0$, and strictly monotone {increasing} if, for all $(t,x) \in \mathcal{C}$, $V_t(t,x) >0$, $\sigma_t(t,x) \geq 0$, and $f_t(t,x) \geq 0$. 

Second, we establish a Lemma proving that $V$ is decreasing (increasing) over time when both $\sigma(\cdot,x)$, and $f(\cdot,x)$ are decreasing (increasing) and at least one strictly so.
     \begin{lemma}\label{prop:convexproblemmonotonicity}
Consider a convex optimal stopping problem \eqref{eq:valuefunction}.
        \begin{enumerate}[topsep=3pt,itemsep=3pt]
            \item If $t \to \sigma(t,x)$ and $t \to f(t,x)$ are nonincreasing, at least one being strictly decreasing, then \eqref{eq:valuefunction} is \textbf{strictly monotone decreasing}.
		
		    \item If $t \to \sigma(t,x)$ and $t \to f(t,x)$ are nondecreasing, at least one being strictly increasing, and $T =\infty$,\footnote{If $T<\infty$, the result would also hold, provided that the payoff of the DM at the boundary is large enough.} then \eqref{eq:valuefunction} is \textbf{strictly monotone increasing}.
        \end{enumerate}
	 \begin{proof}
        To prove point 1, we show that $V_t\leq0$ and $V_t(u,x)<0$ for all $(u,x)\in \mathcal{C}$ when both $\sigma$ and $f$ are nonincreasing in $t$, and at least one is strictly decreasing. The proof of point 2 follows the exact same steps (with reverse inequalities).
        % that $V$ is nondecreasing and strictly increasing in the continuation region when $T =\infty$, both $t \to \sigma(t,x)$ $t \to f(t,x)$ are nondecreasing, and at least one is strictly increasing follows the exact same steps.
        
		For all $t \in [0, T]$, define $A_t = \int_{0}^t a(s,X_t) ds$ with $a(s,x) = \frac{\sigma(t'-t +s,x)^2}{\sigma(s,x)^2}$ and $V_t = \inf\left\{ s \geq 0 \, :\, A_s > t \right\}$, the (generalized) inverse of $A_t$. Let $Y = \{Y_t = X_{V_t}\}_{t\geq0}$ be the strong Feller process defined by the infinitesimal generator $\mathcal{L}^Y = \frac{1}{a(t,x)} \mathcal{L}^X$, and note that $Y$ and $X^{(t',x)}$ are identically distributed (see, e.g., Theorem 8.5.7 in \cite{oksendal2013stochastic}).  In particular, it follows from the optional sampling theorem that, for all stopping time $\tau$ with finite expectation, the distribution of $X^{(t,x)}_{\tau}$ dominates the distribution of $Y_{\tau}$ in the convex order. Therefore, 
        \begin{align*}
             e^{-rt}V(t,x) & = \underset{\tau \in \mathcal{T}}{\sup } \, \mathbb{E}_{(t,x)} \left[ e^{-r \tau} g(X_\tau) + \int_t^\tau e^{-r s} f(s,X_s) d s \right] \\
             & \geq \underset{\tau \in \mathcal{T}}{\sup } \, \mathbb{E}_{(t,x)} \left[ e^{-r \tau} g(Y_\tau) + \int_t^\tau e^{-r s} f(s,Y_s) d s \right]\\
             & \geq \underset{\tau \in \mathcal{T}}{\sup } \, \mathbb{E}_{(t,x)} \left[ e^{-r \tau} g(Y_\tau) + \int_t^\tau e^{-r s} f(s+ t-t',Y_s) d s \right]
             % \\
             % & 
             = e^{-rt}V(t',x)
        \end{align*}
        using the law of iterated expectations, the convexity of $g$ and $f$ in $x$, and the fact that $t \to f(t,x)$ is nonincreasing in $t$.
        
		Finally, if $(t,x) \in \mathcal{C}$, the optimal stopping time is strictly positive, and, thus, one of the two inequality must be strict (since at the optimal stopping time in either problem puts mass on both sides of the intersection point by Proposition \ref{prop:convextsectionsstopping}). 
  %This concludes the proof.
	\end{proof}
	\end{lemma}

		% The result is an immediate consequence of Proposition \ref{prop:convexproblemmonotonicity} and Theorem \ref{corollary:convexstoppingproblemmonotone}, provided that the stopping boundaries are locally bounded away from $\left\{ \underline{x}, \bar{x} \right\}$, which follows from Lemma \ref{lemma:boundedcontinuationregioninsamplingproblems} below.
  Finally, we conclude the proof by combining Lemma \ref{prop:convexproblemmonotonicity} with Corollary \ref{corollary:globalmonotonestopping}.
	\end{proof}

 \begin{lemma}\label{lemma:propertiesofsigma}
	   % (i) $\sigma \in \mathcal{C}^{1,2,\alpha}$. (ii) For all $x \in (0,1)$, $t\to \sigma(t,x)$ is strictly decreasing if the support of $F$ contains more than 2 elements.
      The learning speed $\sigma$ is in  $\mathcal{C}^{1,2,\alpha}$. Moreover, for all $x \in (0,1)$, $t\to \sigma(t,x)$ is strictly decreasing if the support of $F$ contains more than 2 elements.
	\end{lemma}
\begin{proof}
    This lemma follows immediately from the proof of Proposition 3.8 and Corollary 3.10 in \cite{ekstrom2015bayesian}.
\end{proof}

    \subsection{Omitted proofs for Section \ref{subsec:arithmeticandgeometricBM}}\label{app:aBMandgBM}
  \begin{proposition}\label{prop:aBMandgBMconvexity}
        If $f$ and $g$ are convex in $x$, then the value function associated with \eqref{eq:valuefunctionAG} is convex in $x$.\footnote{Proposition \ref{prop:aBMandgBMconvexity} can easily be extended to any process $X = \left\{X_t\right\}_{t\geq 0}$ linear in the initial argument $X_0$. For example, the same argument guarantees convexity when the underlying state process is a Brownian bridge.}
    \end{proposition}
    \begin{proof}%[\textbf{Proof of Proposition \ref{prop:aBMandgBMconvexity}}]
        We prove the convexity of the value function in \eqref{eq:valuefunction} when the state follows a (nonstationary) geometric Brownian motion \eqref{eq:gBM}. The proof for the arithmetic Brownian motion is similar, hence omitted.

        Observe that the value function is given by
        \begin{align*}
             V(t,x) & = \underset{\tau}{\sup } \, \mathbb{E}_{(t,x)}\left[ \int_{t}^{\tau} e^{-rs} f(s, X_s)ds + e^{-r\tau} g\left(X_{\tau}\right) \right] \\
             & = \underset{\tau}{\sup } \, \mathbb{E}_{(t,x)}\bigg[ \int_{t}^{\tau} e^{-\int_t^s r(u)\, \mathrm{d}u} f\left(s, x e^{\int_t^s \left( \mu(u) -\frac{\sigma^2(u)}{2} \right) \, \mathrm{d}u + \int_t^{s} \sigma(u) dB_u }\right)ds \\
             & \qquad \qquad + e^{-\int_0^{\tau} r(u)\, \mathrm{d}u } g\left(x e^{\int_t^{\tau} \left( \mu(s) -\frac{\sigma^2(s)}{2} \right) ds + \int_t^{\tau} \sigma(t) dB_s }\right) \bigg].
        \end{align*}
        Since $f$ and $g$ are convex in $x$, the integrand is convex in $x$. Moreover, taking expectation preserves convexity. Finally, the supremum of convex functions is convex. Thus $V(t,x)$ is convex in $x$.
    \end{proof}

    \begin{proof}[\textbf{Proof of Corollary \ref{corollary:alignedproblemmonotonicity}}]
        If $f_t(t,x), -r_t(t), \mu_t(t)g_x(x), \sigma_t(t) \leq 0$ for all $(t,x) \in [0,T)\times \mathcal{X}$, with at least one inequality strict, implies $V_t < 0$ in the continuation region and $f_t(t,x), -r_t(t), \mu_t(t,x)g_x(x), \sigma_t(t) \geq 0$ for all $(t,x) \in [0,T)\times \mathcal{X}$, with at least one strictly, implies $V_t >0$ in the continuation region, the result follows from Corollary \ref{corollary:globalmonotonestopping}. So, we prove the first implication when $X$ follows \eqref{eq:aBM}. The proofs of the other cases are similar, hence omitted. 

        Suppose that $f_t(t,x), -r_t(t), \mu_t(t)g_x(x) \leq 0$ for all $(t,x) \in [0,T)\times \mathcal{X}$, with at least one inequality strict. By Lemma \ref{lemma:HJB}, $V$ is the unique $L^p$-solution of \eqref{eq:HJB}:
        \begin{align*}
            \max\left\{ f(t,x) -r(t) V(t,x) + \mu(t) V_x(t,x) + \frac{\sigma^2(t)}{2} V_{xx}(t,x), g(x) - V(t,x) \right\}= 0.
        \end{align*}
        Thus, for all $\epsilon >0$,
        \begin{align*}
            \max\left\{ f(t+\epsilon,x) -r(t+\epsilon) V(t,x) + \mu(t+\epsilon) V_x(t,x) + \frac{\sigma^2(t+\epsilon)}{2} V_{xx}(t,x), g(x) - V(t,x) \right\} \leq 0.
        \end{align*}
        By Lemma \ref{lemma:HJB}, $V(t+\epsilon, x)$ is an $L^p$-solution of the above equation. So, by a comparison principle (Lemma \ref{lemma:comparisonprinciple}), $V(t+\epsilon,x) \leq V(t,x)$ on $[0,T)\times \mathcal{X}$, which implies that the continuation region is weakly decreasing over time. Moreover, on $\mathcal{C}$,
        \begin{align*}
            f(t+\epsilon,x) -r(t+\epsilon) V(t,x) + \mu(t+\epsilon) V_x(t,x) + \frac{\sigma^2(t+\epsilon)}{2} V_{xx}(t,x), g(x) - V(t,x) <0.
        \end{align*}
        One then easily sees from the proof of Lemma \ref{lemma:comparisonprinciple} that $V(t,x)>V(t+\epsilon, x)$ on $\mathcal{C}$, and thus $V_t <0$ on $\mathcal{C}$. This concludes the proof.
    \end{proof}

    \section{Appendix: Stochastic deadlines}\label{app:randomdeadlines}
    
    \paragraph{} Formally, to define stochastic deadlines, we need to enlarge the filtered probability space to  
    %To analyze the problem formally, we first enlarge the probability space to allow for stochastic deadlines. Consider the enlarged filtered probability space 
    $(\Omega\times[0,1], \mathcal{F}\times \mathcal{B}([0,1]), \{\mathcal{F}_t\}_{t \geq 0}, \mathbb{P} \otimes \lambda)$, where $\lambda$ is the Lebesgue measure on $[0,1]$. The extended probability space allows for the randomization device needed for deadlines to be stochastic. 
    %The product measure is well-defined %. This follows from the Hahn-Kolmogorov theorem %(often stated as the first part of Fubini's theorem), see, for example,
    %(Theorem 1.27 in \cite{kallenberg2006foundations}). 
    All the objects defined on $\Omega$ are extended to $\Omega \times [0,1]$ in the obvious way. %In particular, we will work with the enlarged filtration defined as the $\mathbb{P}\otimes \lambda$-completion of $\mathcal{F}_t \otimes \mathcal{B}([0,1])$. This filtration satisfies the usual conditions. Furthermore, we will say that a process is $\mathcal{F}_t$-adapted if it is adapted to the trivial extension of $\mathcal{F}_t$: $\mathcal{F}_t \otimes \{[0,1], \emptyset)\}$.
	\begin{definition}\label{def:randomdeadlines}
		A map $\delta: \Omega \times [0,1] \to [0,\infty]$ is a stochastic deadline if it is a $\mathcal{F}_t \otimes \mathcal{B}([0,1])$-stopping time. We will denote by $\mathcal{D}$ the set of all stochastic deadlines.
	\end{definition}
	The probability that the deadline $d$ arrives before time $t$ is given by the optional stochastic measure $F^{\delta}$ on $\mathbb{R}$ defined by:
	\begin{align*}
		F^{\delta}(t, \omega) = \int_{0}^{1} \mathbbm{1}_{\{ \delta(\omega, u) \leq t \}} \, \mathrm{d}u, \quad \mathbb{P}\text{-a.s..}
	\end{align*}
    In this sense, a stochastic deadline is a distribution over stopping times. Next, observe that, for any stochastic deadline, we can compute the DM's payoff as follows.
	\begin{proposition}\label{prop:payoffsrandomdeadline}
		For all functions $g$ and $\gamma$, all $\delta \in \mathcal{D}$, and all $\tau \in \mathcal{T}$,
		\begin{align*}
			\mathbb{E} \left[  g(X_\tau) \chi_{\left\{ \tau \leq \delta \right\}} + \gamma(X_{\delta}) \chi_{\left\{ \tau > \delta \right\}} \right] = \mathbb{E} \left[  (1- F^{\delta}_\tau) g( X_\tau) + \int_{0}^\tau \gamma(X_t) d  F^{\delta}_t\right].
		\end{align*}
	\end{proposition}

    \begin{proof}
		Let $\delta \in \mathcal{D}$ and $\tau \in \mathcal{T}$. Recall that $F^{\delta}_t = \int_0^1 \mathbbm{1}_{\{\delta(\omega, u) \leq t \}} \, \mathrm{d}u$. Then
		\begin{align*}
		\mathbb{E}& \left[  g( X_\tau) \chi_{\left\{ \tau \leq \delta \right\}} + \gamma( X_{\delta(\omega,u)}) \mathbbm{1}_{\left\{ \tau > \delta \right\}} \right] \\
		& = \mathbb{E} \left[ g(X_\tau) \int_{0}^1 \mathbbm{1}_{\left\{ \delta(\omega, u)\geq \tau \right\}} \, \mathrm{d}u +\int_0^1 \gamma( X_{\delta(\omega,u)}) \mathbbm{1}_{\left\{ \tau > \delta(\omega,u) \right\}} \, \mathrm{d}u \right] \\
		& = \mathbb{E}\left[ g(X_\tau) \int_{0}^1 \mathbbm{1}_{\left\{ F^{\delta(\omega,u)}_\tau \leq u \right\}} \, \mathrm{d}u +\int_0^1 \gamma( X_{\delta(\omega,u)}) \mathbbm{1}_{\left\{ \tau > \delta(\omega, u) \right\}} \, \mathrm{d}u \right] \\
		& = \mathbb{E} \left[  (1- F^{\delta}_\tau) g(X_\tau) + \int_{0}^\tau \gamma(X_{\delta}) d  F^{\delta}_t\right],
		\end{align*}
		where we used Fubini's theorem and the independence of $\tau$ and $\mathcal{B}([0,1])$ for the first equality and proposition 4.9 of chapter 0 in \cite{revuz2013continuous} for the last one.
	\end{proof}
    For Poisson deadlines with intensity $\alpha$, $1-F^{\delta}_t = e^{-\int_0^t \alpha(s,X_s)ds}$. So, we can reformulate \eqref{eq:valuefunctiondeadline} to fit our framework:
	\begin{align*}
		V^{\delta}(t,x) = \underset{\tau\in \mathcal{T}(t)}{\sup }\,  \mathbb{E}_{(t,x)} \left[  e^{-\int_{t}^{\tau} (\alpha(s,X_s) +r)ds} g\left(X_{\tau}\right)  + \int_{t}^{\tau} e^{-\int_{t}^{t+s} (\alpha(u, X_u) +r)\, \mathrm{d}u} \gamma(X_s)  \alpha(s,X_s) ds\right],
	\end{align*}
	subject to \setlength{\abovedisplayskip}{0pt}
\setlength{\belowdisplayskip}{4pt} 
	\begin{align*}
		X_{t+s} & = X_t + \int_{t}^{t+s} \sigma( X_u) \, \mathrm{d}B_u.
	\end{align*} 

    Finally, we prove the following Lemma.
  \begin{lemma}\label{lemma:monotonicitystoppingwithdeadline}
        The stopping problem \eqref{eq:valuefunctiondeadline} is strictly monotone decreasing
        \begin{enumerate}[topsep=3.5pt,itemsep=3pt]
            \item if $\alpha(t,x)$ is strictly decreasing in $t$ and $\gamma(x) \geq V^{\delta}(t,x)$ for all $t\in T$; or %a \vee b$; or %\geq \max_{x\in [0,1]} g(x)$ ; or
            \item if $\alpha(t,x)$ is strictly increasing in $t$ and $\gamma(x) \leq V^{\delta}(t,x)$ for all $t\in T$ %\frac{b}{a+b}$. %g(x)$  for all $x\in [0,1]$.
            % \item if $a(t,x)$ is strictly increasing in $t$ and $\gamma(x) \leq g(x)$  for all $x\in [0,1]$.
        \end{enumerate}
        The stopping problem \eqref{eq:valuefunctiondeadline} is strictly monotone increasing
        \begin{enumerate}[topsep=3.5pt,itemsep=3pt]
            \item[3.] if $\alpha(t,x)$ is strictly increasing in $t$ and $\gamma(x) \geq V^{\delta}(t,x)$ for all $t\in T$; or
            % g(x)$  for all $x\in [0,1]$; or
            \item[4.] if $\alpha(t,x)$ is strictly decreasing in $t$ and $\gamma(x) \leq V^{\delta}(t,x)$ for all $t\in T$ %\frac{b}{a+b}$.% g(x)$  for all $x\in [0,1]$.
        \end{enumerate}
    \begin{proof}
        Under conditions 1. and 2. of the Lemma, it is easily seen that the stopping problem is strictly monotone decreasing if $V_t <0$ in the continuation region. Similarly, under conditions 3. and 4. of the Lemma, it is also easily seen that the stopping problem is strictly monotone increasing if $V_t >0$ in the continuation region. So, there remain only to show that $V_t <0$ under condition 1. or 2., and that $V_t >0$ under condition 3. or 4.. 
        
        Suppose that condition 1. holds and recall that the DM's problem is given by
        \begin{align*}
    		V^{\delta}(t,x) =& \underset{\tau\in \mathcal{T}(t)}{\sup }\,  \mathbb{E}_{(t,x)} \left[  e^{-r\tau} g\left(X_{\tau}\right) \mathbbm{1}_{ \{\tau \leq \delta \}} +  e^{-r\delta} \gamma(X_{\delta}) \mathbbm{1}_{ \{\tau > \delta \}} \right],\\
            & \text{subject to } X_{t+s} = X_t + \int_{t}^{t+s} \sigma(u, X_u) \, \mathrm{d}B_u.
    	\end{align*}
        Observe that $\delta$ is the first tick of a Poisson clock with random arrival rate $\alpha(t,X_t)$. In particular, let $\bar{t}>\underline{t}$. $\bar{\delta}(\omega) = \delta(\omega) - \bar{t}$ denote the arrival time of the deadline when the game starts at time $\bar{t}$, and $\underline{\delta}(\omega) =  \delta(\omega) - \underline{t}$ denote the arrival of the deadline when the game starts at time $\underline{t}$. Then, for the random time change $A_t = \int_0^t \frac{\alpha( \underline{t} +t, X_{\underline{t}+t})}{\alpha(\bar{t}+t, X_{\underline{t} +t})}\mathrm{d}t$, one sees that $(\bar{\delta}_{A_t}, \bar{X}_t)$ and $(\underline{\delta}_{t}, \underline{X}_t)$, where $\bar{X}$ and $\underline{X}$ are the state process starting at times $\bar{t}$ and $\underline{t}$ respectively,  are identically distributed. Since $A_t > t$ for all $t$ and all paths $X$, there exists a probability space such that $\underline{\delta} < \bar{\delta}$ and $\bar{X} = \underline{X}$ almost surely. But, then, on that probability space, by Snell envelope theorem,
        \begin{align*}
            V^{\delta}(\bar{t},x) & = \underset{\tau\in \mathcal{T}(\bar{t})}{\sup }\, \mathbb{E}_{(\bar{t},x)} \left[ e^{-r(\tau-\bar{t})}g(X_{\tau})\mathbbm{1}_{\{ \tau < \underline{\delta} +\bar{t} \}} + e^{-r\underline{\delta}}V(\underline{\delta} +\bar{t}, X_{\underline{\delta}+\bar{t}})\mathbbm{1}_{\{ \tau \geq \underline{\delta} +\bar{t} \}} \right].
        \end{align*}
        But, by condition 1., for all $(t,x) \in \mathcal{Y}$, $\gamma(x) \geq V(t,x)$. It follows that
        \begin{align}\label{eq:proofstrictmonotonicitydeadline}
            V^{\delta}(\bar{t},x) & = \underset{\tau\in \mathcal{T}(\bar{t})}{\sup }\, \mathbb{E}_{(\bar{t},x)} \left[ e^{-r(\tau-\bar{t})}g(X_{\tau})\mathbbm{1}_{\{ \tau < \underline{\delta} +\bar{t} \}} + e^{-r\underline{\delta}}V(\underline{\delta} +\bar{t}, X_{\underline{\delta} +\bar{t} })\mathbbm{1}_{\{ \tau \geq \underline{\delta} +\bar{t} \}} \right] \notag \\
            & \leq \underset{\tau\in \mathcal{T}(\bar{t})}{\sup }\, \mathbb{E}_{(\bar{t},x)} \left[ e^{-r(\tau-\bar{t})}g(X_{\tau})\mathbbm{1}_{\{ \tau < \underline{\delta} +\bar{t} \}} + e^{-r\underline{\delta}}\gamma(X_{\underline{\delta} +\bar{t}})\mathbbm{1}_{\{ \tau \geq \underline{\delta} +\bar{t} \}} \right] \\
            & = \underset{\tau\in \mathcal{T}(\underline{t})}{\sup }\, \mathbb{E}_{(\underline{t},x)} \left[ e^{-r(\tau-\underline{t})}g(X_{\tau})\mathbbm{1}_{\{ \tau < \underline{\delta} +\underline{t} \}} + e^{-r\underline{\delta}}\gamma(X_{\underline{\delta} + \underline{t}})\mathbbm{1}_{\{ \tau \geq \underline{\delta} +\underline{t} \}} \right] \notag \\
            & = V^{\delta}(\underline{t}, x). \notag
        \end{align}
        Moreover, for all almost surely strictly positive $\tau$, $\mathbb{P}\left( \underline{\delta} < \tau \right)>0$, and, by condition 1., for all $(t,x)$ in the continuation region, $\gamma(x) > V(t,x)$. So, if $(\bar{t}, x) \in \mathcal{C}$, the inequality \eqref{eq:proofstrictmonotonicitydeadline} is strict. This concludes the proof for case 1..

        The proofs for cases 2., 3., and 4., are similar, hence omitted. 
    \end{proof}
        \end{lemma}
    \noindent Together with Corollary \ref{corollary:globalmonotonestopping}, Lemma \ref{lemma:monotonicitystoppingwithdeadline} delivers \textbf{Proposition \ref{prop:randomdeadlinesboundaries1}.}

    \section{Appendix: Control and Stopping}\label{app:proofoftheoremcontrolledstopping}

        \textbf{Preliminaries:} Following the proof of Lemma \ref{lemma:HJB}, we see that the value function \eqref{eq:valuefunctioncontrol} is the unique $L^p$-solution of 
    \begin{align}\label{eq:HJBc}
        \begin{cases}
            \max \!\bigg\{ \underset{a \in \mathbb{A}}{\max} \,  v_t(t,x) + \mathcal{L}^{a,(t,x)}v(t,x) \!-\!r(a,t,x)v(t,x) \!+\! f(a, t,x), \\
            \qquad \qquad \qquad \qquad \qquad \qquad \qquad \qquad g(x) \!-\!v(t,x) \bigg\} = 0 \text{\small{ a.e. in }} \medmath{[0,T)\times \mathcal{X}}, \\
            v(T,x) = g(x) \text{ on } \mathcal{X},
        \end{cases}
    \end{align}
    where, $\mathcal{L}^{a,(t,x)}f(t,x) = \mu(a, t, x) f_{x}(t,x) +  \frac{\sigma(a, t, x)^2}{2} f_{xx}(t,x).$

    \begin{proof}[\textbf{Proof of Theorem \ref{theorem:localmonotonestoppingcontroldiffusion}}]
		(i) The weak monotonicity of the continuation region (points 1. and 3.) follows immediately from the local monotonicity of the value function. 
  
        (ii) There remains to prove strict monotonicity when the optimal stopping and control problem is locally strictly monotone decreasing on $\left( \underline{t}, \bar{t} \right)$. The argument for a locally strictly monotone increasing optimal stopping problem is identical, hence omitted. %From point 1, since the optimal stopping problem is locally monotone decreasing, so is the continuation region. There remains to show that the boundaries are strictly monotone when they are in $\mathcal{X}$. 
        From 1., the continuation region is monotone decreasing. So, there only remains to show that the boundaries are strictly monotone (when they are in $\mathcal{X}$).
		
		The proof is by contradiction. Suppose that there exists $\underline{t}, \bar{t} \in [0,T)$ and $B \in \mathcal{X}$ such that $(\underline{t}, B)$ and $(\bar{t}, B) \in \partial \mathcal{C}$. For the rest of the proof, we assume that we are looking at an ``upper boundary'' of the continuation region, i.e., there exists $\epsilon>0$ such that for all $(\underline{t},\bar{t}) \times (B-\epsilon, B) \subset \mathcal{C}$. The proof for the ``lower boundary'' is identical.
		
		Consider the rectangular domain $\mathcal{Y} = [\underline{t}, \bar{t}) \times (B-\epsilon,B)$ with parabolic boundary $\partial \mathcal{Y} = \left( [\underline{t},\bar{t}] \times \left( \left\{ B-\epsilon \right\} \cup \left\{ B\right\} \right)   \right) \cup\left( \left\{\bar{t}\right\} \times (B-\epsilon,B) \right) \subset \mathcal{C}$. By Theorem 1 in \cite{durandard2022smoothness}, we know that the value function $V$ is the unique $L^p$-solution of the boundary value problem
		\begin{align*}
    		\begin{cases}
        		\underset{a \in A}{\max }\, \big\{ v_t(t,x) + \frac{\sigma(a,t,x)^2}{2} v_{xx}(t,x) + \mu(a,t,x) v_x(t,x) \\ 
                \qquad \qquad \qquad  -r(a,t,x)v(t,x) + f(a,t,x)\big\} = 0 \text{ if }  (t,x) \in  \mathcal{Y}, \\
        		v(t,x) = V(t,x) \text{ if } (t,x) \in \partial \mathcal{Y}.
    		\end{cases}
		\end{align*}
		By Theorem 2 and Corollary 2 in \cite{durandard2022smoothness}, $V_t \in \mathcal{C}^0\left( \bar{\mathcal{Y}} \right)$. Theorem 1 in \cite{durandard2023existence} and the Envelope Theorem then guarantee that $V_t(t,x) \in {W}^{1,2,p}_{loc}\left( \mathcal{Y}\right)$ and solves
		\begin{align*}
		\begin{cases}
            v_{tt}(t,x) + \frac{\sigma(a(t,x), t,x)^2}{2} v_{txx}(t,x) + \mu(a(t,x), t,x) v_{tx}(t,x) - r(a(t,x), t,x)v_t(t,x) \\
		    \quad =-\sigma(a(t,x), t,x) {\sigma_t( a(t,x), t,x)} v_{xx}(t,x) - \mu_t(a(t,x), t,x) v_{x}(t,x) \\
            \qquad \qquad \qquad \qquad + r_t(a(t,x), t,x)v(t,x) - f_t(a(t,x), t,x) \text{ if }  (t,x) \in  \mathcal{Y}, \\
		  v_t(t,x) = V_t(t,x) \text{ if } (t,x) \in \partial \mathcal{Y},
		\end{cases}
		\end{align*}
        where $a(t,x)$ is a Lispchitz selection that maximizes $-V_{t}(t,x)$ given by Assumption \ref{assumption:Lipschizselection}. Therefore, on $\mathcal{Y}$,
		\begin{align*}
            v_{tt}(t,x) + & \frac{\sigma(a(t,x), t,x)^2}{2} v_{txx}(t,x) + \mu(a(t,x), t,x) v_{tx}(t,x) - r(a(t,x), t,x)v_t(t,x)\\ 
            & = - \sigma(a(t,x), t,x) {\sigma_t( a(t,x), t,x)} v_{xx}(t,x) - \mu_t(a(t,x), t,x) v_{x}(t,x) \\
            & \qquad \qquad \qquad \qquad + r_t(a(t,x), t,x)v(t,x) - f_t(a(t,x), t,x) \\
    		& \geq 0.
		\end{align*}
        The inequality follows from condition \eqref{eq:controlledDOV} in Definition \ref{definition:monotoneenvironmentscontrol}. Then, by the Hopf's boundary Lemma for strong solutions (Lemma 1.39 in \cite{wang2021nonlinear}), $v_{tx}(t,B^{-}) >0$ for all $t\in (\underline{t}, \bar{t})$, which implies that $V_x(t, B^-)$ is strictly increasing for $t\in (\underline{t}, \bar{t})$. On the other hand, for all $t\in (\underline{t}, \bar{t})$, $V(t,B) = g(x)$ and $V \in W^{1,2,p}_{loc}\left( [0,T)\times\mathcal{X}\right)$ implies that $V_x(t,B) = g_x(x)$, which is constant on $(\underline{t}, \bar{t})$: a contradiction. 
  
		%The inequality follows from condition (iii) in Definition \ref{definition:monotoneenvironments}. Then, by Hopf's boundary Lemma (Lemma 1.23 in \cite{wang2021nonlinear}), $v_{tx}(t,B) >0$ for all $t\in (\underline{t}, \bar{t})$ such that $v_{tx}(t,B)$ exists. But, for all $t\in (\underline{t}, \bar{t})$, $V(t,B) = g(x) \Rightarrow v_t(t,B) = V_t(t,B) = \frac{\partial}{\partial t} g(x)= 0$ and therefore $v_{tx}(t,B) = 0$. So if there exists $\left( t,B \right) \in \left( \underline{t}, \bar{t} \right) \times \left\{ B \right\}$ such that $v_{tx}(t,B)$ is well defined, we have a contradiction. This follows from Theorem 2.1 in \cite{wang1992regularity}, since $V \in \mathcal{C}^{1,2, \alpha}\left( [\underline{t}, \bar{t}) \times \left(B-2\epsilon, B\right)\right)$ by Lemma \ref{lemma:HJB}.
		
		So, the upper boundary of the continuation is not flat, which concludes the proof. 
    \end{proof}

    \pagebreak

    \section{Online Appendix}\label{app:online}

    \subsection{When are the boundaries continuous?}\label{subsec:continuousboundaries}

    In the mathematical literature, the stopping boundaries' continuity has received enormous interest both from the probabilistic stopping literature \citep{villeneuve1999exercise, chen2007mathematical, de2015note, de2019lipschitz, de2022stopping} and from the literature on partial differential equations \citep{caffarelli1977regularity,petrosyan2012regularity}. We show that, in monotone decreasing environments, a small strengthening of Condition \ref{condition:singlecrossing} guarantees the boundaries' continuity.
	% \begin{assumption}[SSC]
 % \label{assumption:strongsinglecrossing}
	% 	For all $t\in [0,T)$, the mapping $x \to f(t,x) + \left(\mathcal{L}^{(t,x)} -r\right) g(x)$ is nondecreasing on $\left(\underline{x}, x^c\right]$ and there exists a unique $x^{-}(t) \in \left[\underline{x}, x^c\right]$ such that, for all $x \in \left(\underline{x}, x^c\right]$,$x < x^-(t) \Rightarrow f(t,x) + \left(\mathcal{L}^{(t,x)} -r(t,x)\right) g(x) <0$ and $x > x^-(t) \Rightarrow f(t,x) + \left(\mathcal{L}^{(t,x)} -r\right) g(x) >0.$
        
	% 	Similarly, for all $t\in [0,T)$, the mapping $x \to f(t,x) + \left(\mathcal{L}^{(t,x)} -r(t,x)\right) g(x)$  nonincreasing on $\left[x^c, \bar{x}\right)$ and there exists a unique $x^{+}(t) \in \left[x^c, \bar{x}\right]$ such that, for all $x \in \left[x^c, \bar{x}\right)$, $x > x^+(t) \Rightarrow f(t,x) + \left(\mathcal{L}^{(t,x)} -r(t,x)\right) g(x) <0$ and $x > x^+(t) \Rightarrow f(t,x) + \left(\mathcal{L}^{(t,x)} -r(t,x)\right) g(x) <0.$
	% \end{assumption}
	\begin{condition}[SSC]
    \label{condition:strongsinglecrossing}
		For all $t\in [0,T)$, $f(t,x) + \left(\mathcal{L}^{(t,x)} -r(t,x) \right) g(x)$ is nondecreasing on $\left(\underline{x}, x^c\right]$ and nonincreasing on $\left[x^c, \bar{x}\right)$. Moreover,  for all $t\in [0,T)$, there exists a unique $x^{-}(t) \in \left[\underline{x}, x^c\right]$ such that, $f(t,x) + \left(\mathcal{L}^{(t,x)} -r(t,x)\right) g(x) <0$ if $x \in \left(\underline{x}, x^-(t)\right)$ and $f(t,x) + \left(\mathcal{L}^{(t,x)} -r\right) g(x) >0$ if $x\in (x^-(t), x^c] $. Similarly, for all $t\in [0,T)$ there exists a unique $x^{+}(t) \in \left[x^c, \bar{x}\right]$ such that,  $f(t,x) + \left(\mathcal{L}^{(t,x)} -r(t,x)\right) g(x) >0$ if $x \in \left[x^c, x^+(t)\right)$ and $f(t,x) + \left(\mathcal{L}^{(t,x)} -r(t,x)\right) g(x) <0$ if $x \in \left(x^+(t), \bar{x}\right).$
	\end{condition}

    \begin{proposition}\label{prop:continuousboundaries}
		Let $\underline{t} = \inf\left\{ t\geq 0 \, : \, \underline{b}(t) > -\infty  \right\}$ and $\bar{t} = \inf\left\{ t\geq 0 \, : \, \underline{b}(t) > -\infty  \right\}$. Suppose that the optimal stopping problem is monotone decreasing and that Condition C \ref{condition:strongsinglecrossing} holds. Then the stopping boundaries $t\to \underline{b}(t)$ and $t\to \bar{b}(t)$ are continuous on $[\underline{t},T)$ and on $[\bar{t}, T)$, respectively.
	\end{proposition}
	The proof of Proposition \ref{prop:continuousboundaries} is technical. It follows the ideas in \cite{de2015note}. Moreover, the argument behind Proposition \ref{prop:continuousboundaries} only works in monotone decreasing environment, as the proof relies on the fact that $X$ moves towards the boundary when starting close to it. In monotone increasing environments, the process $X$ moves away from the boundary as time passes. Intuitively, the diffusion does not ``see'' the discontinuities of the boundaries in the increasing environment.
    
    We start with an easy lemma.
    \begin{lemma}\label{lemma:nostoppingaroundxc}
		Suppose that Condition C\ref{condition:strongsinglecrossing}(SSC) holds, then 
		\begin{align*}
		\left\{ (t,x) \in [0,T)\times\mathcal{X} \, : \, x \in \left( x^-(t), x^+(t) \right) \right\} \subset \mathcal{C}.
		\end{align*}
	\end{lemma}

    \begin{proof}
		By Lemma \ref{lemma:HJB}, $V$ is the $L^p$-solution of \hyperlink{link:HJB}{(HJB)}. Therefore, $V$ is a subsolution of 
        \begin{align*}
            \left( \mathcal{L}^{(t,x)} -r(t,x) \right) V(t,x) + f(t,x) =0 \text{ in } [0,T)\times \mathcal{X}.
        \end{align*}
        Under Condition C\ref{condition:strongsinglecrossing}, it can only be the case if $V >g$ on
        \begin{align*}
            \left\{ (t,x) \in [0,T)\times \mathcal{X} \, : \, x \in \left( x^-(t), x^+(t) \right) \right\}.
        \end{align*}
        The conclusion follows.
	\end{proof}
        
    Lemma \ref{lemma:nostoppingaroundxc} implies that
	\begin{align}\label{eq:stoppinginnegativecontinuation}
	   \mathcal{S} \subset \left\{ (t,x) \in [0,T)\times \mathcal{X} \, : \, f(t,x) + \left(\mathcal{L}^{(t,x)} -r(t,x)\right) g(x) < 0 \right\}.
	\end{align}
    With that in mind, we are now ready to prove Proposition \ref{prop:continuousboundaries}.

    \begin{proof}[\textbf{Proof of Proposition \ref{prop:continuousboundaries}}]
    Since the optimal stopping problem is monotone decreasing, by Theorem \ref{theorem:localmonotonestopping}, the stopping region is given by 
		\begin{align*}
			\mathcal{S} = \left\{ \left(t, x\right) \in \mathbb{R}_+ \times \mathcal{X} \, : \,  x \not \in \left( \underline{b}(t), \bar{b}(t) \right) \right\},
		\end{align*}
		where $\underline{b}: \mathbb{R}_+ \to \mathcal{X}$ is c\`{a}dl\`{a}g {nondecreasing} and $\bar{b}: : \mathbb{R}_+ \to \mathcal{X}$ is c\`{a}dl\`{a}g {nonincreasing}, with $\underline{b}(T) \leq x^c\leq \bar{b}(T)$. So we only need to prove that the boundaries are left-continuous too.
		
		The proof is by contradiction. Suppose that there exists $\bar{t} \in [0,T)$ such that $\bar{b}$ or $\underline{b}$ is discontinuous at $t$. For the rest of the proof, we will assume that $\bar{b}$ is discontinuous at $\bar{t}$. The proof for $\underline{b}$ is identical. Since $\bar{b}$ is c\`{a}dl\`{a}g  nonincreasing, $\bar{b}(\bar{t}^-) > \bar{b}(\bar{t})$. Let $a,b \in \mathcal{X}$ be such that $(a,b) \subset (\bar{b}(\bar{t}), \bar{b}(\bar{t}^-))$ and consider the rectangular domain $\mathcal{Y} = [\underline{t}, \bar{t}) \times (a,b)$ with parabolic boundary $\partial \mathcal{Y} = \left( [\underline{t},\bar{t}] \times \left( \left\{ a \right\} \cup \left\{ b\right\} \right)   \right) \cup\left( \left\{T\right\} \times (a,b) \right)$. Using Lemma \ref{lemma:nostoppingaroundxc} and \eqref{eq:stoppinginnegativecontinuation}, we can choose $\underline{t}$ and $(a,b)$ such that
		\begin{align}\label{eq:proofcontinuousboundarieslowerbound}
			f(s,x) + \left(\mathcal{L}^{(t,x)} -r(t,x)  \right) g(x) < -\delta \text{ on } \mathcal{Y}
		\end{align}
		for some $\delta >0$ small.
		
		Moreover, by Lemma \ref{lemma:HJB}, $V$ is the unique $L^p$-solution of the boundary value problem
		\begin{align*}
		\begin{cases}
			v_t(t,x)+ \mathcal{L}^{(t,x)}v(t,x) - r(t,x)v(t,x)  + f(t,x) = 0 \text{ if }  (t,x) \in  \mathcal{Y}, \\
			v(t,x) = V(t,x) \text{ if } (t,x) \in \partial \mathcal{Y}.
		\end{cases}
		\end{align*}
		Let $\phi \in \mathcal{C}_c^{\infty}\left( [a,b] \right)$ such that $\phi(x) \geq 0$ and $\int_{a}^{b} \phi(x) dx =1$. Then, multiplying the above equality by $\phi$ and integrating, we have
		\begin{align*}
			\int_{t}^{\bar{t}} \int_{a}^{b} &\left( V_t(s,x) + f(s,x) + \left(\mathcal{L}^{(s,x)} -r(s,x)  \right) V(s,x)\right) \phi(x) dx ds =0.
		\end{align*}
		for all $t \in [\underline{t}, \bar{t})$.
		By Fubini's theorem,
		\begin{align*}
			\int_{t}^{\bar{t}} \int_{a}^{b}  V_t(t,x) \phi(x) dx ds & = \int_{a}^{b} \int_{t}^{\bar{t}} V_t(t,x) ds\phi(x) dx \\
			& = \int_{a}^{b} \left( V(\bar{t},x) - V(t,x) \right) \phi(x) dx \\
			& < \int_{a}^{b} \left( g(x) - g(x) \right) \phi(x) dx =0.
		\end{align*}
		where the inequality follows from the fact that $V(t,x) > g(x)$ for all $(t,x) \in \mathcal{Y}$, as it is a subset of the continuation region. Therefore, for all $t \in [\underline{t}, \bar{t})$,
		\begin{align}\label{eq:proofcontinuousboundariescontradictions}
			\int_{t}^{\bar{t}} \int_{a}^{b} &\left( f(s,x) + \left(\mathcal{L}^{(s,x)}  -r(s,x)  \right) V(s,x)\right) \phi(x) dx ds > 0. 
		\end{align}
		Moreover, integrating by parts, for all $t \in [\underline{t}, \bar{t})$,
		\begin{align*}
			\int_{t}^{\bar{t}} \int_{a}^{b} &\left( f(s,x) + \left(\mathcal{L}^{(s,x)} -r(s,x)  \right) V(s,x)\right) \phi(x) dx ds \\
			& = \int_{t}^{\bar{t}} \int_{a}^{b} \left( \left( f(s,x) -r(s,x) V(s,x)\right) \phi(x) + V(s,x) \mathcal{L}^{(s,x)}_*\phi(x)  \right) dx ds;
		\end{align*}
		where $\mathcal{L}_*$ is the adjoint of $\mathcal{L}$: for all $\psi \in \mathcal{C}^{1,2}\left( [0,T]\times \mathcal{X} \right)$,
		\begin{align*}
		\mathcal{L}^{(t,x)}_* \psi(t,x) = \frac{\partial^2}{\partial x^2} \left( \frac{\sigma^2(t,x)}{2} \psi(t,x) \right) - \frac{\partial}{\partial x} \left( \mu(t,x) \psi(t,x) \right).
		\end{align*} 
		For $s\in [\underline{t},\bar{t})$, 
		\begin{align*}
			\int_{a}^{b} & V(s,x) \mathcal{L}^{(s,x)}_*\phi(x) dx \\
			& = \int_{a}^{b} \mathbbm{1}_{\{ \mathcal{L}^{(s,x)}_*\phi(x) <0 \}} V(s,x) \mathcal{L}^{(s,x)}_*\phi(x) dx  + \int_{a}^{b}  \mathbbm{1}_{\{ \mathcal{L}^{(s,x)}_*\phi(x) \geq 0 \}}  V(s,x) \mathcal{L}^{(s,x)}_*\phi(x) dx \\
			& \leq C\left| s-\bar{t} \right| \int_{a}^{b} \mathbbm{1}_{\{ \mathcal{L}^{(s,x)}_*\phi(x) \geq 0 \}} \mathcal{L}^{(s,x)}_*\phi(x) dx  + \int_{a}^{b} g(x) \mathcal{L}^{(s,x)}_*\phi(x) dx \\
			&  = C\left| s-\bar{t} \right| \int_{a}^{b} \mathbbm{1}_{\{ \mathcal{L}^{(s,x)}_*\phi(x) \geq 0 \}} \mathcal{L}^{(s,x)}_*\phi(x) dx  + \int_{a}^{b} \phi(x) \mathcal{L}^{(s,x)} g(x) dx.
		\end{align*}
		The inequality follows from the two inequalities $V(t,x) \geq g(x)$ and $V(s,x) \leq  g(x) + C\left| s-\bar{t} \right|$ for some $C>0$ (since $t\to V(t,x)$ is Lipschitz, as $V$ belongs to $W^{1,2,p}\left( \mathcal{Y} \right)$). The last equality is obtained by integration by parts. Let
		\begin{align*}
			\bar{\phi}(t) = \int_{t}^{\bar{t}}\int_{a}^{b} \mathbbm{1}_{\{ \mathcal{L}^{(s,x)}_*\phi(x) \geq 0 \}} \mathcal{L}^{(s,x)}_*\phi(x) dxds \geq 0, \quad t \in [\underline{t}, \bar{t}).
		\end{align*}
		Then, for all $t \in [\underline{t}, \bar{t})$,
		\begin{align*}
			\int_{t}^{\bar{t}} \int_{a}^{b} &\left( f(s,x) + \left(\mathcal{L}^{(s,x)} -r(s,x)  \right) V(s,x)\right) \phi(x) dx ds \\
			& \leq \int_{t}^{\bar{t}} \int_{a}^{b}\left( f(s,x) + \left(\mathcal{L}^{(s,x)} -r(s,x)  \right) g(x)\right) \phi(x) dx ds +\bar{\phi} C\left| t-\bar{t} \right|^2 \\
			& \leq - \delta (\bar{t} -t) +\bar{\phi} C\left| t-\bar{t} \right|^2,
		\end{align*}
		where we used \eqref{eq:proofcontinuousboundarieslowerbound} and that $\phi$ integrates to 1 to obtain the second inequality. But then there exists $t \in [\underline{t}, \bar{t})$ such that
		\begin{align*}
			\int_{t}^{\bar{t}} \int_{a}^{b} &\left( f(s,x) + \left(\mathcal{L}^{(s,x)} -r(s,x)  \right) V(s,x)\right) \phi(x) dx ds \leq 0,
		\end{align*}
		which contradicts \eqref{eq:proofcontinuousboundariescontradictions}.
    \end{proof}

\end{document}